\documentclass[a4paper,11pt]{iopart}
\pdfoutput=1 

\pdfoutput=1

\usepackage{graphicx,sidecap}
\usepackage{bm}
\usepackage{amssymb}
\usepackage{cite}
\usepackage{mathrsfs}
\usepackage{amsmath}
\usepackage{bm}
\usepackage{xcolor}
\usepackage{hyperref}
\usepackage{enumerate}
\usepackage[toc,title]{appendix}
\usepackage{verbatim}
\usepackage[margin=30pt, bf, font=footnotesize, center, justification=justified]{caption}[2004/07/16]
\usepackage[T1]{fontenc} 
\usepackage{braket}
\usepackage{mathdots}
\usepackage{stackrel}
\usepackage{bbold}
\usepackage{mathtools}
\usepackage{subfigure}
\usepackage[T1]{fontenc}               
\usepackage[left=3cm,
            right=2.5cm,
            top=2.5cm,
            bottom=3cm
            ]{geometry}                
\usepackage{mathrsfs}
\usepackage{appendix}
\usepackage{fancyhdr}
\usepackage{amsmath,                   
            amssymb,                   
            amsthm}                    

\hypersetup{colorlinks,bookmarksopen,bookmarksnumbered,
citecolor=red,
linkcolor=blue,
pdfstartview=green,
urlcolor=teal}

\catcode`@=11 
\renewcommand\tableofcontents{%
  \section*{\contentsname}%
  \@starttoc{toc}%
}
\catcode`@=12

\begin{document}

\title[Symmetry resolved entanglement in 2d systems via dimensional reduction
]
{Symmetry resolved entanglement in two-dimensional systems via dimensional reduction
}

\vspace{.5cm}

\author{Sara Murciano$^1$, Paola Ruggiero$^2$ and Pasquale Calabrese$^{1,3}$}
\address{$^1$SISSA and INFN Sezione di Trieste, via Bonomea 265, 34136 Trieste, Italy.}
\address{$^2$DQMP, University of Geneva, 24 Quai Ernest-Ansermet, CH-1211 Geneva, Switzerland}
\address{$^3$International Centre for Theoretical Physics (ICTP), Strada Costiera 11, 34151 Trieste, Italy}

\vspace{.5cm}

\begin{abstract}
We report on the calculation of the symmetry resolved entanglement entropies in two-dimensional many-body systems of free bosons and fermions 
by \emph{dimensional reduction}. 
When the subsystem is translational invariant in a transverse direction,  this strategy allows us 
to reduce the initial two-dimensional problem into decoupled one-dimensional ones in a mixed space-momentum representation. 
While the idea straightforwardly applies to any dimension $d$, here we focus on the case $d=2$ and derive explicit expressions for two lattice models possessing a $U(1)$ symmetry, i.e., free non-relativistic massless fermions and free complex (massive and massless) bosons. 
Although our focus is on symmetry resolved entropies, some results for the total entanglement are also new. 
Our derivation gives a transparent understanding of the well known different behaviours between massless bosons and fermions in $d\geq2$:
massless fermions presents logarithmic violation of the area which instead strictly hold for bosons, even massless. 
This is true both for the total and the symmetry resolved entropies. 
Interestingly, we find that the equipartition of entanglement into different symmetry sectors holds also in two dimensions at leading order in subsystem size; 
we identify for both systems the first term breaking it.
All our findings are quantitatively tested against exact numerical calculations in lattice models for both bosons and fermions. 
\end{abstract}

\maketitle

\newpage

\tableofcontents

\section{Introduction}

One of the most fascinating aspects of the entanglement entropy in the ground states of extended quantum systems is that it scales with the area of a subsystem rather than its volume, as it happens, instead, for generic eigenstates in the middle of the spectrum. This feature is known as {\it area law} \cite{eisert-2010}.
The area law is a well established concept for gapped (massive) systems \cite{eisert-2010,bombelli, sredniki, intro1, intro2, intro3}. 
On the other hand, if the correlations  are long-ranged, i.e., the system is massless,  area law may be violated like in the prototypical example of
one-dimensional (1d) conformal invariant systems \cite{cc-04, cc-09, hlw-94, vidal} for which there is a multiplicative logarithmic correction to it.
In higher dimensions, massless systems behave rather differently depending on the fine details of the model. 
It is impossible to mention all aspects of the problem, but the most striking aspect is that 
while free massless non-relativistic fermions show logarithmic violations of the area law \cite{widom1,widom2, widom3, widom4, widom5,fz-07,  mint2,ryu,s-12,cmv-12}, 
in free massless bosons it strictly holds \cite{suggest,widom1,areaLawplenio1, areaLawplenio2, areaLawplenio3, CH2d}.
While in general it is not known how the entanglement scales in interacting massless bosons and fermions, there are indications that 
the structure found for free systems should be robust also against interactions, see e.g. Refs. \cite{fendley,twist,konik,kdmsa-13,sjfm-13,im-13,p-16,mt-13,dp-15,mfs-09,wws-16,s-10,dsy-12}. 

\begin{figure}[t]
\centering
\subfigure
{\includegraphics[width=0.45\textwidth]{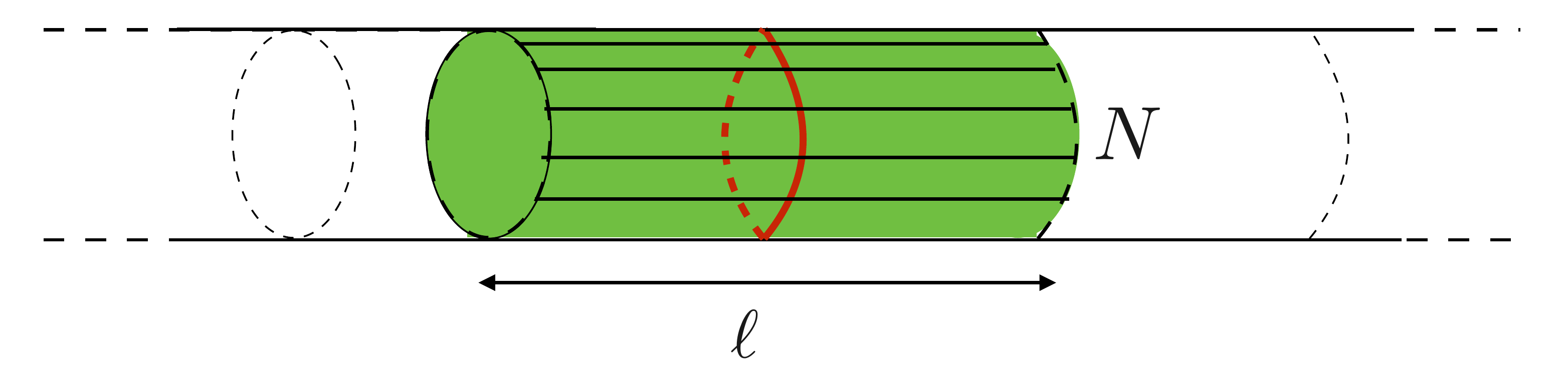}}
\subfigure
{\includegraphics[width=0.25\textwidth]{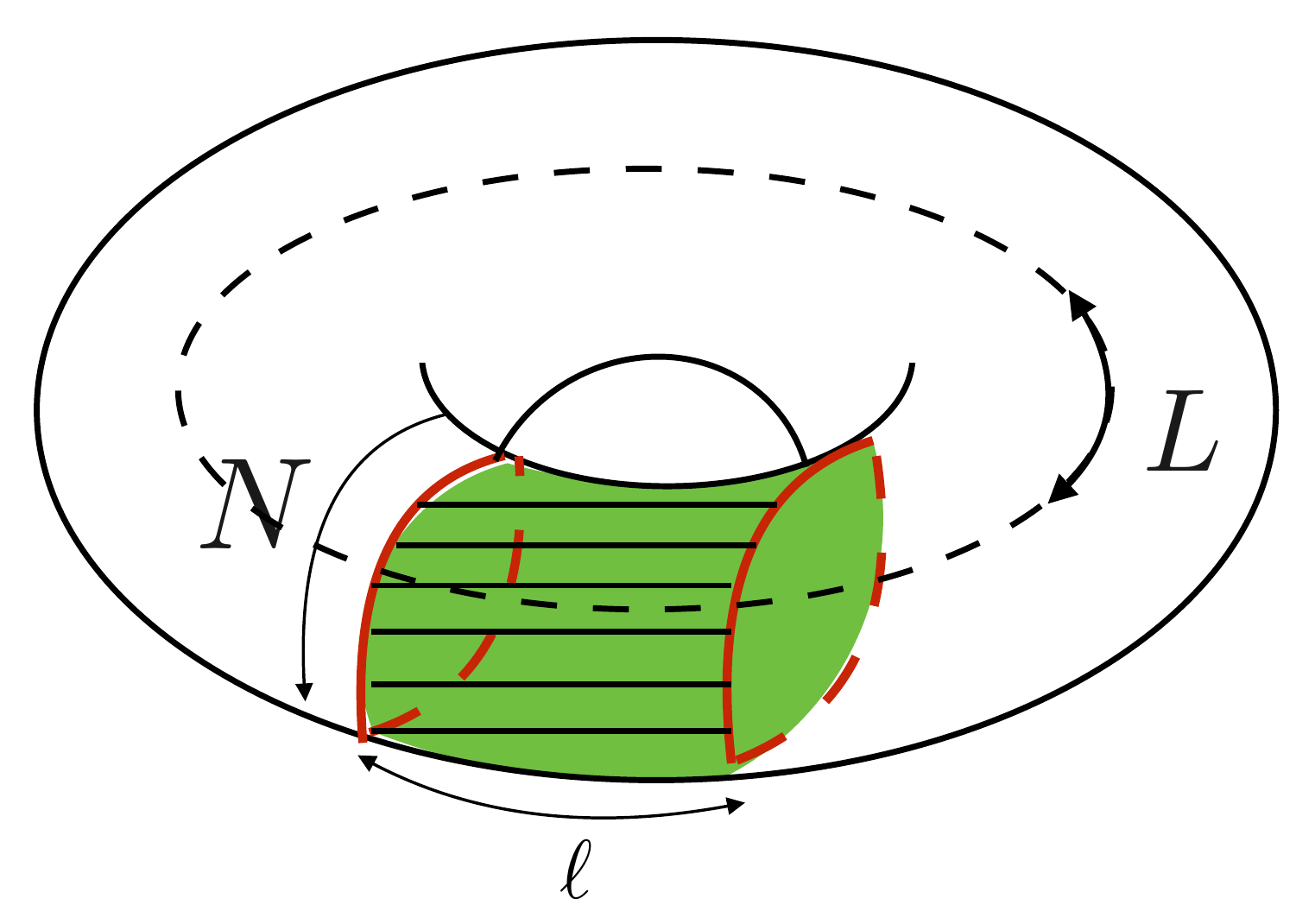}}
\caption{The geometries of the 2d systems we study in this paper: along the longitudinal $x$-direction the system is either infinite  (left) or finite with length $L$ (right).
In both cases, periodic boundary conditions are imposed along the transverse $y$-direction of size $N$. 
The geometry is then either an infinite cylinder (left)  or a torus (right). 
The entangling region is always a periodic strip of length $\ell$ along the $x$-axis,  as highlighted in green.
}\label{fig:cartoon}
\end{figure}

 A natural and transparent way to see this fundamental difference between free bosons and fermions is {\it dimensional reduction}, a strategy 
for the computation of the entanglement entropy first suggested in \cite{suggest} and since then exploited in many different circumstances. 
The idea is very simple: 
if the subsystem of a free two-dimensional (2d) model is translational invariant in one compact direction (that we call transverse,  say along the $y$-axis), 
we can perform Fourier transform in this direction and reduce the problem to the sum of 1d ones, for which exact results are known.
Two examples of  geometries for which the dimensional reduction works are shown in Figure \ref{fig:cartoon} and they are the only ones we will consider in this paper. 
Actually, this technique can be straightforwardly applied in generic dimensions $d$ (with $d-1$ compact ones), but we focus here in 2d for clarity of the presentation  
(the only difference in the final result is just the sum over many transverse components). 

The main goal of this paper is to apply dimensional reduction to the computation of the symmetry resolved entanglement \cite{lr-14,goldstein, goldstein1, goldstein2, xavier}
entropies for 2d free fermions and bosons with a $U(1)$ symmetry, exploiting known results in 1d for bosons \cite{MDC-19-CTM} and fermions \cite{riccarda,SREE2dG}.
These quantities account for the entanglement within the different symmetry sectors (see section \ref{sec:mainSREE} for precise definitions). 
While the role of symmetries is crucial in the study of many-body systems, the importance of symmetry resolved entanglement measures has been only 
recently understood, and it has been underlined also from an experimental point of view \cite{fis}.

The paper is organised as follows.
In section \ref{sec:main} we will do a brief recap of the needed 1d  results.
Sections \ref{sec:fermions} and \ref{sec:bosons} are the core of the paper, where we derive our results for total and symmetry resolved entropies for fermions and bosons, 
respectively.
Step by step, we benchmark our analytic results against exact numerical computations. 
We draw our conclusions in Section \ref{concl}. 
Details about the numerical techniques are provided in Appendix~\ref{app:a}. 
Appendix~\ref{app:b} provides a further application of the dimensional reduction to an anisotropic free fermion model.

\section{One-dimensional recap}\label{sec:main}
In this section, we provide an overview of the the results about one-dimensional models that we will need for the dimensional reduction in the following sections.  
For free fermions these are based on  Toeplitz determinants and Fisher-Hartwig techniques, while for free bosons on corner transfer matrix.

\subsection{R\'enyi and Entanglement Entropies}\label{sec:mainREE}
Given a bipartition of a system in a pure state $\ket{\psi}$ into $A \cup B$,  the reduced density matrix (RDM) $\rho_A$ of the subsystem $A$ is defined by tracing over the degrees of freedom of the subsystem $B$, i.e.  
\begin{equation}
\label{eq:RDMdefinition}
\rho_A=\mathrm{Tr}_B \rho, 
\end{equation}
where $\rho =\ket{\psi} \bra{\psi}$ is the density matrix of the entire system. A measure of the bipartite entanglement is given by the R\'enyi entanglement entropies $S_n$, 
defined as
\begin{equation}
\label{eq:defRenyi}
S_n\equiv\dfrac{1}{1-n}\log \mathrm{Tr}\rho^n_A,
\end{equation}
whose limit $n \to 1$ is the von Neumann entanglement entropy, i.e., $S_1=-{\rm Tr} \rho_A \log \rho_A$. 
These entanglement entropies have been investigated for a large number of extended quantum systems in several different physical situations and with many 
different techniques (see e.g. Refs. \cite{intro1,intro2,intro3} as reviews). 

Below we only report some results that we will use in the following sections. 
We start by considering the one-dimensional tight binding model, i.e., the free spinless fermions described by the Hamiltonian
\begin{equation}
\label{eq:Hamiltonian1dFF}
H_{FF}=-\frac{1}{2}\sum_{i} \left( c^{\dagger}_{i+1} c_i+c^{\dagger}_{i} c_{i+1}\right)+\sum_i\mu c^{\dagger}_{i}c_{i},
\end{equation}
where $\mu$ is the chemical potential, $c_i$ and $c^\dag_i$ the ladder operators of the fermions obeying standard anticommutation relations $\{ c_i, c_j^{\dagger} \}= \delta_{ij}$.
We only focus on the ground state here.  When $|\mu|<1$ the theory is gapless.
The Jordan Wigner transformation maps the model to the spin-1/2 XX chain in a magnetic field.
We consider the  subsystem $A$  to be an interval made of $\ell$ consecutive sites. 
For large $\ell$, the asymptotic scaling  of the entanglement entropies is given by \cite{korepin,ce-10}
\begin{equation}
\label{eq:1dresults}
S_n=\dfrac{1+n^{-1}}{6}\log \left( 2\ell \sin k_F \right)+\Upsilon_n,
\end{equation}
where $k_F=\arccos (\mu/2)$ is the Fermi momentum and 
\begin{equation}
\label{eq:UpsilonR}
\Upsilon_n= \frac{n+1}{n} \displaystyle \int_0^{\infty} \frac{dt}{t}\left[\frac{1}{1-n^{-2}} \left(\frac{1}{n \sinh t/n}-\frac{1}{\sinh t} \right)\frac{1}{\sinh t} -\frac{e^{-2t}}{6}\right].
\end{equation}
The leading logarithmic term in Eq. \eqref{eq:1dresults} is universal and follows from conformal field theory \cite{cc-04, cc-09, hlw-94, vidal};  
in contrast, the non-universal constant $\Upsilon_n$ has been derived using the Fisher-Hartwig conjecture \cite{korepin}. 
For a finite system of length $L$ with PBC's, the same form also holds replacing $\ell$ with $\frac{L}{\pi} \sin \frac{\pi\ell }{L}$ \cite{cc-04}.

Exact results are available also for free bosonic systems on the lattice, i.e., for  the harmonic chain.
In this case, one exploits  the Baxter corner transfer matrix (CTM) approach \cite{baxter}. 
A chain of oscillators of mass $M = 1$ with frequency $\omega_0$, coupled together by springs with elastic constant $k$ 
(which, without loss of generality, we assume to be $k=1-\omega_0$), is described by the Hamiltonian
\begin{equation}
\label{eq:hamHC1d}
H_{B}=\dfrac{1}{2}\sum_{i} p^2_i +\omega_0^2q^2_{i} +k(q_{i+1}-q_{i})^2,
\end{equation}
where the $p_i$ and $q_i$ satisfy the canonical commutation relations $[q_i,q_j] = [p_i,p_j] = 0$ and $[q_i,p_j] = i\delta_{ij}$.
A canonical transformation of variables $(p_i, q_i)$ allows us to have only one relevant system parameter in Eq. (\ref{eq:hamHC1d}), i.e., $\omega^2_0/k$ or, 
equivalently, $(1-k)^2/k$ \cite{solvable}. 

In Refs. \cite{Gaussian,nishino,nishino1} it has been shown that a harmonic chain is related to a two-dimensional classical Gaussian model.
Such correspondence is at the basis of the CTM approach which allows us to write the RDM as the partition function of the two-dimensional classic model.
For the case of an infinite subsystem size (namely, a semi-infinite line), this correspondence allows us to express $\rho_A$ as (up to a prefactor) 
\cite{peschel1,Peschel, Gaussian}
\begin{equation}
\rho_A \sim e^{-\mathcal{H}_{CTM}},
\label{ctm}
\end{equation}
where $\mathcal{H}_{CTM}$ is an effective Hamiltonian which, due to the gaussian nature of the model, can be diagonalised. This means that all the eigenvalues of the RDM can be determined exactly. In particular, in Ref.~\cite{Gaussian} it was shown that
 \begin{equation}
 \mathcal{H}_{CTM}=\displaystyle \sum_{j=0}^{\infty} \epsilon(2j+1)\beta^{\dagger}_j \beta_j, \qquad \epsilon =\dfrac{\pi I(\sqrt{1-\kappa^2})}{I(\kappa)},
 \label{HCTM}
 \end{equation}
 where $I(\kappa)$ is the complete elliptic integral of the first kind, $\kappa$ can be defined in terms of the system parameters as $\kappa/(1-\kappa)^2=k/\omega_0^2$ 
 and $\beta_j, \beta_j^{\dagger}$  are bosonic ladder operators.  
 From Eq. (\ref{eq:defRenyi}), we easily read off the R\'enyi entropies of a harmonic chain as \cite{cc-04}
 \begin{equation}
 \label{eq:defctm}
 S_n=\sum_{j=0}^{\infty} n\log [1-e^{-(2j+1) \epsilon}]-\sum_{j=0}^{\infty}\log [1-e^{-(2j+1)n\epsilon}].
 \end{equation}
In the critical regime in which $\omega_0 \to 0$ (i.e., $\epsilon \to 0$), we recover the field theory result \cite{cc-04}
 \begin{equation}
 \label{eq:cftresult}
 S_n\simeq\dfrac{1+n^{-1}}{12}\log \xi. 
\end{equation}  
where $\xi \sim \omega_0^{-1}$ is the correlation length (the inverse gap) of the system. 
Although all the previous formulas are valid for a semi-infinite line, we have that for a finite subsystem of length $\ell$, as long as $\ell \gg \xi$, the clustering of the 
RDM implies that it becomes the product of two independent ones at the two boundaries \cite{albaES} and hence that the R\'enyi entropies are just the double of the one above for a single boundary, with exponentially suppressed 
corrections in $\ell/\xi$ \cite{ccd-08,cd-09}.  

In the following, we will mainly be interested in systems with a $U(1)$ internal symmetry. We then consider a complex bosonic theory which on the 
lattice is a chain of complex oscillators (we dub complex or double harmonic chain). 
The latter is the sum of two real harmonic chains in the variables $(p^{(1)},q^{(1)})$ and $(p^{(2)},q^{(2)})$, i.e.
\begin{equation}
\label{eq:complex}
H_{CB}(p^{(1)}+ip^{(2)},q^{(1)}+iq^{(2)})=H_{B}(p^{(1)},q^{(1)})+H_{B}(p^{(2)},q^{(2)}).
\end{equation}
For the double chain the entanglement Hamiltonian is the sum of two $\mathcal{H}_{CTM}$ of the form \eqref{HCTM}. 
For each of them we introduce the associated ladder operators, $\beta_{1,i}$ and $\beta_{2,i}$. Since these operators commute, the RDM factorises as 
\begin{equation} \label{b1b2}
\rho_A= \rho_A^{\beta_1}\otimes \rho_A^{\beta_2},
\end{equation}
where we denoted the RDM for $\beta_{1,i}$ and $\beta_{2,i}$ with $\rho_A^{\beta_1}$ and $\rho_A^{\beta_2}$ respectively.
Therefore the R\'enyi entropies for a complex chain are just the double of those for a real chain.

\subsection{Symmetry Resolved Entanglement Entropies}\label{sec:mainSREE}
Let us focus now on systems with a $U(1)$ symmetry and with an associated conserved charge denoted by ${Q}$. 
Moreover, we assume it is possible to write ${Q}$ as the sum of local operators, i.e., ${Q}=\sum_i {Q}_i$, where ${Q}_i$ is the local contribution.  Taking the trace over $B$ of $[\rho,{Q} ] = 0$, and defining ${Q}_A=\sum_{i\in A} {Q}_i$,
we find that $[\rho_A,{Q}_A] = 0$. This implies that $\rho_A$ is block-diagonal and each block corresponds to a different charge sector labelled by the eigenvalue $q$ of ${Q}_A$
\begin{equation}
\rho_A= \oplus_q \Pi_q \rho_A, 
\end{equation}
where $\Pi_{q}$ is the projector onto the subspace of states of region $A$ with charge $q$
(with a slight abuse of notation we just wrote $\Pi_q \rho_A$ instead of $\Pi_q \rho_A \Pi_q$, being sure that will not generate any confusion).
Such decomposition defines the (normalised) RDM in the $q$-sector, $\rho_A (q)\equiv \Pi_q \rho_A/{\rm Tr}( \Pi_q \rho_A)$.
The generalised moments
\begin{equation}
\label{eq:znq}
\mathcal{Z}_n(q)\equiv \mathrm{Tr} (\Pi_q\rho^n_A),
\end{equation}
are useful to write down the \emph{symmetry resolved} R\'enyi  and von Neumann entropies, 
meaning the entropies associated to each $\rho_A(q)$ that are given by
\begin{equation}
\label{eq:SREE1}
S_n(q)\equiv \dfrac{1}{1-n}\log {\rm Tr} (\rho_A(q))^n= \dfrac{1}{1-n}\log \left[ \dfrac{\mathcal{Z}_n(q)}{\mathcal{Z}^n_1(q)}\right], \qquad S_{1}(q)=\lim_{n\rightarrow 1} S_n(q).
\end{equation}
Notice that ${\cal Z}_1(q)$ is the probability $p(q)$ of finding $q$ in a measurement of ${Q}_A$, i.e., $p(q)={\rm Tr} ( \Pi_q \rho_A)$. 
The total von Neumann entanglement entropy $S_1$ and the symmetry  resolved ones satisfy the relation \cite{nc-10,fis}
\begin{equation}
\label{eq:SvN}
S_{1}=\displaystyle \sum_q p(q) S_{1}(q)- \displaystyle \sum_q p(q) \log p(q).
\end{equation}
Eq.~(\ref{eq:SvN}) has a clear physical interpretation. The first contribution is known as configurational entanglement entropy ($S^c$) and depends on the entropy of each charge sector, weighted with its probability. The second contribution is the fluctuation entanglement entropy ($S^f$) which is due, as the name says, to the fluctuations of the charge within the subsystem.  
The fluctuation entropy $S^f$, also known as number entropy, has been studied in many different context \cite{fis,wv-03,SREE2d,delmaestro2,kusf-20,kusf-20b}. 
More complicated formulas can also be written for the R\'enyi entropies  \cite{crc-20}.

Finally, we can define the (normalised) charged moments of $\rho_A$ as
\begin{equation}
\label{eq:firstdef}
Z_n(\alpha)\equiv\mathrm{Tr}\rho_A^ne^{i {Q}_A \alpha},
\end{equation}
which, importantly, are related to the generalised moments (\ref{eq:znq}) by Fourier transform, i.e., \cite{goldstein}
\begin{equation}
\label{eq:defF}
\mathcal{Z}_n(q)=\displaystyle \int_{-\pi}^{\pi}\dfrac{d\alpha}{2\pi}e^{-iq\alpha}Z_n(\alpha).
\end{equation}
Recently, such relation allowed to derive interesting results about the different symmetry-resolved contributions for CFTs, 
free gapped and gapless systems of bosons and fermions, integrable spin chains and disordered systems. 
As for the total entropy, we will review only the results relevant for our purposes, while a plethora of others can be found in the literature 
\cite{goldstein, goldstein1, goldstein2, xavier,riccarda,SREE2dG,crc-20,MDC-19-CTM,lr-14,ccgm-20,SREE,xhek,cms-13,d-16,matsuura,CFH,CFH2,ch-rev,ssr-17,neg2,clss-19}.

In the one-dimensional tight binding model, the generalised Fisher-Hartwig conjecture has been used to obtain the asymptotic behaviour of the symmetry resolved entropies 
at leading and subleading orders \cite{riccarda,SREE2dG}. 
The charged moments are 
 \begin{equation}
\label{eq:1d}
\log Z_n(\alpha)=\dfrac{ik_F \ell}{\pi}\alpha-\left[\frac{1}{6}\left(n-\frac{1}{n} \right) +\frac{2}{n}\left(\frac{\alpha}{2\pi} \right)^2\right]\log 2\ell \sin k_F +\Upsilon(n,\alpha)+o(1),
\end{equation}
 where  $\Upsilon(n,\alpha)$ is a real and even function of $\alpha$ defined as
 \begin{equation}
\label{eq:UpsilonR2}
\Upsilon(n,\alpha)= ni\displaystyle \int_{-\infty}^{\infty}dw\left[\tanh (\pi w)-\tanh(\pi n w +i\alpha/2)\right]\log \frac{\Gamma \left(\frac{1}{2}+iw \right)}{\Gamma \left(\frac{1}{2}-iw \right)}.
\end{equation}
Taking the Fourier transforms and expanding for large $\ell$, one gets for the symmetry resolved entropies at the leading orders \cite{riccarda}
\begin{equation}
S_n(q)= S_n
- \frac{1}{2} \ln \left( \frac{2}{\pi} \ln ( 2\ell \sin k_F) \right)  +\frac{\ln n}{2(1-n)} +o(1).
\label{Snsr}
\end{equation}
The fact that up to order $O(1)$ the symmetry resolved entropies do not depend on $q$ has been dubbed {\it equipartition of entanglement} \cite{xavier}.
The first term breaking equipartition appears at order $O(1/(\log \ell)^2)$ \cite{riccarda}.

The same quantities have been also investigated for off-critical quantum bosonic chains through the Baxter's CTM  for the bipartition in two 
semi-infinite systems \cite{MDC-19-CTM} (and generalised to finite subsystems in \cite{ccgm-20}). 
The approach of the previous subsection can, in fact, be adapted to the computation of the symmetry resolved entropies in the non-critical complex harmonic chain 
which (in contrast with its real analogue) possesses a $U(1)$ symmetry. 
The starting point is to write the charge operator ${Q}_A$ in terms of the $\beta_1$'s and $\beta_2$'s ladder operators associated to the bosons of the two (real) chains (cfr. Eq.~\eqref{b1b2}), as \cite{MDC-19-CTM}
 \begin{equation} \label{chargebosonA}
 {Q}_A=\sum_{j=0}^{\infty}\beta_{1,j}^{\dagger}\beta_{1,j}-\beta_{2,j}^{\dagger}\beta_{2,j}.
 \end{equation}
For the charged moments, we need to compute ${\rm Tr} ( \rho_A^n e^{i {Q}_A\alpha} )$, but using the form in Eq.~\eqref{chargebosonA} for ${Q}_A$, the trace factorises as 
\begin{equation}
Z_n(\alpha)={\rm Tr}\rho_A^n e^{i {Q}_A\alpha}= {\rm Tr}[(\rho_A^{\beta_1})^n e^{iN_A^{\beta_1}\alpha}]\times [{\rm Tr}(\rho_A^{\beta_2})^n e^{-iN_A^{\beta_2}\alpha}],
\end{equation}
where $N_A^{\beta_1}=\sum_{j\in A} \beta^{\dagger}_{1,j}\beta_{1,j}$ and 
$N_A^{\beta_2}=\sum_{j\in A} \beta^{\dagger}_{2,j}\beta_{2,j}$. The two factors are equal, except for the sign of $\alpha$.  
Therefore, the charged moments for the complex harmonic chain are  \cite{MDC-19-CTM}
\begin{equation}
\label{eq:second}
\log Z_{n}(\alpha)=\sum_{j=0}^{\infty} 2n\log [1-e^{-(2j+1) \epsilon}]-\sum_{j=0}^{\infty}\log [1-e^{-(2j+1)n\epsilon+i\alpha}]-\sum_{j=0}^{\infty}\log [1-e^{-(2j+1)n\epsilon-i\alpha}].
\end{equation}
In the critical region $\epsilon\to0$, Eq. (\ref{eq:second}) reduces to
\begin{equation}
\label{eq:crit1}
\log Z_n(\alpha)\simeq\left[ \dfrac{1}{n}\left( \dfrac{\alpha}{2\pi}\right)^2- \dfrac{|\alpha|}{2\pi n}+\dfrac{1}{6 n}-\dfrac{n}{6} \right] \log \xi+O(1).
\end{equation}
A Fourier transform allows to get the generalised moments and, eventually, the symmetry resolved entropies. We only report the final result, which reads
\begin{equation}\label{eq:resv}
S_n(q)= \frac{2}{1-n}\sum_{k=1}^{\infty}\Big[n\log (1-e^{-2\epsilon  k})-\log (1-e^{-2n\epsilon  k}) \Big] 
+ \dfrac{1}{1-n}\log \dfrac{\Phi_q(e^{-n \epsilon } )}{(\Phi_q(e^{- \epsilon } ))^n} ,
\end{equation}
with
\begin{equation}
\Phi_q(u) =
\sum_{k=0}^\infty (-1)^k u^{k^2 + k+|q|(2 k + 1)}\,.
\label{Phi2}
\end{equation}
While in general there is no entanglement equipartition for these bosonic chains, in the critical limits $\epsilon\to0$ one has 
\begin{equation}
S_n(q)=\frac{1}{1-n}\log \frac{\mathcal{Z}_n(q)}{\mathcal{Z}_1^n(q)}=S_n(q=0)+ \frac{n \epsilon^2 q^2}2 +O(\epsilon^3),
\label{Sncr}
\end{equation}
and equipartition is recovered at the leading order in $\epsilon$.
Again, all the previous formulas are valid for $A$ being a semi-infinite line. The results for a finite interval are obtained exploiting the clustering of the RDM, 
following, e.g., Refs. \cite{albaES,ccgm-20}.

\section{Two-dimensional Free Fermions} \label{sec:fermions}
In this section we compute the R\'enyi entropies and the symmetry-resolved entropies in the ground state of a two-dimensional free fermionic system. 
For the total entropies, our results confirm the known logarithmic violation of the area law \cite{widom2,widom3,widom4,widom5,mint2},
which generalises also to the symmetry resolved analogue. 

\subsection{The model and the bipartition} 
Let us consider a quadratic fermionic system on a two-dimensional square lattice with isotropic hopping between nearest-neighbour sites. It is described by the following Hamiltonian
\begin{equation}
\label{eq:Hamiltonian}
H_{FF}=-\frac{1}{2}\sum_{\braket{\mathbf{i},\mathbf{j}}}(c^{\dagger}_{\mathbf{i}}c_{\mathbf{j}}+c^{\dagger}_{\mathbf{j}}c_{\mathbf{i}})+\mu\sum_{\mathbf{i}}c^{\dagger}_{\mathbf{i}}c_{\mathbf{i}},
\end{equation}
where $\mu$ is the chemical potential for the spinless fermions $c_{\mathbf{i}}$, with $\mathbf{i}= (i_1, i_2 )$ a vector identifying a given lattice site, 
and $\braket{\mathbf{i},\mathbf{j}}$ stands for nearest neighbours. 
Specifically, we consider  a set of $N$ coupled identical parallel chains, hence $N$ is the finite length along one direction (say the $y$-axis).
In the other direction, say the $x$-axis, the system is either infinite or finite with length $L$. 
PBC's are imposed along the $y$-axis. 
The subsystem $A$ is a (periodic) strip of length $\ell$ along the $x$-axis, (see Figure\,\ref{fig:cartoon}).

Given the special geometry we consider, we can take the Fourier transform  along the transverse $y$ direction.
The partial Fourier transforms $\tilde{c}_{j_1,r}$ and its inverse are 
\begin{equation}
\tilde{c}_{j_1,r}=\displaystyle {\frac{1}{\sqrt{N}}} \sum_{j=0}^{N-1}c_{j_1,j}e^{-2\pi i j r /N}, \qquad 
c_{j_1,j_2}=\frac{1}{\sqrt{N}}\displaystyle \sum_{r=0}^{N-1}\tilde{c}_{j_1,r}e^{2\pi i j_2 r /N}, 
\end{equation}
leading to the  Hamiltonian in mixed space-momentum representation
 \begin{equation}
\label{eq:Hamiltonian1}
H_{FF}=\sum_{r=0}^{N-1}H_{k^{(r)}_y}.
\end{equation} 
The operator $H_{k^{(r)}_y}$ is the Hamiltonian in the $k^{(r)}_y=\frac{2\pi r}{N}$ transverse  momentum sector: 
\begin{equation}
\label{eq:Hamiltonian2}
H_{k^{(r)}_y}=-\frac{1}{2}\sum_{i=1}^L \left( \tilde{c}^{\dagger}_{i,r}\tilde{c}_{i+1,r}+\mathrm{h.c.}\right)+\sum_i\mu_{r} \tilde{c}^{\dagger}_{i,r}\tilde{c}_{i,r},
\end{equation}
where 
\begin{equation} 
 \mu_{r}=\mu-\cos k^{(r)}_y,
\label{mur} 
\end{equation} 
and $L$ is the length of the chain along the $x$-axis. 
In this way, the Hamiltonian is mapped to a sum of $N$ independent one-dimensional chains with chemical potential $\mu_{r}$ depending on the 
transverse momentum $k^{(r)}_y$. 

We focus on the critical regime of the whole 2d system, which is in attained for $0<\mu < 2$.
In terms of the one dimensional systems, this constraint on $\mu$ means that all transverse modes with $|\mu_{r}|<1$ are critical, 
while the others are not. This inequality is satisfied for 
\begin{equation} \label{OmegaN}
r \in \Omega_{\mu}=\left[ 0,\frac{\arccos(\mu-1)N}{2\pi } \right[ \; \cup \;  \left]N \left( 1-\frac{\arccos(\mu-1)}{2\pi } \right), N-1 \right].
\end{equation}
The inner extremes of the intervals are not part of $\Omega_\mu$.
The case $\mu=0$ deserves particular attention: when dealing with a finite number of chains, also the mode $r=0$ has to be removed from $\Omega_\mu$. 
This difference is irrelevant in the limit $N \to \infty$ when the fraction of critical chains is simply given by $\frac{\arccos (\mu -1)}{\pi}$.
 
 Since the Hamiltonian is a sum of different sectors, the ground state density matrix factorises and so does the RDM
 \begin{equation}
\rho_A= \bigotimes_{r =1}^N \rho^A_{k_y^{(r)}}=\bigotimes_{r \in \Omega_{\mu}} \rho^A_{k_y^{(r)}} .
\label{facRDM}
 \end{equation}
In the last equality, we stress that the only relevant modes are the ones corresponding to critical 1d chains. The blocks corresponding to non-critical chains are projectors on the 1d vacuum state, i.e. without fermions. As a consequence, hereafter, we only take into account the gapless modes, which belong to $\Omega_{\mu}$.
The RDM $\rho^A_{k_y^{(r)}}$ of the 1d subsystem associated to the $r$-th mode can be written as \cite{correlation,pe-09,solvable}
\begin{equation}
 \label{eq:rho}
  \rho^A_{k_y^{(r)}}=\det C_{k^{(r)}_y} \exp \left( \sum_{i,j} [\log( C^{-1}_{k^{(r)}_y} -1)]_{i,j}\tilde{c}^{\dagger}_{i,r} \tilde{c}_{j,r}\right),
\end{equation}
 where the matrix $ C_{k^{(r)}_y} \equiv \braket{\tilde{c}^{\dagger}_{i,r} \tilde{c}_{j,r}}$  is the correlation matrix restricted to the $r$-th subsystem $A$. 
 The entanglement entropy is easily expressed in terms of the eigenvalues of such correlation matrix (see appendix \ref{app:a} for further details). 
 
  We start by considering the model in the thermodynamic limit in the longitudinal ($x$) direction, i.e., $L \to \infty$ (Figure~\ref{fig:cartoon}, left panel). 
 For the ground-state of $N$ infinite chains the correlation matrix $\mathbf{C}$ of the whole (two-dimensional) subsystem can be written as
 \begin{equation}
 \label{eq:correlation}
 \mathbf{C}= \oplus_r C_{k^{(r)}_y} ,
 \end{equation}
 where  $C_{k^{(r)}_y}$ reads
 \begin{equation}
 \label{eq:corelation1}
  C_{k^{(r)}_y}(i,j)=\dfrac{\sin k_r^F (i-j)}{\pi (i-j)}, \qquad k_r^F=\arccos \mu_{r},
 \end{equation}
as a function of the Fermi momentum $k_r^F$ of each $r$-th chain ($\mu_{r}$ is given in Eq. \eqref{mur}).
This is due to the factorisation of the Hilbert space into the different modes, which corresponds to a block diagonal structure of the correlation matrix: each block is associated 
to a transverse mode and, as a consequence, to a given 1d ground state. 

\subsection{R\'enyi and Entanglement Entropies}\label{sec:fermiREE}

From the structure of the Hamiltonian in Eq. ($\ref{eq:Hamiltonian1}$), the R\'enyi entropies can be computed by invoking the one-dimensional results discussed in section (\ref{sec:mainREE}): the entanglement entropy is additive on tensor products and therefore decomposes as 
 \begin{equation}
 \label{eq:entropy1}
 S^{2d}_n\left( \bigotimes_{r \in \Omega_{\mu}}\rho^A_{k_y^{(r)}} \right)=\sum_{r \in \Omega_{\mu}}S^{1d}_{n,r}, \qquad S^{1d}_{n,r}=\dfrac{1}{6}\Big(1+\dfrac1n\Big)\log (2\ell\sin k_r^F)
 +\Upsilon_n +o(1),
 \end{equation}
 where $\Upsilon_n$ is in Eq. (\ref{eq:UpsilonR}).

\begin{figure}[t]
\centering
\subfigure
{\includegraphics[width=0.49\textwidth]{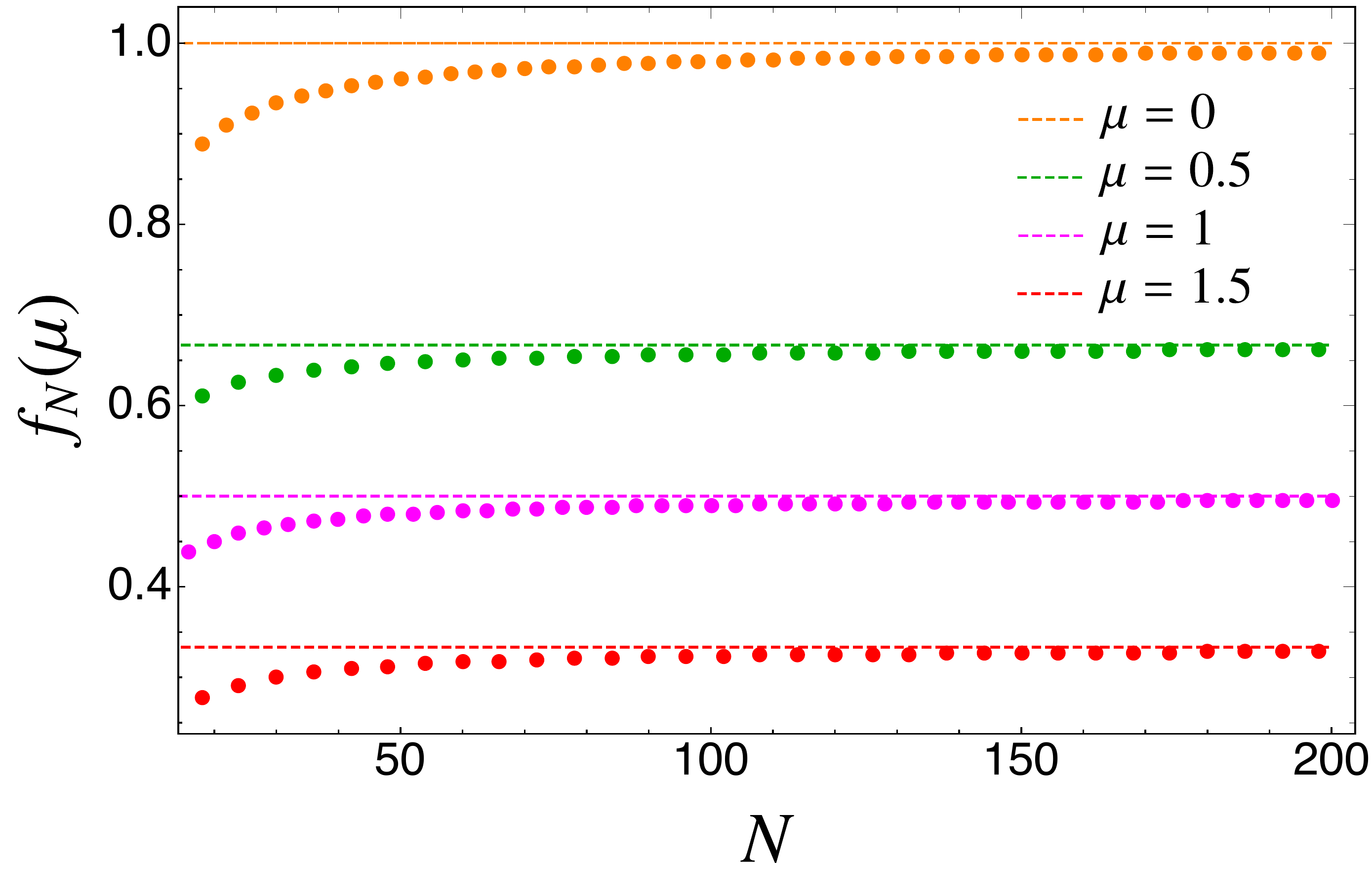}}
\subfigure
{\includegraphics[width=0.495\textwidth]{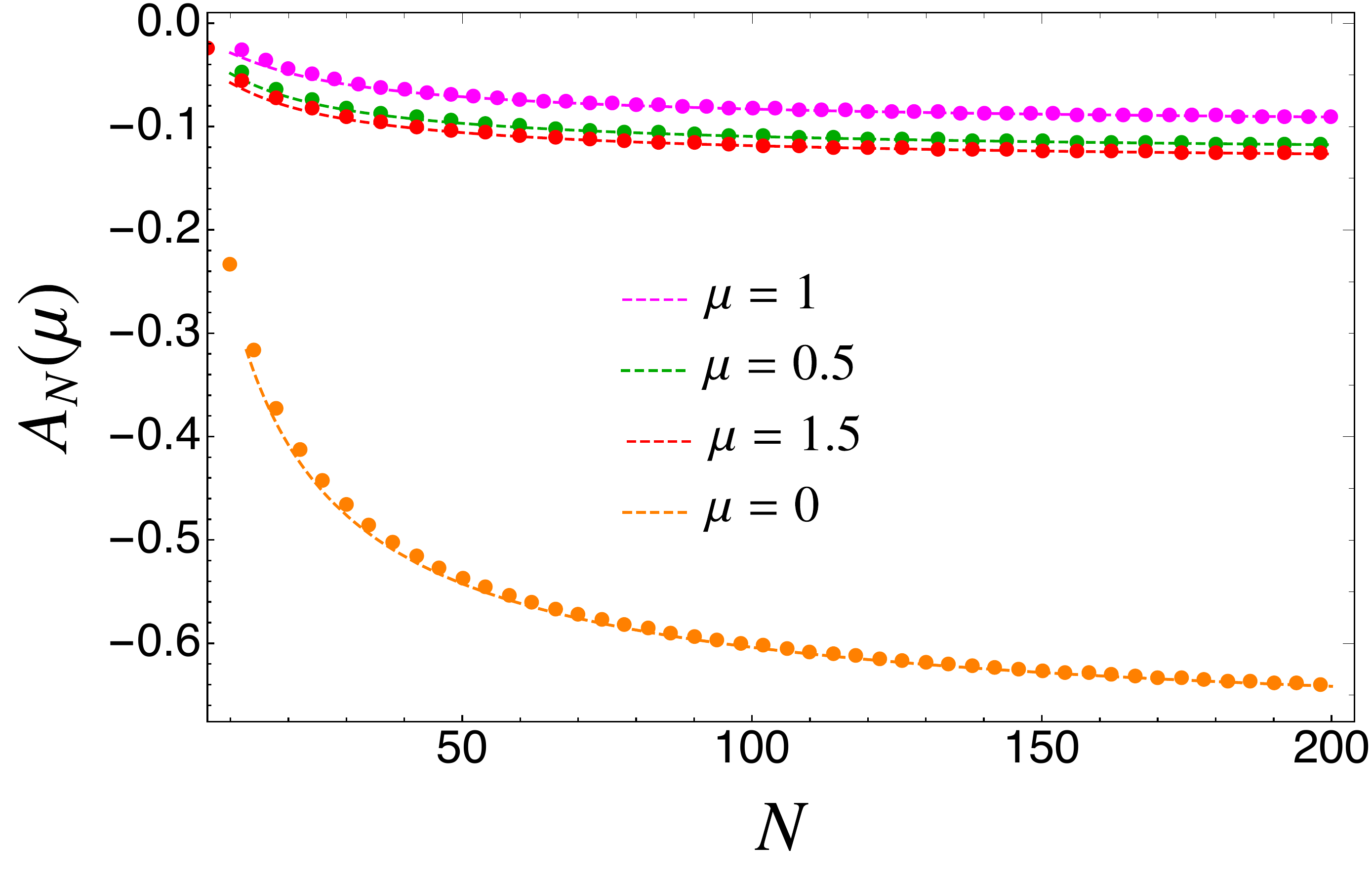}}
\caption{Left panel: The fraction of critical modes, $f_N(\mu)$, is plotted as a function of length $N$ of the transverse direction for four different values of chemical potential $\mu$. The curves approach the constant value reported in Eq.\,(\ref{eq:fN}) and plotted as dashed lines.  Right panel: The function $A_N(\mu)$ in Eq. \eqref{ANm} as a function of $N$ for four different values of chemical potential $\mu$.
For all $\mu$, the curves  approach $A_\infty(\mu)$ reported as dashed lines including the correction up to $O(1/N)$, which are clearly important to have 
a good match even for $N$ as large as $200$.
}\label{fig:ANm}
\end{figure}

Note that in our setting, the Fermi momentum of each transverse mode-chain  can be explicitly written down as 
 \begin{equation}
 \label{eq:fermi}
 \sin k_r^F=\sqrt{1-\left(\mu -\cos \left(\dfrac{2\pi r}{N} \right) \right)^2}.
 \end{equation}
Plugging this relation into Eq. (\ref{eq:entropy1}) we get 
 \begin{equation}
\label{eq:try1}
S^{2d}_n =\dfrac{f_N(\mu)N}{6}\Big(1+\dfrac1n\Big)\log (2\ell)+f_N(\mu)N\Upsilon_n+\dfrac{1}{12}\Big(1+\dfrac1n\Big)
\sum_{r\in \Omega_{\mu}}\log \left[ 1-\left(\mu-\cos\Big(\dfrac{2\pi r}{N}\Big)\right)^2\right], 
\end{equation}
where $f_N(\mu)$ denotes the fraction of critical modes (i.e., the number of modes belonging to $\Omega_{\mu}$ divided by $N$).
%
It is useful to define also the quantity 
\begin{equation}
A_N(\mu) = \dfrac{1}{2N}\sum_{r\in \Omega_{\mu}}\log \left[ 1-\left(\mu-\cos\left(\dfrac{2\pi r}{N}\right)\right)^2\right], 
\label{ANm}
\end{equation}
so that we have  
\begin{equation}\label{eq:finiten}
S^{2d}_n  =\dfrac{1}{6}\Big( 1+\dfrac1n \Big) \Big( f_N(\mu)\log \, 2\ell+ A_N(\mu) \Big)N +Nf_N(\mu)\Upsilon_n,
\end{equation}
As aforementioned, when $N \to \infty$ the prefactor of the logarithmic term simply becomes 
\begin{equation}\label{eq:fN}
 Nf_{\infty}(\mu) =N \frac{\arccos(\mu -1)}{\pi}.
\end{equation} 
In the left panel of Figure \ref{fig:ANm} we report $f_N(\mu)$ as function of $N$ for a few values of $\mu$, showing the approach to $N\to\infty$.

In the right panel of Figure \ref{fig:ANm} we report a similar plot for $A_N(\mu)$, as function of $N$  for four different values of $\mu$.
As $N$ increases, it approaches an asymptotic value that can be explicitly calculated.
In fact, in the limit of large $N$, the sum in Eq. \eqref{eq:try1} turns into 
\begin{equation}
\label{eq:sumtointegral}
\begin{split}
&\frac{1}{2}\sum_{r \in \Omega_{\mu}}\log \left| 1-\left(\mu-\cos\left(\dfrac{2\pi r}{N}\right)\right)^2\right| \rightarrow \\
&\dfrac{N}{ 2\pi} \displaystyle \int_{0}^{\arccos(\mu-1)} dx\, \log \left( 1-\left( \mu - \cos (x) \right)^2 \right)-\frac{1-\log (2\pi \sqrt{\mu(2-\mu)})+\log N}{2},
\end{split}
\end{equation}
where we have subtracted the (divergent) contribution from the upper extreme of integration, corresponding to the modes $r= \{ f_{\infty}(\mu)N, (1- f_{\infty}(\mu))N  \}$, which are excluded from the sum in the left hand side (see Eq.~\eqref{OmegaN}).
The explicit computation of the integral gives
\begin{multline}
\label{eq:integral}
\dfrac{N}{ 2\pi} \displaystyle \int_{0}^{\arccos(\mu-1)} dx\, \log \left( 1-\left( \mu - \cos (x) \right)^2 \right)=\\
-\dfrac{N}{2\pi} \left[\pi \log  (1+4\mu+2\mu^2-2(1+\mu)\sqrt{\mu^2+2\mu})+ \right. 
\arccos (\mu -1) \log (4(1+\mu+\sqrt{\mu^2+2\mu}))+\\
+\left. {\rm Im}(\mathrm{Li}_2(e^{2i\arccos(\mu-1)})+2\mathrm{Li}_2(e^{i\arccos(\mu-1)}(1+\mu+\sqrt{\mu^2+2\mu})))\right],
\end{multline}
where $\mathrm{Li}_2$ is the dilogarithmic function  $\mathrm{Li}_2\equiv\sum_{k=1}^{\infty}\frac{z^k}{k^2}$.
Once again, the case $\mu=0$ deserves particular attention because also the divergence coming from the lower extreme of integration in (\ref{eq:sumtointegral}) 
has to be subtracted (i.e., the limits $\mu \to 0$ and $N \to \infty$ do not commute). 
Thus, one has to carefully perform a Taylor expansion of the integrand around both extremes of integration.
The final result is
\begin{equation}\label{eq:mu0}
 A_N(0) \to  A_\infty(0)=-\log 2- \frac{2-2\log \pi+2\log N}{N}.
\end{equation}
The logarithmic correction for small values of $N$ is evident in Figure\,\ref{fig:ANm} for all values of $\mu$, but it is more pronounced for $\mu=0$, as clear from 
the analytic expressions.
Hence the total entropy for large $N$ is 
\begin{equation}
\label{eq:integral2}
S^{2d}_n =\dfrac{f_{\infty}(\mu) N}{6}\Big(1+\dfrac1n\Big) \log 2\ell+ \dfrac{N}{6} \Big(1+\dfrac1n\Big)  A_\infty(\mu) +f_{\infty}(\mu) N\Upsilon_n ,
\end{equation}
which we recall is valid at order $o(\ell^0)$ and $O(N^0)$. 
We will see that to have a good agreement with numerical data at finite but large $\ell$, it is needed to keep the $\log N$ contribution in $A_N(\mu)$.

When both $\ell$ and $N$ are large, it is useful to look at the special case of the subsystem $A$ being a square strip with $N=\ell$, when Eq. \eqref{eq:integral2} 
is rewritten as 
\begin{equation}
\label{eq:generalise}
S^{2d}_n=\frac{f_{\infty}(\mu) }{6}\Big(1+\dfrac1n\Big) \ell \log 2\ell + \ell \frac{1}{6}\Big(1+\dfrac1n\Big) A_\infty(\mu) +f_{\infty}(\mu) \ell\Upsilon_n +O(\ell^0).
\end{equation}
(Actually, any choice of $N$ and $\ell$ proportional to each other, $N= a \ell$, would be equivalent, with just an overall factor $a$, 
but for simplicity let us just think to $a=1$.) 
Let us briefly comment Eq. (\ref{eq:generalise}). 
It shows the expected logarithmic correction to the area law and our derivation gives a clearer understanding of such behaviour: 
it is a simple consequence of the fact that we are dealing with an extensive a number of critical chains, i.e. proportional to $N=\ell$, whose entropy obeys a logarithmic scaling so that each of them contributes 
proportionally  to $\log \ell$ to the total entropy.
Moreover, it also agrees with the result obtained by the application of Widom conjecture  (see, e.g.,\cite{widom2, widom3, widom4}) 
that  provides an explicit formula for the prefactor of the leading term of the entanglement entropy of free fermions in any dimension, i.e., 
$S_1=C \ell^{d-1} \log\ell +O(\ell^{d-1})$, with $C$ given by
\begin{equation}
\label{eq:widom}
C=\dfrac{1}{12(2\pi)}\displaystyle \int_{\partial \Lambda} \displaystyle \int_{\partial \Gamma(\mu)} | n_x \cdot n_p| dS_x dS_p, 
\end{equation}
where $\Lambda$ is the considered subsystem with volume normalised to one, $\Gamma(\mu)$ is the volume in momentum space enclosed by the Fermi surface, 
$n_p$, $n_x$ are the unit normals to the boundaries of these volumes and the integration is carried over the surface of both domains. 
In the case of interest for this paper, given the compactification along the $y$ direction, the Fermi surface is defined by the solutions of $S_p=0$ 
(with $S_p=\mu-\cos k_x-\cos k_y$). By performing the line integrals, Eq. (\ref{eq:widom}) becomes
\begin{equation}
\label{eq:prova}
C =\dfrac{\arccos(\mu -1)}{3\pi},
\end{equation}
in agreement with  Eq. \eqref{eq:generalise}. 
We stress that while the leading terms in the two approaches are identical, the dimensional reduction provides an
explicit prediction also for the subleading term proportional to $\ell$, as in Eq. \eqref{eq:generalise}, which cannot be derived by Widom conjecture.

\subsubsection{Some generalisations.}

The same approach is straightforwardly adapted to the computation of the entanglement entropies  in the case of  Dirichlet (open) boundary conditions (DBC's) along 
the transverse direction ($y$-axis), i.e., imposing  $c_{j_1,0}=c_{j_1,N}=0$.
Although these boundary conditions break the translational invariance in the transverse direction, one can use the Fourier sine transform (rather than the standard one). 
The only final difference is that the set of modes $\Omega_{\mu}$ in \eqref{OmegaN} corresponding to critical chains will now start from $r=1$ (instead of $r=0$).
The same strategy applies when the total system is a finite block of $L$ sites along the $x$-direction, with PBC's (see the right panel in Figure\,(\ref{fig:cartoon})). 
In this case, the only difference is that the scaling of the one-dimensional R\'enyi entropies for a system with PBC's reads 
\begin{equation}
\label{eq:fs}
S^{1d}_{n,j}=\dfrac{1+n}{6n}\log \Big(\frac{2L}{\pi}\sin \Big(\frac{\pi\ell}{L} \Big)\sin k_{j}^F\Big)+\Upsilon_n.
\end{equation}
Therefore, we have for any finite $N$ 
\begin{equation}
\label{eq:generalisefs}
S^{2d}_n=\frac{n+1}{6n}\left[f_N(\mu) \log \Big(\frac{L}{2\pi}\sin \Big(\frac{\pi\ell}{L}\Big) \Big) +  A_N(\mu)\right]N+f_N(\mu)N \Upsilon_n,
\end{equation}
and similarly for large $N$ with $f_N(\mu)\to f_{\infty}(\mu)$ and $A_N(\mu)\to A_\infty(\mu)$.

\subsubsection{Numerical checks.}

\begin{figure}[t]
\centering
\subfigure
{\includegraphics[width=0.325\textwidth]{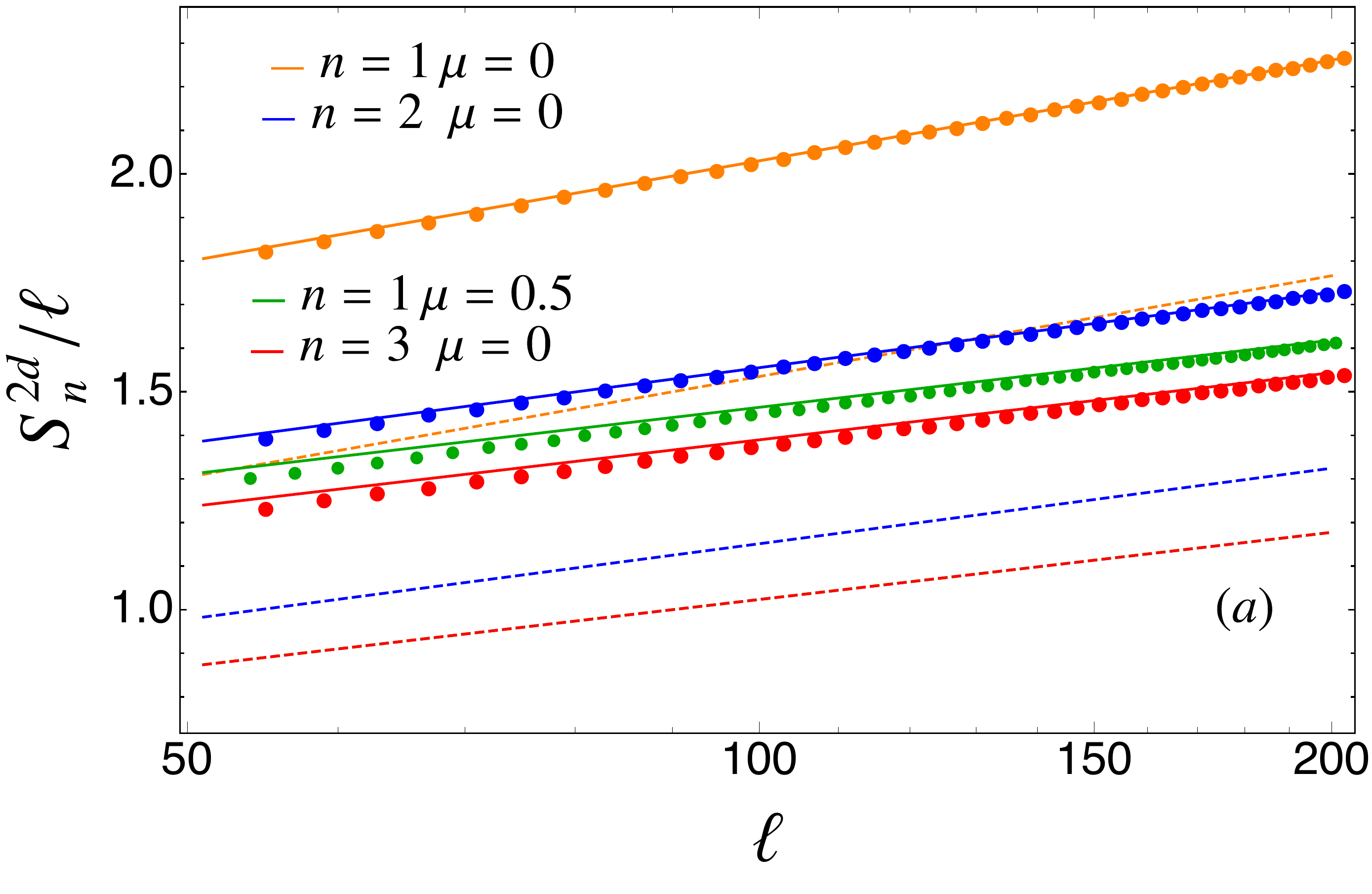}}
\subfigure
{\includegraphics[width=0.325\textwidth]{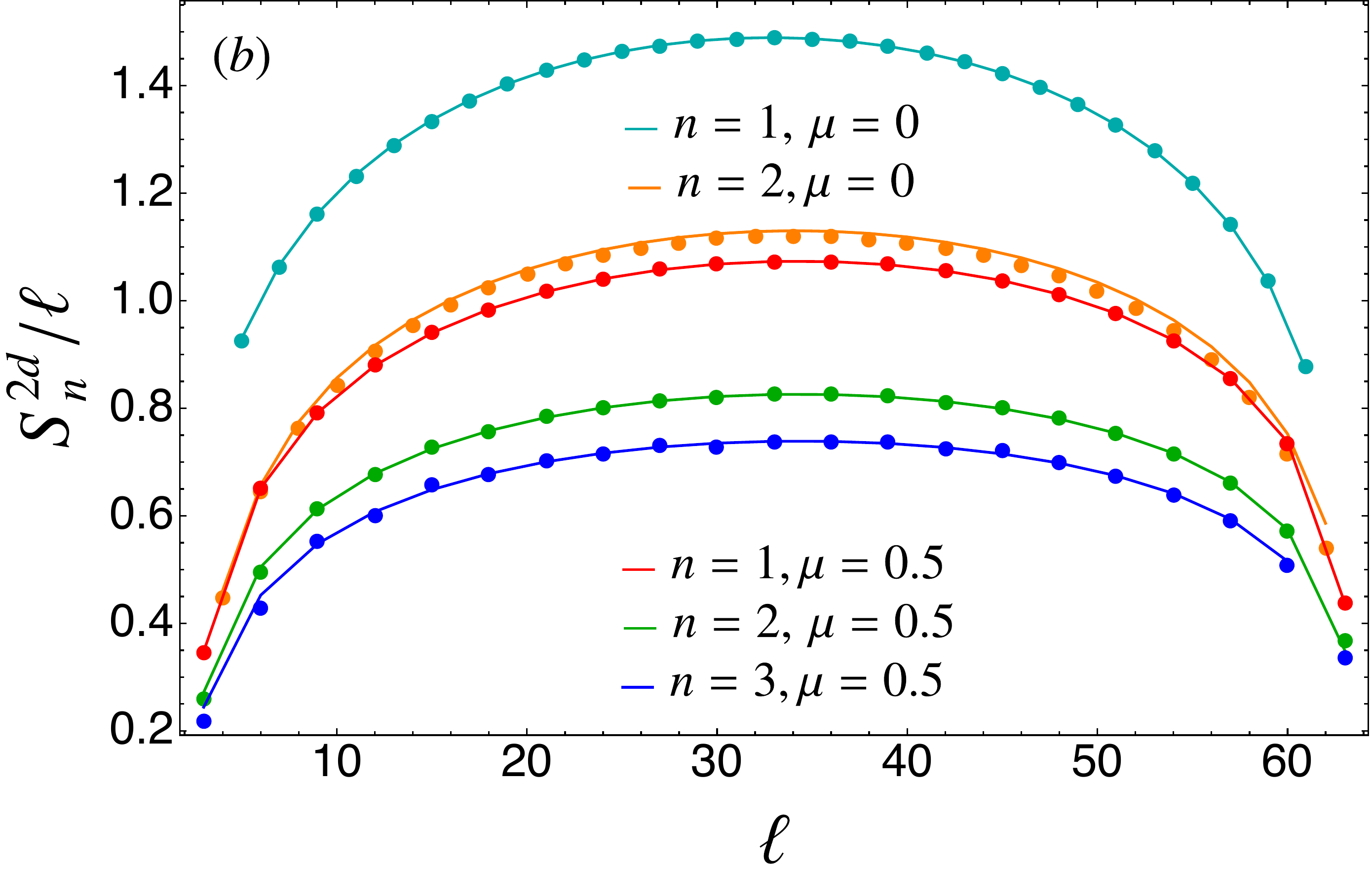}}
\subfigure
{\includegraphics[width=0.325\textwidth]{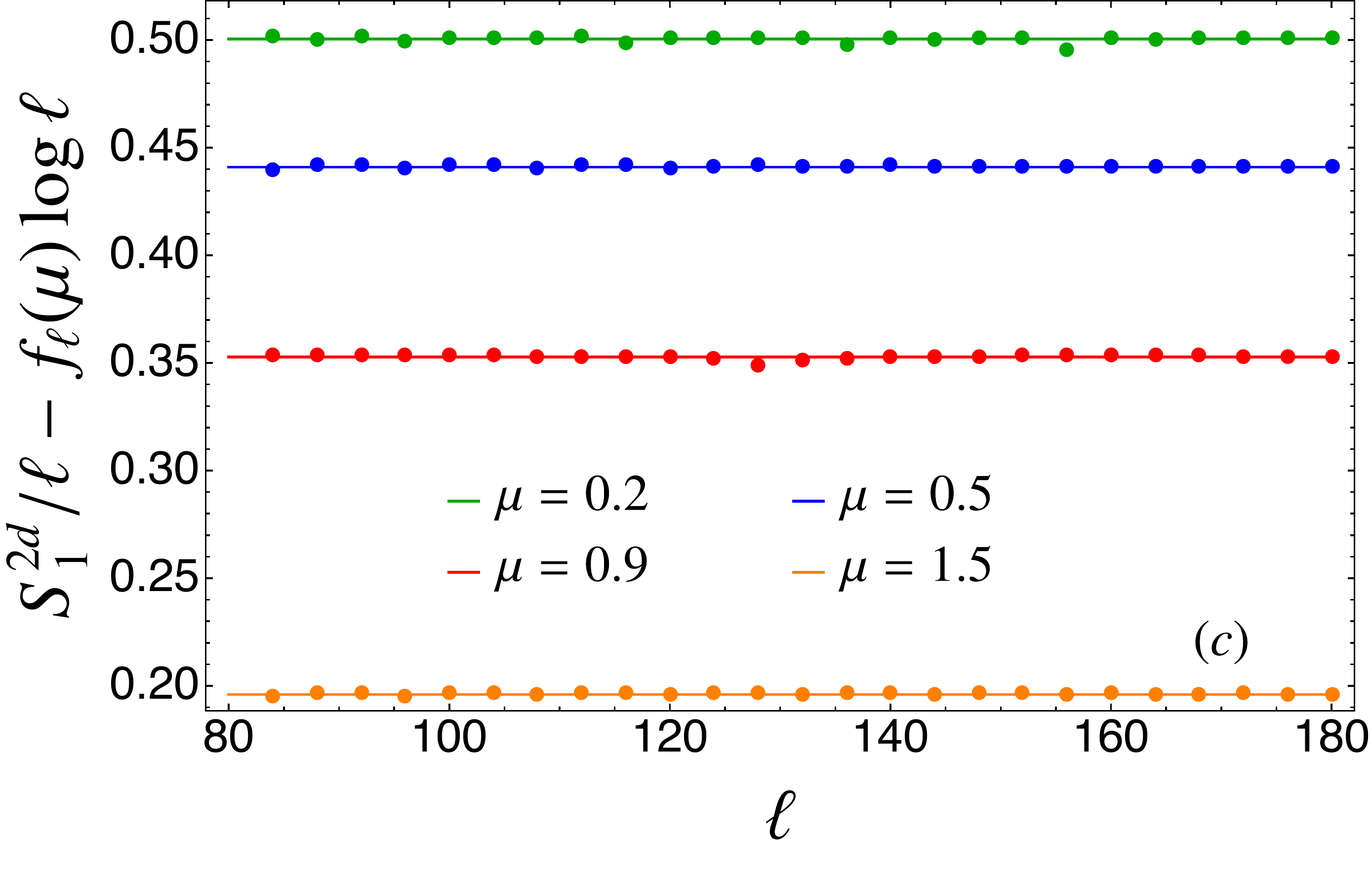}}
\caption{Leading scaling behaviour of the R\'enyi entropies $S^{2d}_n$ of 2d free fermions both for infinite (a) and finite system size $L$ (b) 
in the longitudinal direction.  
In the transverse direction, we fix the periodic size $N$ to equal $\ell$, the subsystem length in the longitudinal direction. 
The numerical results (symbols) for different values of $\mu$ and $n$ are reported as function of $\ell$. 
They match well the theoretical prediction of Eqs.  (\ref{eq:generalise}) and (\ref{eq:generalisefs}); 
the dashed lines in (a) are the leading behaviour $\propto \ell \log\ell$ which is clearly not enough accurate. 
The non-universal coefficient proportional to the area, $2\ell$, in Eq. (\ref{eq:integral}) is well captured by the numerics, as highlighted in (c).
}\label{fig:prima}
\end{figure}

We now benchmark the results for the total entropies against exact numerical calculations obtained by the free-fermion techniques reported in the Appendix \ref{app:a}.
In Figure~\ref{fig:prima} we report the numerical data of the R\'enyi entropies for different values of the  index $n$ and chemical potential $\mu$, 
both for infinite (panel (a)) and finite (panel (b)) system size. 
We fix the transverse direction $N$ to be equal to the longitudinal subsystem length $\ell$, so that the subsystem $A$ is a square with PBC in the transverse direction. We also properly choose the values of $\mu$ and $\ell$ such that $\ell f_{\ell}(\mu)$ is an integer number to eliminate effects due to partial fillings of modes.
The theoretical predictions for the leading scaling in Eqs.  (\ref{eq:generalise}) and (\ref{eq:generalisefs}) are also reported for comparison.
These include both the leading term and the subleading one proportional to the area ($2\ell$) between the subsystem $A$ and the rest of the system.

It is evident that the analytical results correctly describe the data. 
We also report (as dashed lines) the sole leading universal behaviour $\propto \ell \log \ell$:
this universal term alone does not match the data for these values of $\ell$, highlighting the 
importance of the subleading terms $\propto  \ell$ that we calculated analytically here for the first time. 
In the panel (c) of the same figure, we plot the data for the von Neumann entropy where we subtracted the leading term $f_{\ell}(\mu)\ell \log\ell$ to show the 
non-universal subleading terms found in Eq. (\ref{eq:generalise}) alone. 

\begin{figure}
\centering
\subfigure
{\includegraphics[width=0.325\textwidth]{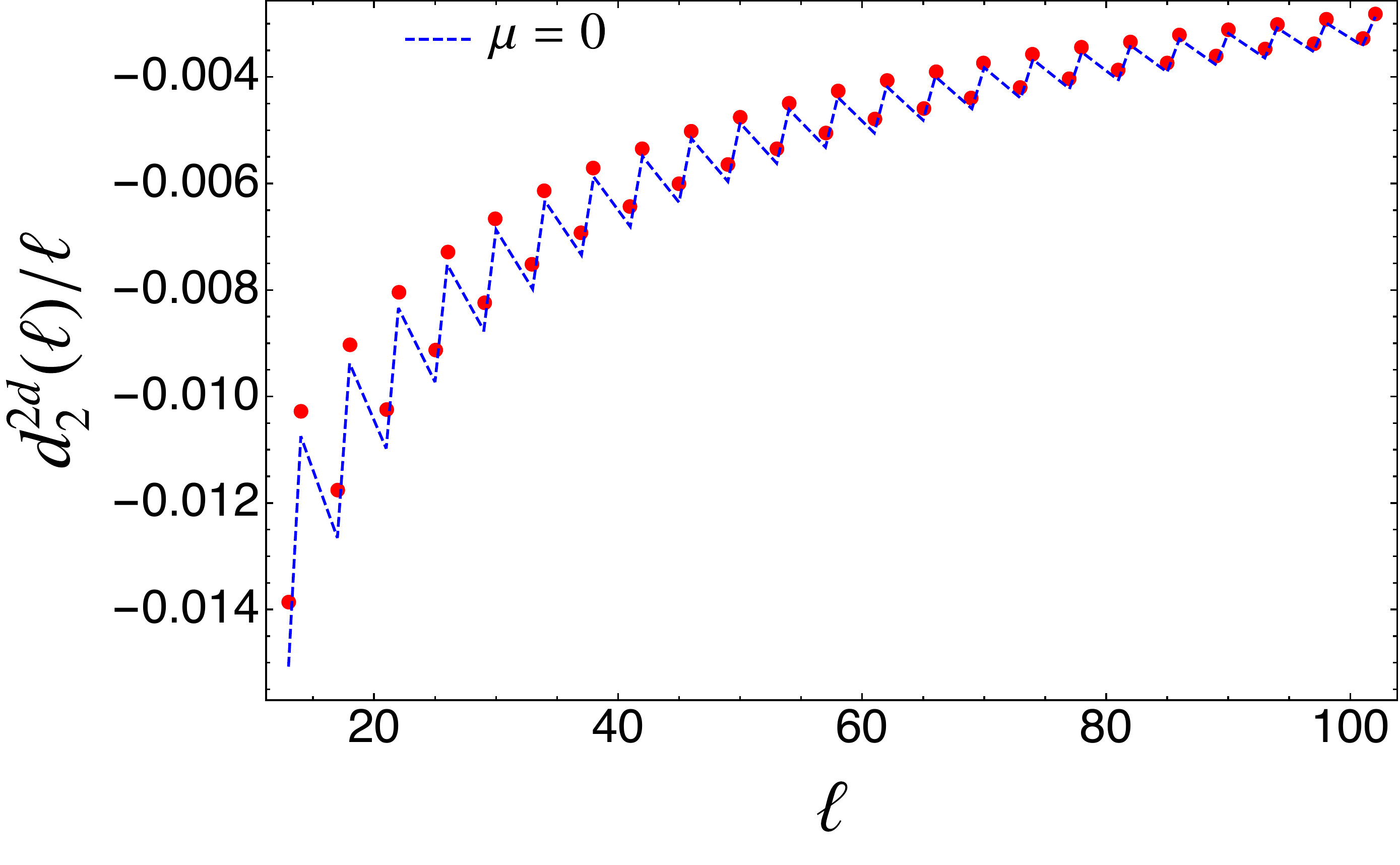}}
\subfigure
{\includegraphics[width=0.325\textwidth]{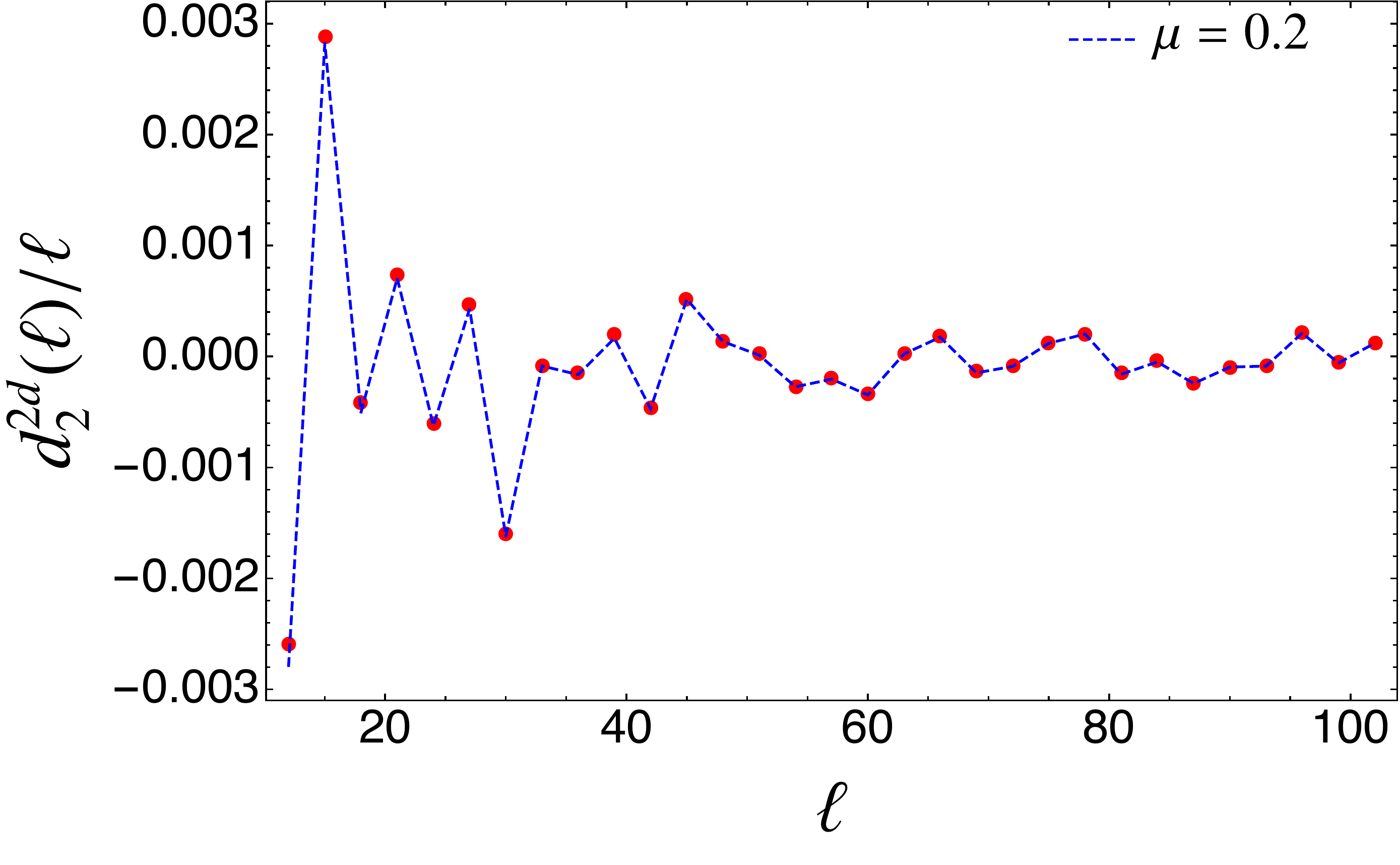}}
\subfigure
{\includegraphics[width=0.325\textwidth]{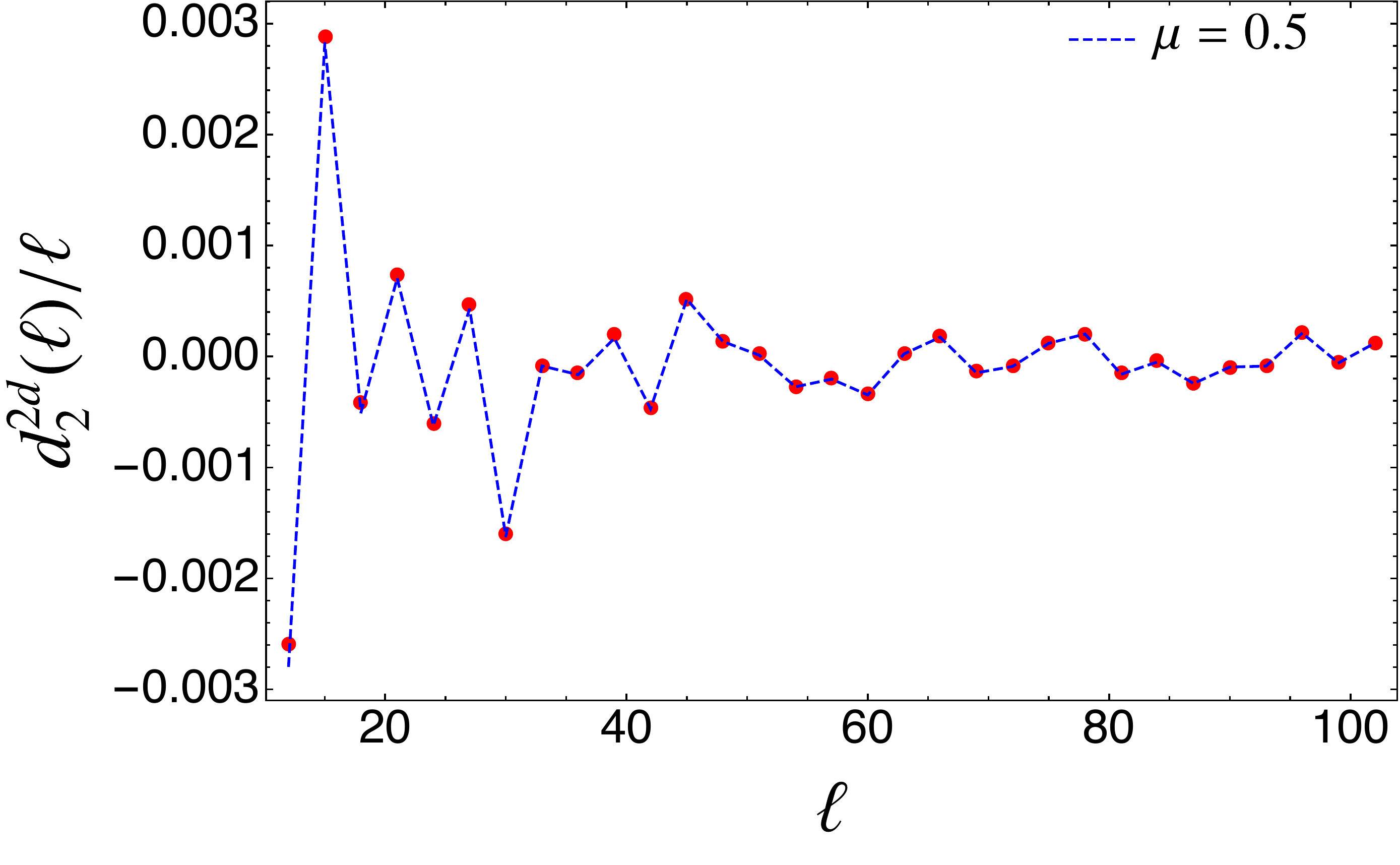}}
\caption{Subleading corrections to scaling $d^{2d}_2(\ell)$ for 
the $2d$ free fermionic model for different values of chemical potential $\mu=0, 0.2, 0.5$ in the three panels. 
The red symbols correspond to numerical values, while the dashed blue lines are the analytical prediction in Eq. (\ref{eq:oscsn}).}\label{fig:oscillations}
\end{figure}

In Figure~\ref{fig:prima}, subleading oscillating corrections for $n\neq 1$ are visible. 
These are easily understood as a consequence of the well studied unusual corrections to the scaling in 1d \cite{sublead,pc2010,ce-10,CMV-11}, which are present for 
generic bipartitions (see e.g. \cite{mint2}). 
Anyhow, in our special case we can exploit the dimensional reduction also to derive exact predictions for these corrections in 2d.    
In 1d, for the tight-binding model, they behave like \cite{pc2010,ce-10} 
\begin{equation}
d^{1d}_n(\ell)\equiv S^{1d}_n(\ell)-S_n^{1d,(0)}(\ell)=f_n \cos (2\ell k_F) |2\ell \sin k_F|^{-2/n}, \label{eq:oscsn1d}
\end{equation}
where $S_n^{1d,(0)}(\ell) $ is the (leading) prediction from the Fisher-Hartwig conjecture in Eq.~(\ref{eq:1dresults}), 
(which contains both the CFT prediction and the non universal constant term $\Upsilon_n$) 
and the amplitude is
\begin{equation}
\label{eq:fn}
f_n=\frac{2}{1-n}\frac{\Gamma^2((1+n^{-1})/2)}{\Gamma^2((1-n^{-1})/2)}.
\end{equation}
By dimensional reduction we have that each chain with Hamiltonian $H_{k^{(r)}_y}$ in Eq. \eqref{eq:Hamiltonian1}
has corrections given by Eq. \eqref{eq:oscsn1d} with the appropriate Fermi momentum. 
Summing over the contributions given by each mode in Eq. (\ref{eq:oscsn1d}), we obtain (for the case $N=\ell$ of the figure)
\begin{equation}
d^{2d}_n(\ell)\equiv S^{2d}_n(\ell)-S_n^{2d,(0)}(\ell)=f_n\sum_{j \in \Omega_{\mu}}\cos (2\ell k_j^F) |2\ell \sin k_j^F |^{-2/n},\label{eq:oscsn}
\end{equation}
where $S_n^{2d,(0)}$ stands for the leading terms in Eq. \eqref{eq:finiten}. 
The accuracy of these subleading corrections is tested against numerical data in Figure \ref{fig:oscillations} for three different values of the chemical potential, 
finding perfect agreement. Several frequencies corresponding to the various $k_j^F$ are clearly visible in the figure.

\subsection{Symmetry Resolved Entanglement Entropies}\label{sec:fermiSREE}
The same dimensional reduction technique can further be used to compute the symmetry resolved entanglement entropies. 
Indeed, from Eq. (\ref{eq:Hamiltonian}) the particle number ${Q}= \sum_{\mathbf{i}}c^{\dagger}_{\bf{i}}c_{\bf{i}}$ is a conserved $U (1)$ charge of the model
in arbitrary dimension.
The strategy is exactly as before: we consider a finite system in the transverse direction with PBC and so reduce to a one-dimensional problem for
the charged  moments and then, via Fourier transform, we get the symmetry resolved entropies.

\subsubsection{Charged moments.}
Because of the factorisation of the RDM \eqref{facRDM} and of the additivity of the conserved charge, we can rewrite
\begin{equation}\label{eq:fact}
\rho_A^n  e^{i {Q}_A \alpha}= \bigotimes_{r \in \Omega_{\mu}}\rho^{n}_{A,k_y^{(r)}}e^{i {Q}^{(r)}_A \alpha} \,,
\end{equation}
where ${Q}_A^{(r)}$ is the charge operator restricted to the $r$-th transverse mode.
This factorisation allows us to rewrite in terms of  the one-dimensional results for the charged moment 
\begin{equation}
 \label{eq:entropy}
 \log Z_n^{2d}(\alpha)=\sum_{r \in \Omega_{\mu}}\log Z^{1d}_{n,r}(\alpha),
 \end{equation}
and, using the explicit 1d result Eq. (\ref{eq:1d}), the  sum is performed as 
\begin{multline}
\label{eq:try2exact}
\log Z^{2d}_n(\alpha)\simeq i \bar{q} \alpha
-\left[\frac{1}{6}\left(n-\frac{1}{n} \right) +\frac{2}{n}\left(\frac{\alpha}{2\pi} \right)^2\right] \left( f_N(\mu)\log 2\ell  +  A_N(\mu)\right)N +Nf_N(\mu) \Upsilon(n,\alpha).
\end{multline}
The first term in Eq. (\ref{eq:try2exact}) is purely  imaginary and it is the average number of particle within $A$, for large $N$ explicitly given by 
\begin{equation}
\label{eq:approx}
\bar{q}= \frac{N \ell  }{\pi^2}\displaystyle \int_0^{\arccos(\mu-1)}dx\arccos (\mu-\cos x) .
\end{equation}
It is extensive in the subsystem volume ($N \ell$), as it should, and at half-filling, $\mu=0$, it reproduces the simple result $\bar{q}=N \ell/2$.

\begin{figure}
\centering
\subfigure
{\includegraphics[width=0.49\textwidth]{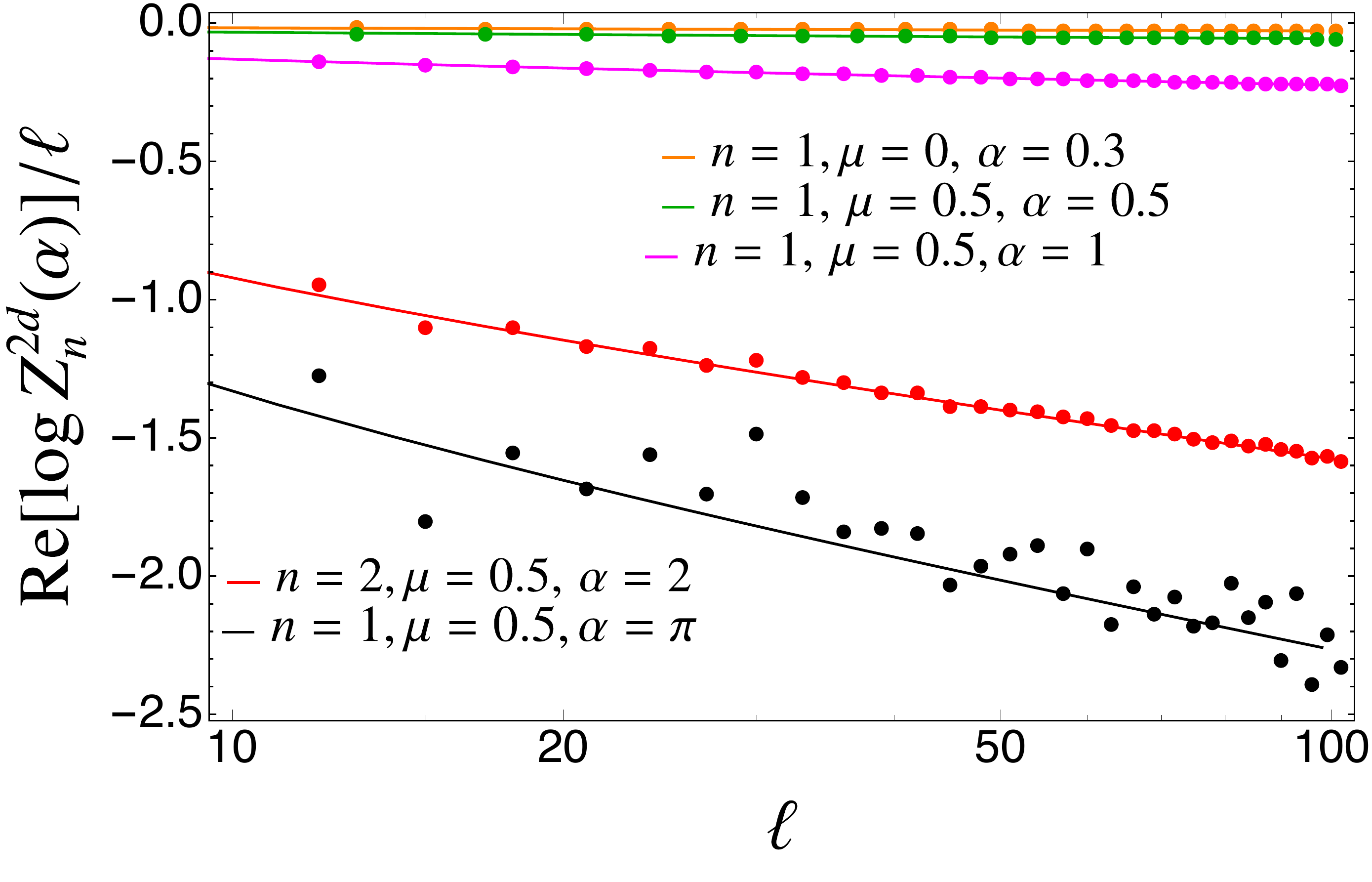}}
\subfigure
{\includegraphics[width=0.49\textwidth]{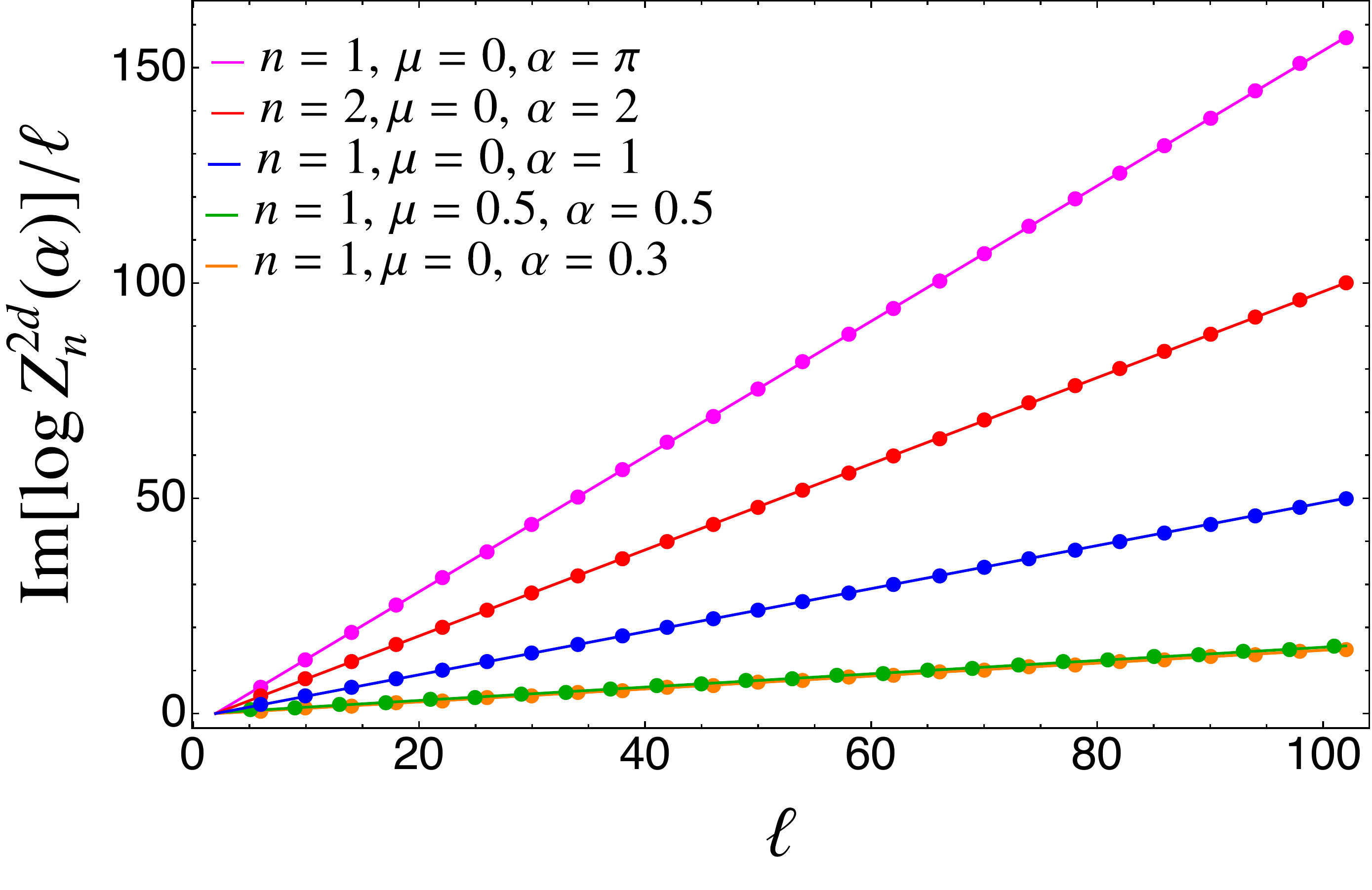}}
\caption{Leading scaling behaviour of the real and imaginary part of the charged moments $\log Z^{2d}_n(\alpha)$ in 2d free fermionic model 
for an infinite cylinder with transverse length  $N=\ell$, equal to the subsystem length in the longitudinal direction. 
The numerical results (symbols) for several values of $\alpha$ and $n$ are reported as function of $\ell$ for different $\mu$'s. 
Different colours represent different choices of the parameters $n, \alpha, \mu$. 
The corresponding analytic predictions (continuous lines), Eqs. (\ref{eq:try2exact}) and (\ref{eq:try2exactfs}), are also reported.}
\label{fig:imrezalphaF}
\end{figure}

In Eq.\,(\ref{eq:try2exact}), it is useful to write $\Upsilon(n,\alpha)$ as
\begin{equation}
\label{eq:simplifyupsilon}
\Upsilon(n,\alpha)=\Upsilon(n)+\gamma(n) \alpha^2 + \epsilon(n,\alpha), \qquad \epsilon(n,\alpha)=O(\alpha^4),
\end{equation}
where
\begin{equation}
\label{eq:upsilonridotta}
\gamma(n)=\frac{ni}{4}\displaystyle \int_{-\infty}^{\infty} dw [\tanh^3(\pi n w )-\tanh(\pi n w)]\log \dfrac{\Gamma(\frac{1}{2}+iw)}{\Gamma(\frac{1}{2}-iw)}.
\end{equation}
In Ref. \cite{riccarda} it has been shown that the quadratic approximation of Eq. (\ref{eq:simplifyupsilon}) is appropriate for many of applications since
$\epsilon(n,\alpha) \ll \gamma(n) \alpha^2$.  In particular, this approximation allows us for an explicit analytic computation of the symmetry resolved moments $\mathcal{Z}_n(q)$. 
Therefore, hereafter we will keep only the terms up to $O(\alpha^2)$ and we rewrite (\ref{eq:try2exact}) in the compact form as: 
\begin{equation}
\label{eq:compact}
\log Z^{2d}_n(\alpha)\simeq\log Z^{2d}_n(0)+i\bar{q}\alpha-\alpha^2( \mathcal{B}_n f_N(\mu)\log 2\ell +\mathcal{C}_n)N ,
\end{equation}
with 
\begin{equation}
\label{eq:istr}
\begin{split}
& \mathcal{B}_n=\dfrac{1}{2\pi^2n},\\
& \mathcal{C}_n=\dfrac{A_N(\mu)}{2\pi^2n}-f_N(\mu)\gamma(n).
\end{split}
\end{equation}

In Figure \ref{fig:imrezalphaF} we report the numerical data both for the real and the imaginary part of $\log Z^{2d}_n(\alpha)$ for different values of $n$ and $\alpha$.
Here the system is an infinite cylinder of circumference $\ell$ and the subsystem $A$ is again a periodic square strip with longitudinal length equal to $\ell$. Here and throughout this section, the values of $\mu$ and $\ell$ are chosen such that $\ell f_{\ell}(\mu)$ is an integer number. Moreover, when $\mu=0$, we focus on the case $\ell$ even.
The theoretical prediction in Eq. (\ref{eq:try2exact}) is also reported for comparison, showing that the analytical result correctly describes the data as long as $|\alpha| < \pi$
(as well known already in 1d, see e.g. \cite{riccarda,SREE2dG}). 
The oscillating corrections to the scaling become relevant when $\alpha$ moves close to $\pm \pi$. 
The reason is that some terms in the generalised Fisher-Hartwig approach become larger 
and Eq. (\ref{eq:1d}) is not a good approximation at the considered  intermediate values of $\ell$ \cite{SREE2dG,riccarda}. 
In the figure it is evident that these oscillations arise also for $n=1$, contrarily  to what happens for $\alpha=0$. 
\begin{figure}
\centering
\subfigure
{\includegraphics[width=0.335\textwidth]{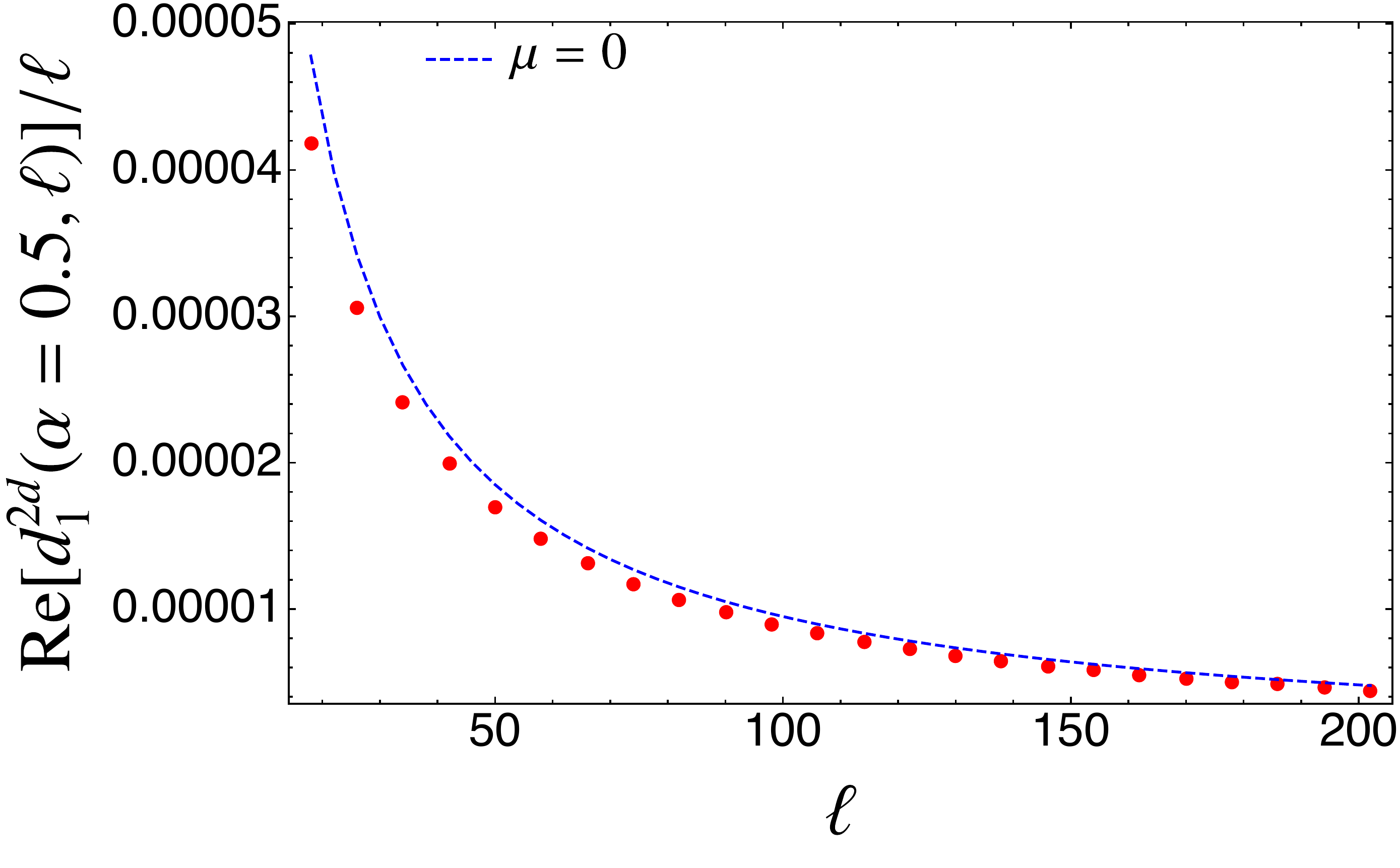}}
\subfigure
{\includegraphics[width=0.321\textwidth]{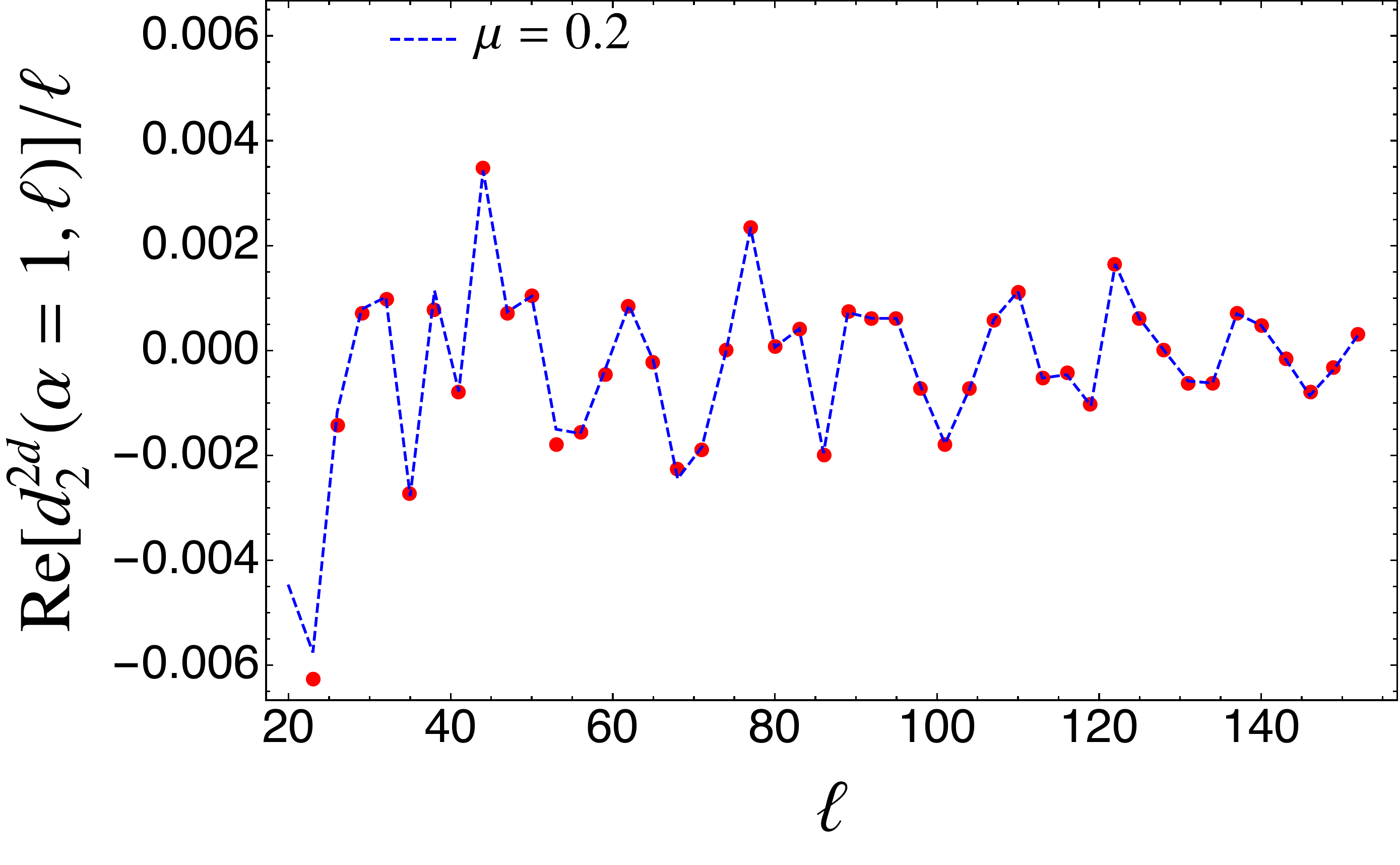}}
\subfigure
{\includegraphics[width=0.321\textwidth]{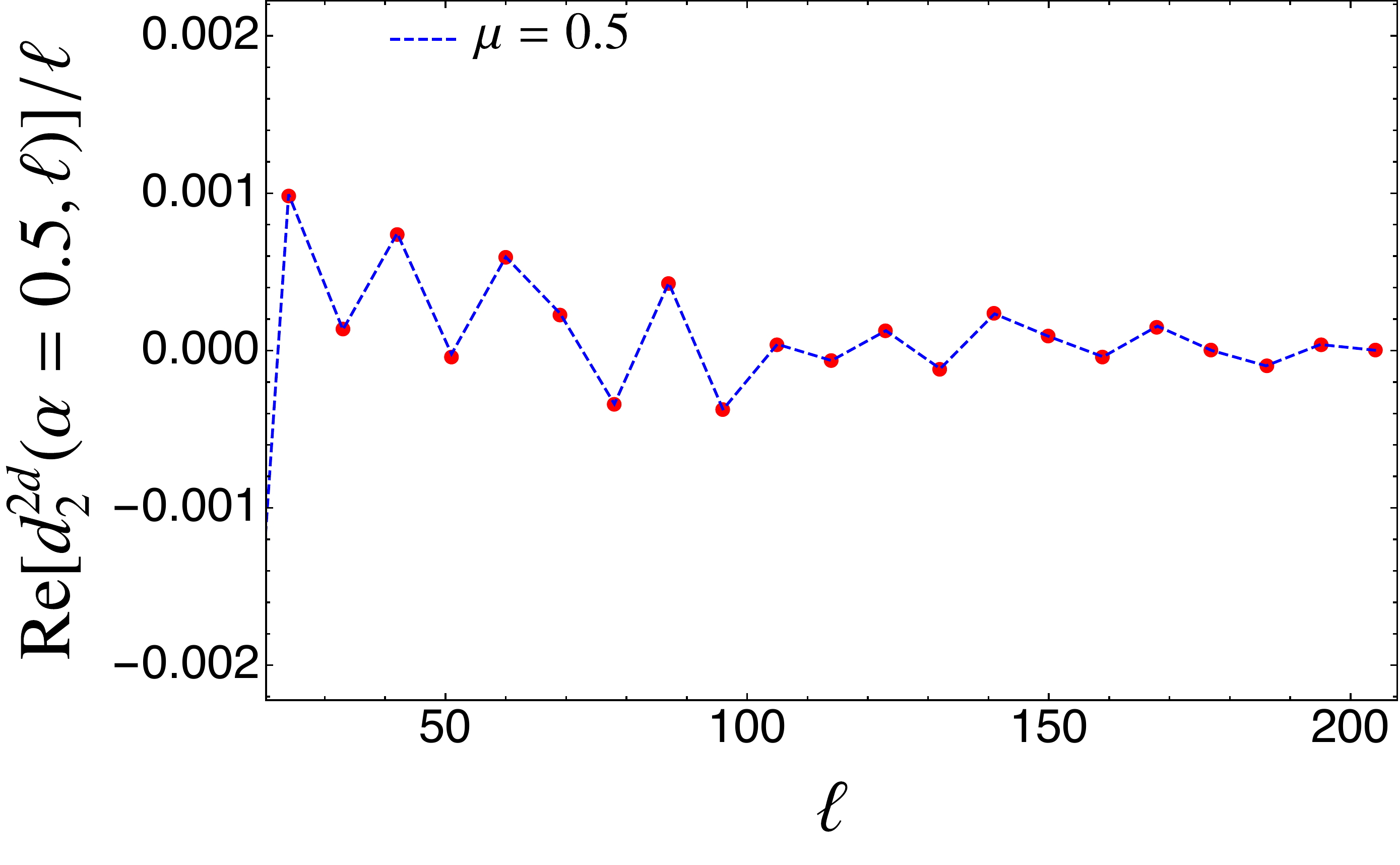}}
\subfigure
{\includegraphics[width=0.34\textwidth]{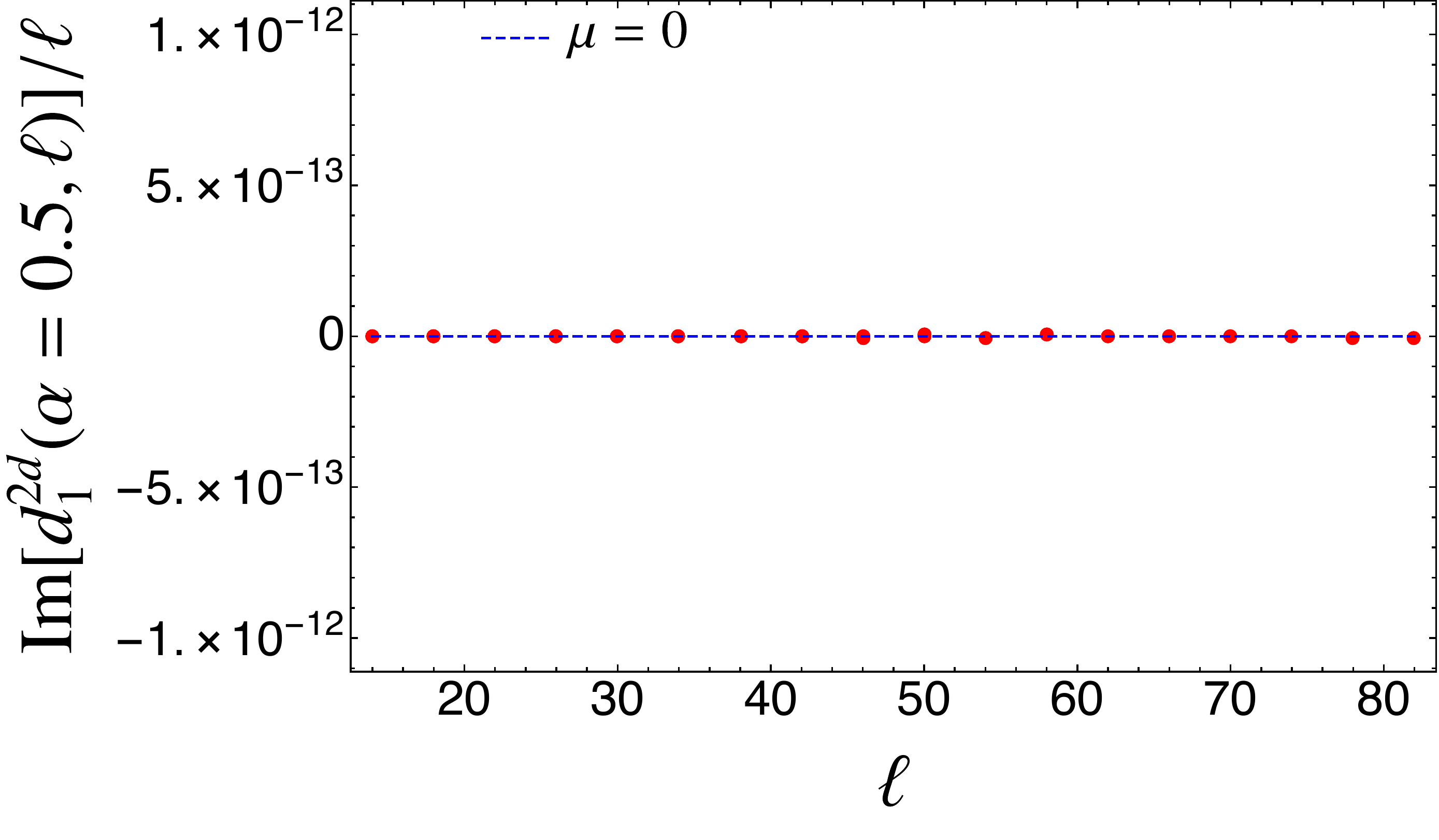}}
\subfigure
{\includegraphics[width=0.313\textwidth]{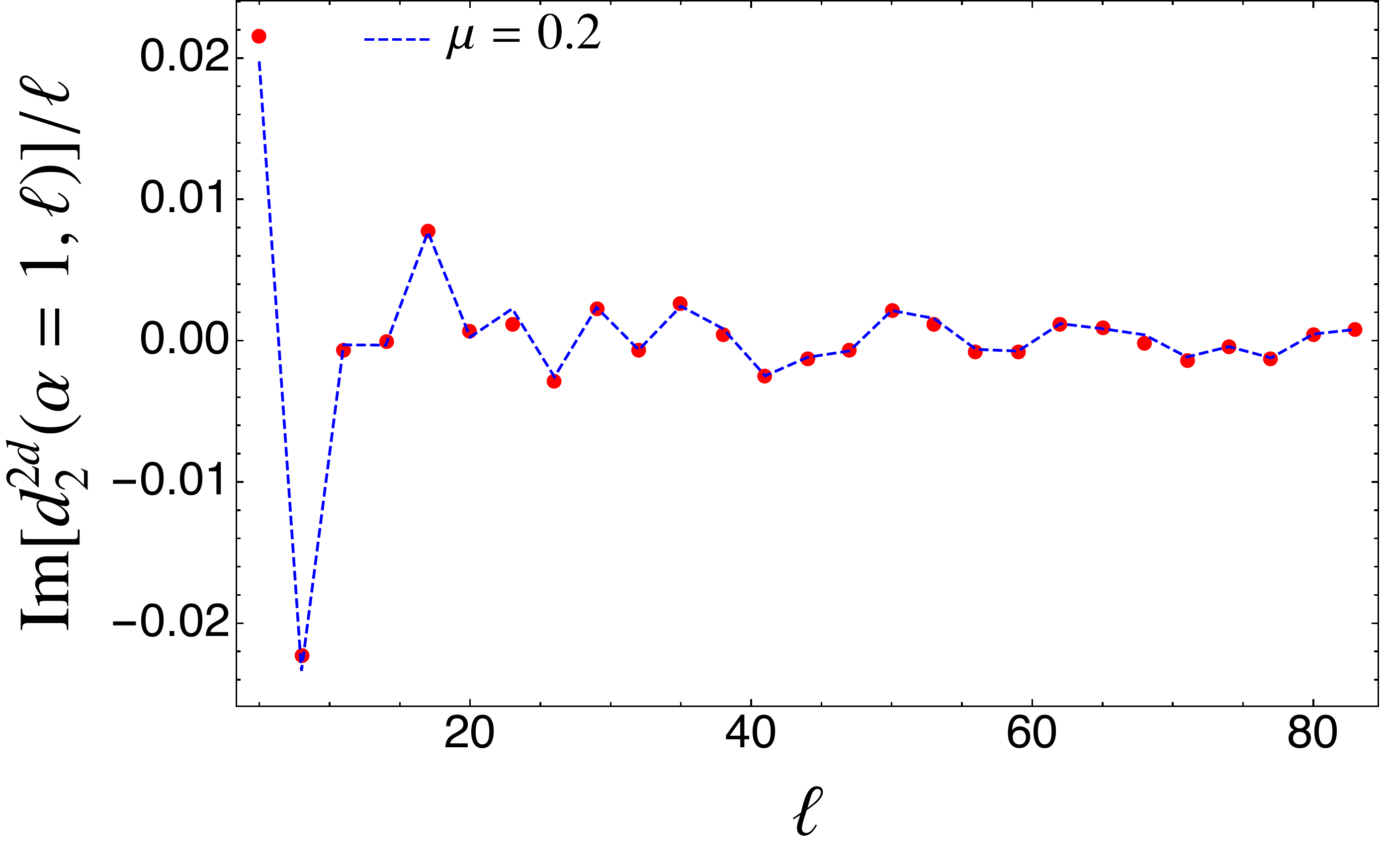}}
\subfigure
{\includegraphics[width=0.33\textwidth]{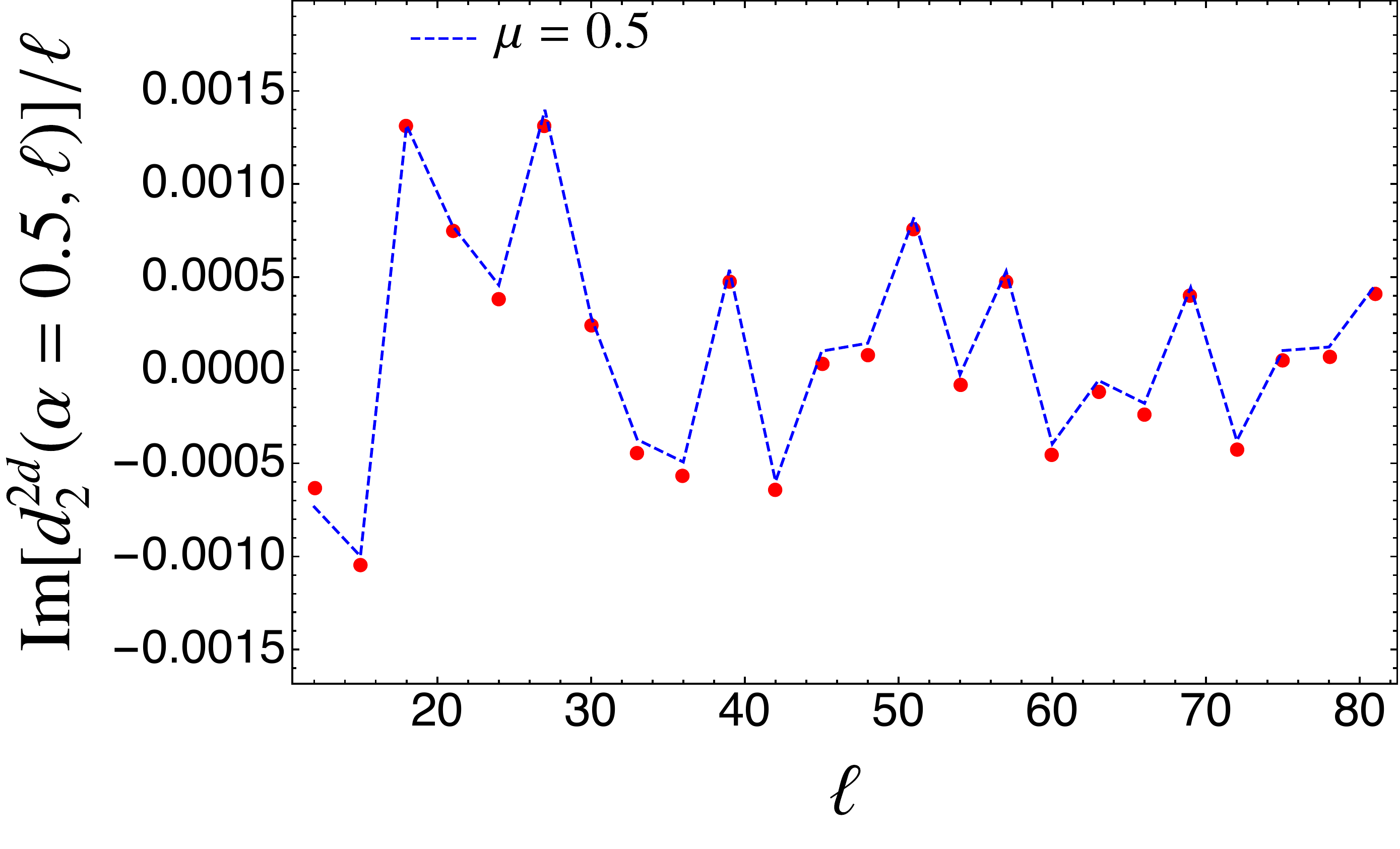}}
\caption{Corrections to scaling $d_n(\ell)$ (cf. Eq. \eqref{eq:osc}) for the 2d free fermionic model. 
Real and imaginary parts are reported in the upper and lower row, respectively, with the three columns corresponding to different values of 
$\mu$ and $\alpha$. The dashed blue lines are the analytical predictions in Eq. (\ref{eq:osc}).}
\label{fig:oscillationsalpha}
\end{figure}

We can also study the subleading oscillatory behaviour exploiting the one-dimensional results \cite{riccarda} (valid for $-\pi<\alpha<\pi$), i.e.,
\begin{multline}
\label{eq:osc1d}
d^{1d,j}_n(\alpha,\ell) \equiv \log Z^{1d}_n(\alpha)-\log Z_n^{1d,(0)}(\alpha) =\\ 
=e^{-2ik_j^F\ell}(2\ell \sin k_j^F)^{-\frac{2}{n}(1-\frac{\alpha}{\pi})}f^{(1)}_n(\alpha) +e^{2ik_j^F\ell}(2\ell \sin k_j^F)^{-\frac{2}{n}(1+\frac{\alpha}{\pi})}f^{(2)}_n(\alpha)+\dots, 
\end{multline}
where $Z_n^{1d,(0)}(\alpha) $ is the (leading) prediction of the generalised Fisher-Hartwig conjecture, Eq. (\ref{eq:1d}), and
\begin{equation}
\label{eq:fna}
f^{(1)}_n(\alpha)=\frac{\Gamma^2(\frac{1}{2}+\frac{1}{2n}-\frac{\alpha}{2\pi n})}{\Gamma^2(\frac{1}{2}-\frac{1}{2n}+\frac{\alpha}{2\pi n})}, \qquad f^{(2)}_n(\alpha)=\frac{\Gamma^2(\frac{1}{2}+\frac{1}{2n}+\frac{\alpha}{2\pi n})}{\Gamma^2(\frac{1}{2}-\frac{1}{2n}-\frac{\alpha}{2\pi n})}.
\end{equation}
In 2d,  the subleading oscillatory behaviour is easily obtained summing the contributions for each mode given by Eq. (\ref{eq:osc1d}), resulting  for $N=\ell$ in 
\begin{equation}
d^{2d}_n(\alpha,\ell)\equiv \log Z^{2d}_n(\alpha)-\log Z_n^{2d,(0)}(\alpha)
=\sum_{j \in \Omega_{\mu}} d^{1d,j}_n(\ell,\alpha), \label{eq:osc}
\end{equation}
which is compared with numerical data in Figure \ref{fig:oscillationsalpha}, finding perfect agreement. 
For $\mu=0$, the oscillatory behaviour of the imaginary part of the charged moments vanishes.

\begin{figure}[t]
\centering
\subfigure
{\includegraphics[width=0.49\textwidth]{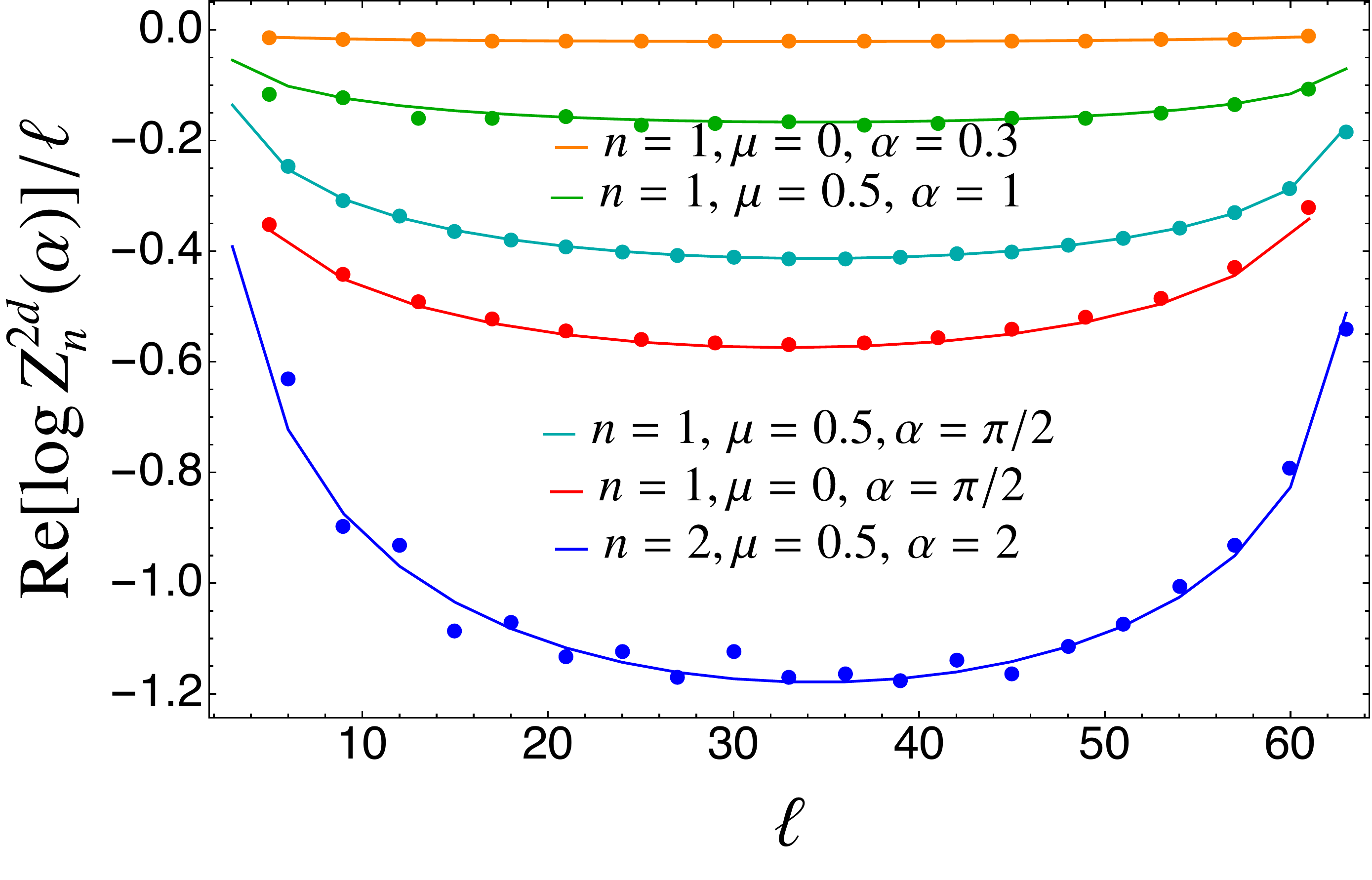}}
\subfigure
{\includegraphics[width=0.49\textwidth]{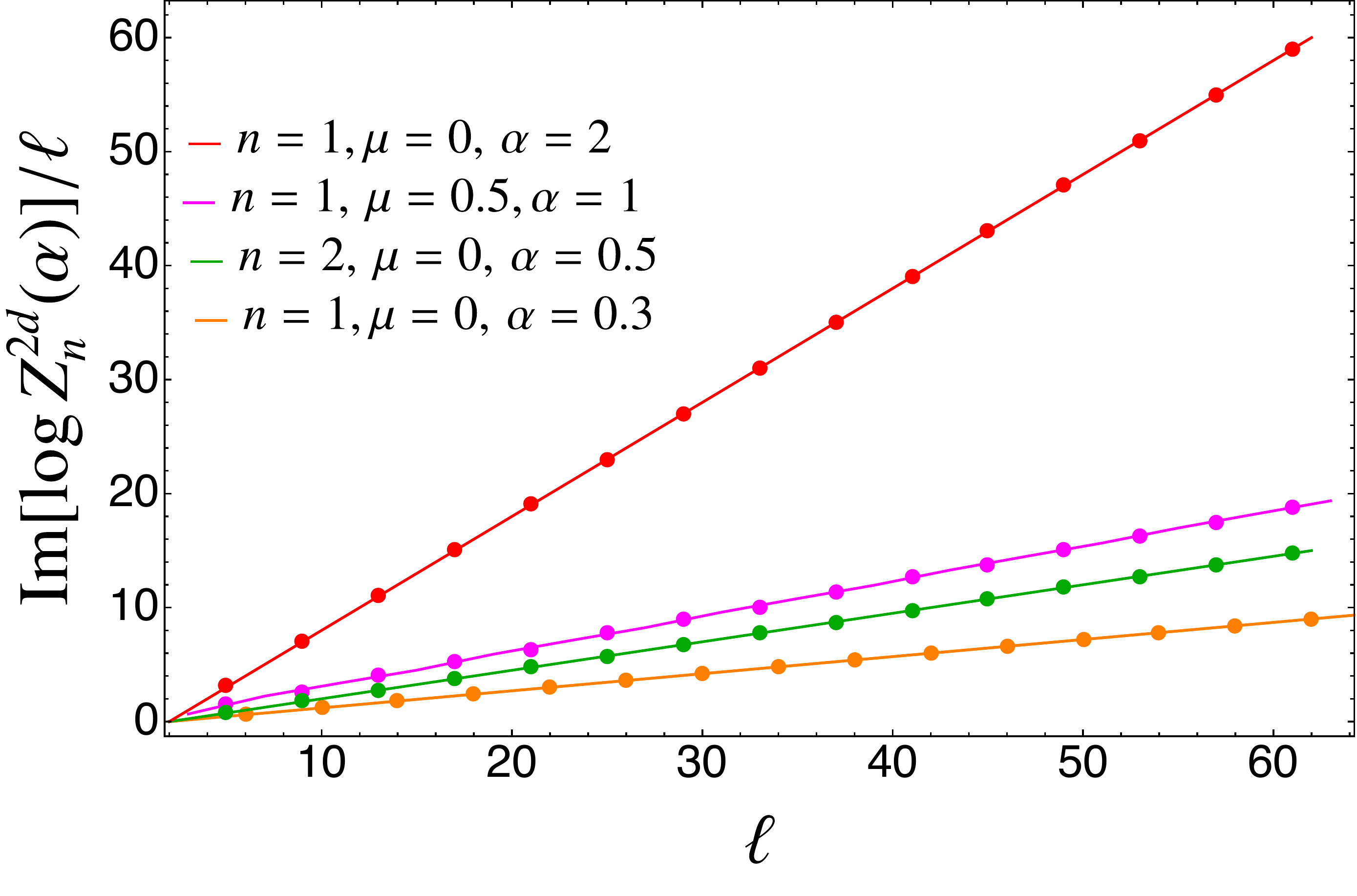}}
\caption{ Leading scaling behaviour of the real and imaginary part of the charged moments $\log Z^{2d}_n(\alpha)$ in 2d free fermionic model for a periodic system 
in both directions. The longitudinal circumference is $L=64$ while the transverse one is equal to $\ell$, the longitudinal subsystem length. 
The numerical results (symbols) for several values of $\alpha$ and $n$ are reported as function of $\ell$ for different $\mu$'s. 
Different colours represent different choices of the parameters $n, \alpha, \mu$. 
The corresponding analytic prediction, Eq. (\ref{eq:try2exactfs}), is also reported as continuous lines.}\label{fig:imrezalphaFS}
\end{figure}

A simple but interesting generalisation of the calculation we just presented concerns the geometry of a torus as depicted in the right of Figure \ref{fig:cartoon}.
The longitudinal size of the system is $L$. %
The charged moments are again obtained by summing up the contribution of the different transverse modes as  critical 1d chains, using the 
finite size form with the chord length.
Summing up the contributions of the  the transverse modes we get
\begin{multline}
\label{eq:try2exactfs}
\log Z^{2d}_n(\alpha)\simeq\\ \simeq 
i\alpha \bar{q}-\left[\frac{1}{6}\Big(n-\frac{1}{n} \Big) +\frac{2}{n}\left(\frac{\alpha}{2\pi} \right)^2\right] 
N\left[ f_N(\mu) \log \Big[\frac{2L}{\pi} \sin\Big(\frac{\pi \ell}{L} \Big)\Big] 
+ A_N(\mu)\right]+ N f_N(\mu)\Upsilon(n,\alpha).
\end{multline} 
The accuracy of this prediction is tested in Figure \ref{fig:imrezalphaFS} against exact numerical calculations. 

\subsubsection{Symmetry resolution.}
We now can compute the Fourier transform $\mathcal{Z}^{2d}_n(q)$ of the charged moments using the leading order terms of $Z_n^{2d}(\alpha)$ 
taking into account the effect of the non-universal pieces.
This Fourier transform is
\begin{equation}
\label{eq:Ftrasform}
\mathcal{Z}^{2d}_n(q) = \displaystyle \int_{-\pi}^{\pi} \dfrac{d\alpha}{2\pi}e^{-iq\alpha}Z^{2d}_n(\alpha)\simeq
Z_n^{2d}(0)\displaystyle \int_{-\pi}^{\pi} \dfrac{d\alpha}{2\pi}e^{-i(q-\overline{q})\alpha-\alpha^2 b_n},
\end{equation}
where the coefficient of the quadratic term is 
\begin{equation}
\label{eq:varianceR}
b_n= \mathcal{B}_n f_N(\mu) \log 2\ell +\mathcal{C}_n N.
\end{equation}
For large subsystem size $\ell$ and/or $N$, we can treat the integral by means of the saddle point approximation and use as domain of integration $[- \infty, + \infty]$, 
getting
\begin{equation}
\label{eq:gaussian}
\mathcal{Z}_n^{2d}(q)\simeq 
Z_n^{2d}(0) e^{-\frac{(q-\overline{q})^2}{4N(\mathcal{B}_n f_N(\mu)\log 2\ell +\mathcal{C}_n )}}\sqrt{\dfrac{1}{4\pi N(\mathcal{B}_n f_N(\mu)\log 2\ell +\mathcal{C}_n)}},
\end{equation}
where $Z_n^{2d}(\alpha)$ is given in Eq. \eqref{eq:try2exact} and we report it again for completeness in a coincise form for $\alpha=0$
\begin{equation}
Z_n^{2d}(0)= \Big((2\ell)^{f_N(\mu)} e^{A_{N}(\mu)} \Big)^{-\frac{1}{6}(n-\frac{1}{n}) N} 
e^{\Upsilon(n) N f_N(\mu)}. 
\end{equation}
In full analogy to the 1d case, the probability distribution functions given by these moments are still Gaussian with mean  $\bar{q}$ and variance 
that for large $\ell$ and $N$ grows as $\sqrt{N \log \ell}$.  
An equivalent result was already obtained in Refs. \cite{SREE2d,SREE2dG} for Fermi gases in arbitrary dimension using the Widom's conjecture. 
The novelty of this formula is an exact prediction for the coefficient $\mathcal{C}_n$ that renormalises the variance at order $O(\ell)$ and, as we will see, 
will play a crucial role for an accurate computation of the symmetry resolved entropies. 

Let us briefly discuss the terms that have been neglected in the derivation of Eq.~(\ref{eq:gaussian}) which are the same as in 1d \cite{riccarda}. 
The main approximation is to ignore  $\epsilon(n,\alpha)$ in Eq.~(\ref{eq:upsilonridotta}) which induces a correction going like $1/(N \log \ell)$. 
The subleading corrections to $Z_n^{2d}(\alpha)$ in Eq.~\eqref{eq:oscsn} only induce power-law corrections and are subdominant compared to one above. 
Finally  the corrections coming from having replaced the extremes of integration $\pm \pi$ with $\pm\infty$ are really small: 
they decay as $e^{- \pi^2 b_n}/b_n$, i.e., exponentially in $N$.

\begin{figure}
\centering
\subfigure
{\includegraphics[width=0.495\textwidth]{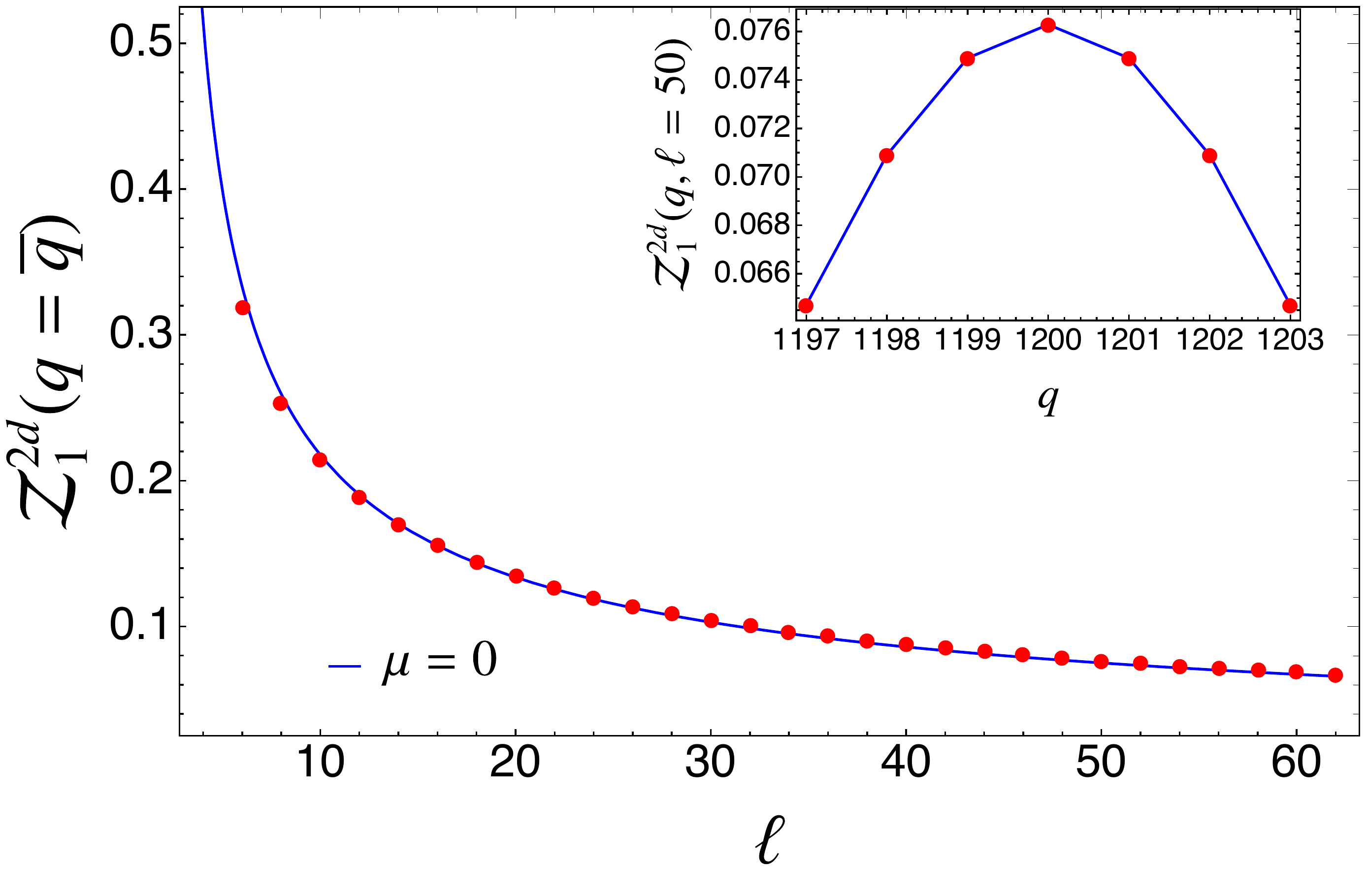}}
\subfigure
{\includegraphics[width=0.495\textwidth]{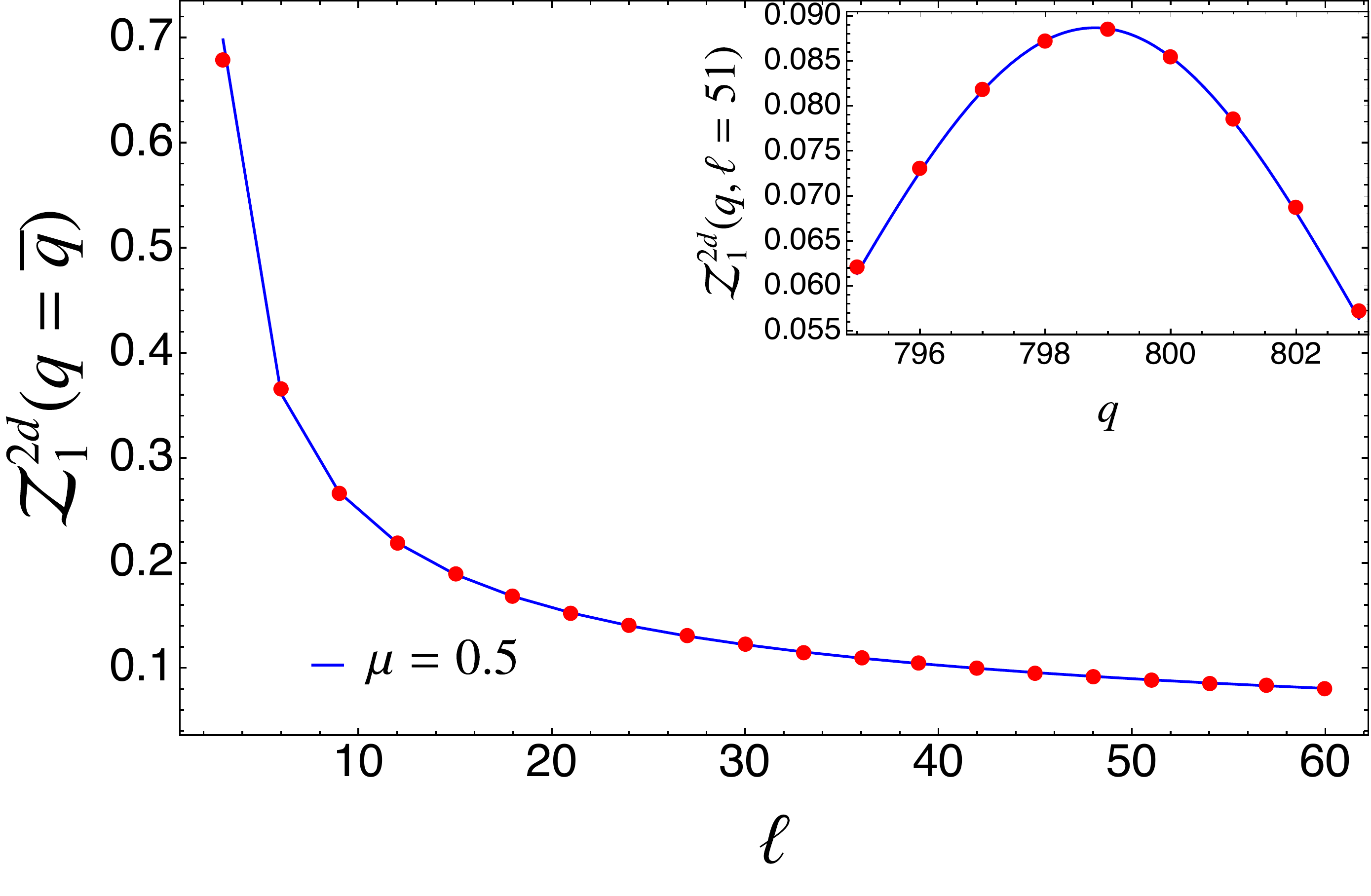}}
\caption{The probability $\mathcal{Z}_1^{2d}(q)$ for 2d free fermions with chemical potential $\mu=0$ (left) and $\mu=0.5$ (right). 
The red symbols are the numerical values and blue lines are the analytical prediction (\ref{eq:gaussian}).
In the main frame $\mathcal{Z}_1^{2d}(q=\bar{q})$ is shown as a function of $\ell$, whereas in the inset we fix $\ell$ and $\mathcal{Z}_1^{2d}(q)$ 
is plotted as a function of $q$. }
\label{fig:znqfermion}
\end{figure}

The accuracy of Eq.~\eqref{eq:gaussian} is checked for different values of $\mu$ in Figure \ref{fig:znqfermion} where we report the numerically calculated Fourier transforms
and the analytical prediction. 
It is evident from the data in the main frames and in the insets that both the $\ell$ and the $q$ dependence of $\mathcal{Z}_n(q)$ is perfectly captured by our approximation.

With these ingredients at our disposal, we are ready to compute the asymptotic behaviour of the symmetry resolved entanglement, given by 
\begin{equation}
\label{eq:def}
S^{2d}_n(q)=\dfrac{1}{1-n} \log \left[ \dfrac{\mathcal{Z}^{2d}_n(q)}{\mathcal{Z}^{2d}_1(q)^n}\right]\simeq \dfrac{1}{1-n}\log \dfrac{Z^{2d}_n(0)}{(Z^{2d}_1(0))^n}\dfrac{e^{-\frac{(q-\overline{q})^2}{4b_n}}}{e^{-\frac{n(q-\overline{q})^2}{4b_1}}}\dfrac{(4\pi b_n)^{-1/2}}{(4 \pi b_1)^{-n/2}}.
\end{equation}
After some simple algebra, we can write 
\begin{multline}
\label{eq:simmresolved}
S^{2d}_n(q)=S^{2d}_n -\dfrac{1}{2} \log \left( \dfrac{2N}{\pi} \left(  f_{\infty}(\mu)\log (2\ell)+ f_{\infty}(\mu) \delta_n + A_\infty(\mu)  \right)\right)+\dfrac{\log n}{2(1-n)}+\\ 
(q-\overline{q})^2\pi^4\dfrac{n}{1-n}\dfrac{ (\gamma(1)-n\gamma(n))}{N[f_{\infty(\mu) \log (2\ell) + f_{\infty}(\mu)\kappa_n} + A_\infty(\mu) ]^2}+\cdots,
\end{multline}
where $S^{2d}_n$ is the total R\'enyi entropy, 
\begin{equation}
\label{eq:deltan}
 \delta_n=-\dfrac{2\pi^2 n  (\gamma(n)-\gamma(1))}{1-n},
\end{equation}
and
\begin{equation}
\label{eq:kappan}
 \kappa_n=-\pi^2(\gamma(1)+n\gamma(n)).
\end{equation}
Eq.~\eqref{eq:simmresolved} provides the leading  behaviour for large $\ell$ and $N$ as well as the non-universal additive constants, 
and a $q$-dependent  subleading correction 
which scales as $N^{-1}(\log \ell)^{-2}$. Such correction provides the first term in the expansion for large $\ell$ and $N$ which depends on the symmetry sector. 
As in the corresponding 1d calculation \cite{riccarda}, it can be calculated from the subleading terms of the variance of $\mathcal{Z}^{2d}_n(q)$, 
in particular the additive non-universal constant ${\cal C}_n$ in Eq. \eqref{eq:istr}. 
So while few leading terms satisfy the equipartition of entanglement, we can precisely identify the first term that breaks it. 
Taking now the limit for $n\rightarrow 1$ of \eqref{eq:simmresolved}, we get the von Neumann entropy
\begin{multline}
\label{eq:vN}
S^{2d}_{1}(q)=S^{2d}_{1} -\dfrac{1}{2} \log \left( \dfrac{2N}{\pi} \left(f_{\infty}(\mu)\log(2\ell) +  f_{\infty}(\mu)\delta_1 + A_\infty(\mu)  \right)\right)-\dfrac{1}{2}+ \\ 
  +(q-\overline{q})^2\pi^4\dfrac{ (\gamma(1)+\gamma'(1))}{N[f_{\infty}(\mu)\log (2\ell) +f_{\infty}(\mu)\kappa_1+ A_\infty(\mu)]^2}+\cdots
\end{multline}

\begin{figure}
\centering
\subfigure
{\includegraphics[width=0.485\textwidth]{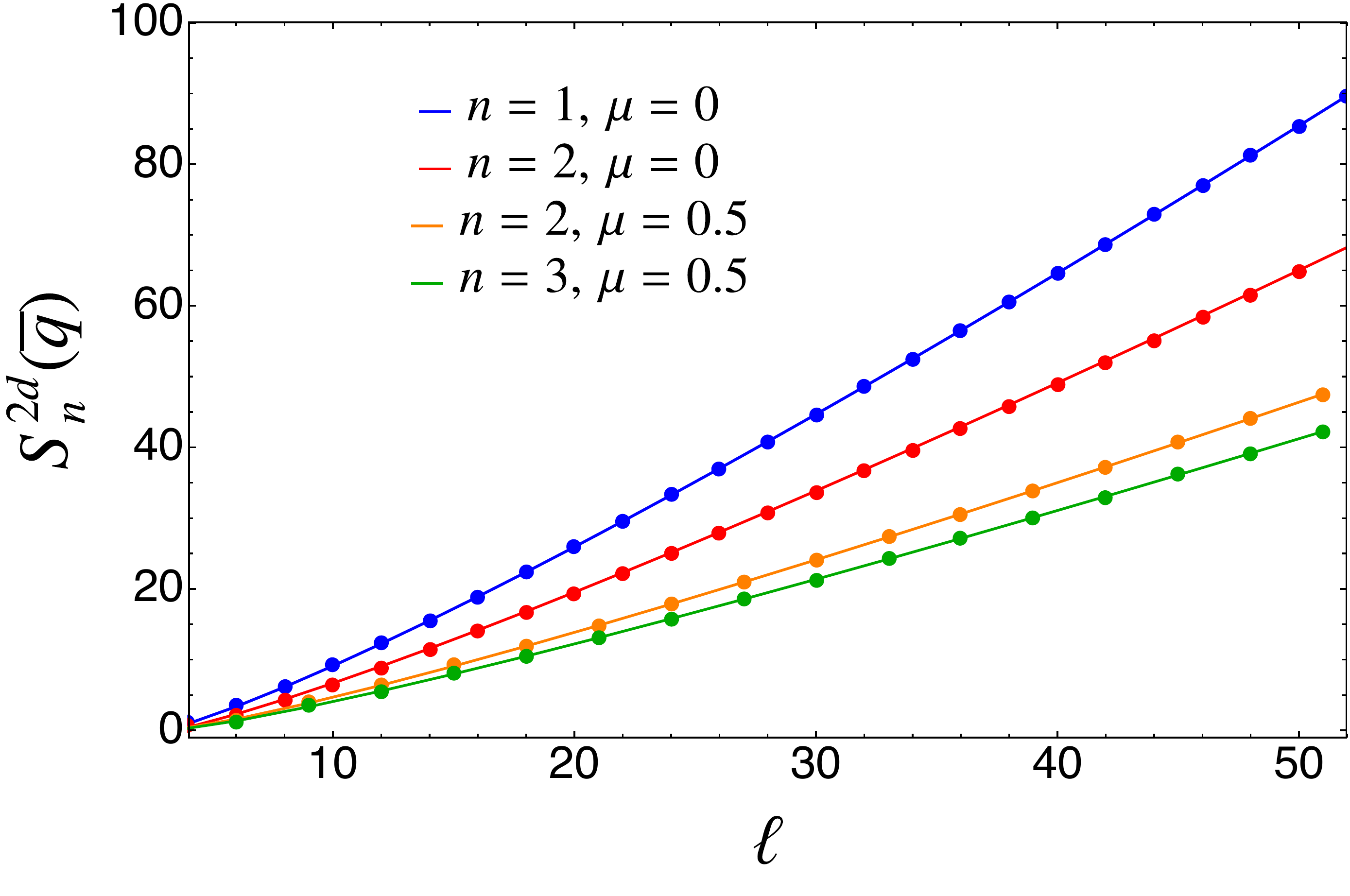}}
\subfigure
{\includegraphics[width=0.485\textwidth]{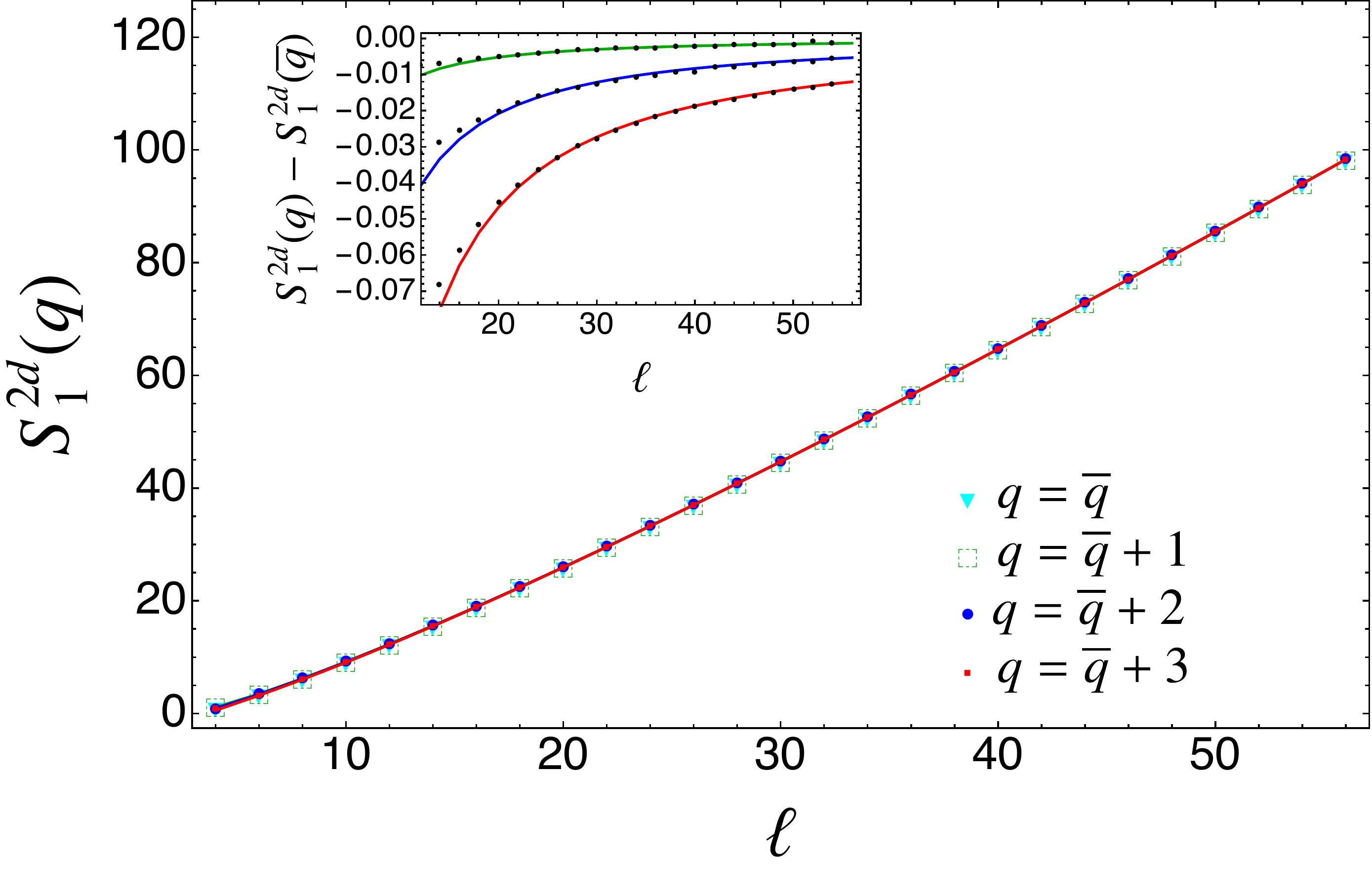}}
\caption{ Symmetry resolved R\'enyi entanglement entropies $S_n^{2d} (q)$ of 2d free fermions for $n=1,2,3$ and different values of $\mu$. We fix the transverse direction $N=\ell$, equal to the length of the subsystem in the longitudinal direction.
In the left panel the numerical data (symbols) of 2d free fermions for $q = \bar{q}$ are compared with the theoretical predictions of Eqs. (\ref{eq:simmresolved}) and (\ref{eq:vN}). 
In the right panel we show four values of $q$ (namely $q-\bar q=0,1,2,3$).
The data are almost coinciding on this scale, so in the inset we report their difference which is perfectly captured by the theoretical prediction.
}
\label{fig:symmresfermions}
\end{figure}

These predictions for the symmetry resolved entanglement are compared with the numerical data in Figure \ref{fig:symmresfermions}.
In the left panel we consider the scaling with $\ell$ of $S_n(\bar q)$ and it is evident that the numerical data perfectly match with the theoretical prediction 
in Eqs. (\ref{eq:simmresolved}) and (\ref{eq:vN}). 
The corrections in $(q-\bar{q})$ are suppressed as $1/(N(\log\ell)^2)$ and the curves in the right panel  seem to be on top of each other on the scale of the plot.
In order to appreciate their distance, in the inset we report the differences with $S_n(\bar q)$ (focusing on $n=1$)
and we show that they are well described by our prediction. The agreement is excellent even for relatively small values of $\ell$ and $N$ of the order of $20$.

In the one-dimensional case, the corrections of order $\log(\log \ell)$ coming from the symmetry resolved entanglement 
exactly cancel in the total entanglement entropy, when summing to the fluctuation entanglement as in Eq.~\eqref{eq:SvN}. 
The same occurs also in 2d. In fact,  the fluctuation entanglement in our case is given by 
\begin{equation}
\label{eq:fluct}
S^{2d,f}=-\displaystyle \int_q \mathcal{Z}_1(q) \log \mathcal{Z}_1(q) \simeq \frac{1}{2} (1+\log 4\pi b_1)= \frac{1}{2}+\frac{1}{2} \log \left( \frac{2}{\pi} N f_{\infty}(\mu)\log \ell \right)+O((\log \ell)^{-1}).
\end{equation}
From this equation, it is clear that the first two leading terms in $S^{2d,f}$ cancel exactly with the corresponding ones in the symmetry resolved entanglement in Eq. (\ref{eq:vN}).

\begin{figure}
\centering
\subfigure
{\includegraphics[width=0.495\textwidth]{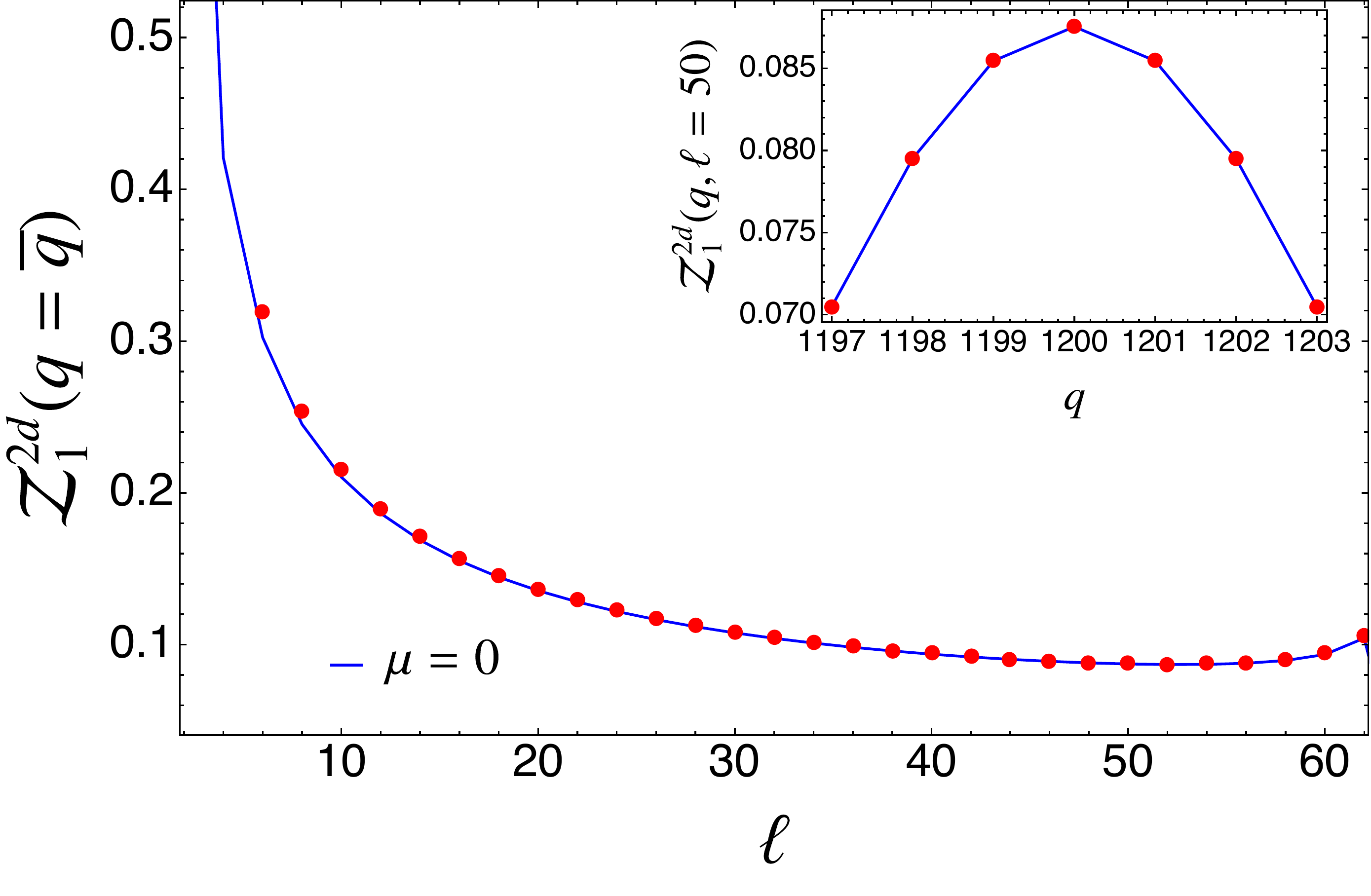}}
\subfigure
{\includegraphics[width=0.495\textwidth]{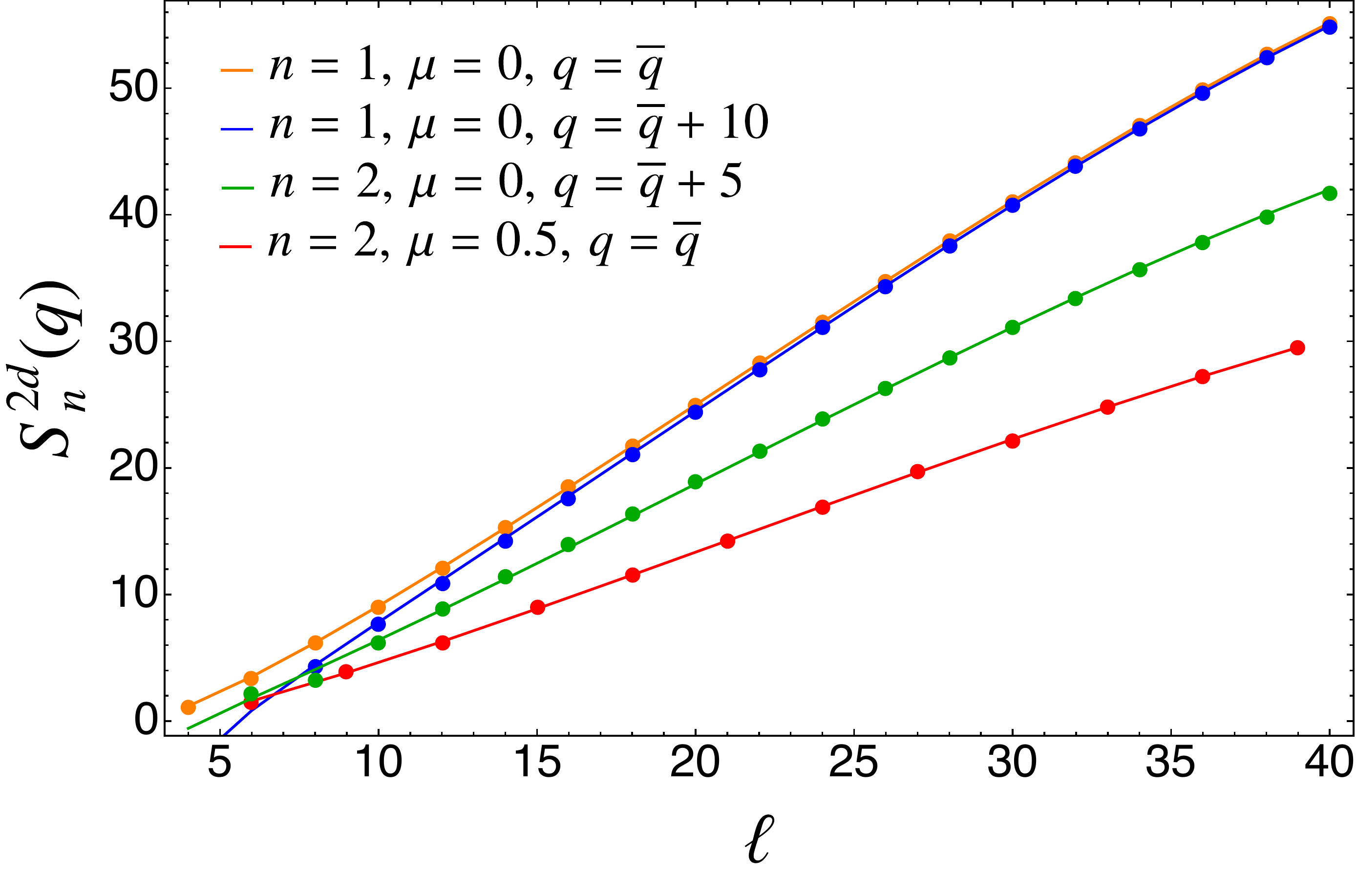}}
\caption{Left panel: $\mathcal{Z}_1^{2d}(q=\bar{q})$ of 2d free fermions for different values of $\mu$ and finite torus of longitudinal length $L=64$. 
We fix the transverse size $N$ equal to the subsystem length $\ell$. 
Red symbols correspond to numerical data. The blue lines show the analytical prediction (\ref{eq:fourierfs}). 
The inset shows $\mathcal{Z}_1^{2d}(q)$ for $\ell=10$ as a function of $q$. 
Right panel: Symmetry resolved R\'enyi entanglement entropies $S_n^{2d} (q)$ of 2d free fermions for $n=1,2$ and different $\mu$'s and $q$: 
the numerical data (symbols) are compared with the theoretical prediction (continuous lines) of Eq. (\ref{eq:simmresolvedfs}). 
As $(q-\bar{q})$ increases, the neglected terms become more relevant, therefore the agreement between numerics and analytical predictions worsens.}
\label{fig:znqfermionfs}
\end{figure}

We quickly discuss now what happens in the case of the torus geometry:  we only report the final results, since the calculations are only a slight modification of the previous ones. The Fourier transform of  Eq. (\ref{eq:try2exactfs}) is 
\begin{equation}
\label{eq:fourierfs}
\mathcal{Z}_n(q) \simeq Z_n^{2d}(0) \frac{e^{-\frac{(q-\overline{q})^2}{4N\left[\frac{1}{2\pi^2 n} \left(f_N(\mu)  \log \left( \frac{2L}{\pi} \sin\left( \frac{\pi \ell}{L} \right) \right) + A_N(\mu)\right)-f_N(\mu)\gamma(n)\right] }}}{\sqrt{4N\pi\left[\frac{1}{2\pi^2 n} \left(f_N(\mu)\log \left( \frac{2L}{\pi} \sin\left( \frac{\pi \ell}{L} \right) \right) + A_N(\mu)\right)-f_N(\mu)\gamma(n)\right]}}
\end{equation}
where 
\begin{equation}
\begin{split}
& Z_n^{2d}(0)=e^{\Upsilon(n) N f_N(\mu) } \left(e^{A_{N}(\mu) }\left(\frac{2L}{\pi}\sin(\frac{\pi \ell}{L})\right)^{f_N(\mu)}\right)^{-\frac{1}{6}(n-\frac{1}{n})N}.
\end{split}
\end{equation}
The symmetry resolved entropies are then easily worked out as 
\begin{multline}
\label{eq:simmresolvedfs}
S^{2d}_n(q)=S^{2d}_n -\dfrac{1}{2} \log \left[ \dfrac{2N}{\pi} \left(f_{\infty}(\mu) \log\left(\frac{2L}{\pi}\sin(\frac{\pi \ell}{L})\right) +f_{\infty}(\mu)\delta_n +A_{\infty}(\mu)  \right)\right]+\dfrac{\log n}{2(1-n)}+\\ 
(q-\overline{q})^2\pi^4\dfrac{n}{1-n}\dfrac{ (\gamma(1)-n\gamma(n))}{N[f_{\infty}(\mu)\log\left(\frac{2L}{\pi}\sin(\frac{\pi \ell}{L})\right) +f_{\infty}(\mu)\kappa_n +A_{\infty}(\mu)]^2}+\cdots.
\end{multline}
These results are tested with numerics in Figure (\ref{fig:znqfermionfs}). 

\section{Two-dimensional  Free Bosons} \label{sec:bosons}

In this section we consider the entanglement entropy and its partition into the different charge sectors for a lattice discretisation of the complex Klein--Gordon theory, 
namely  coupled complex harmonic oscillators on a two-dimensional square  lattice. 
Here we will apply the same strategy of the previous section to recast, once again, our problem into the sum of uncoupled one-dimensional chains.

\subsection{R\'enyi and Entanglement Entropies}\label{sec:bosonsREE}
Let us examine a two-dimensional system of $L\times N$ real coupled oscillators, where $L$ and $N$ are the lengths along the $x$-- (longitudinal) and $y$--direction 
(transverse), respectively (see Figure \ref{fig:cartoon}).
The 2d Hamiltonian describing a real 2d square lattice of harmonic oscillators is 
\begin{equation}
\label{eq:Hamiltonian2db}
H_{B}=\dfrac{1}{2}\displaystyle  \sum_{x=1}^{L}\sum_{y=1}^{N}\left[ p^2_{x,y}+\omega_0^2 q^2_{x,y}+{\kappa}_x(q_{x+1,y}-q_{x,y})^2+{\kappa}_y(q_{x,y+1}-q_{x,y})^2 \right],
\end{equation}
where $q_{x,y}, p_{x,y}$ and $\omega_0$ are coordinate, momentum and self-frequency of the oscillator at site $(x,y)$ while $\kappa_x$ and $\kappa_y$ are the 
nearest-neighbour couplings.  As in the one-dimensional case, the 2d lattice of complex oscillators is 
\begin{equation}
\label{eq:complex2d}
H_{CB}(p^{(1)}+ip^{(2)},q^{(1)}+iq^{(2)})= H_{B}(p^{(1)},q^{(1)})+H_{B}(p^{(2)},q^{(2)}).
\end{equation}
If we define $\mathbf{p}=\frac{p^{(1)}+ip^{(2)}}{\sqrt{2}}$ and  $\mathbf{q}=\frac{q^{(1)}+iq^{(2)}}{\sqrt{2}}$, Eq. (\ref{eq:complex2d}) becomes
\begin{equation}
\label{eq:complex2d1}
\begin{split}
H_{CB}=&\displaystyle  \sum_{x=1}^{L}\sum_{y=1}^{N}\left[\mathbf{p}^{\dagger}_{x,y}\mathbf{p}_{x,y}+\omega_0^2 \mathbf{q}^{\dagger}_{x,y}\mathbf{q}_{x,y}+\right.\\
&\left. +\kappa_x (\mathbf{q}_{x+1,y}-\mathbf{q}_{x,y})^{\dagger}(\mathbf{q}_{x+1,y}-\mathbf{q}_{x,y}) + \kappa_y (\mathbf{q}_{x,y+1}-\mathbf{q}_{x,y})^{\dagger}(\mathbf{q}_{x,y+1}-\mathbf{q}_{x,y})  \right].
\end{split}
\end{equation}
Imposing PBC's along the $y$-direction, we can exploit the translational invariance and use Fourier transform in the transverse direction, to get the mixed 
space-momentum representation
\begin{equation}
\mathbf{q}_{x,y}=\dfrac{1}{\sqrt{N}}\displaystyle \sum_{r=0}^{N-1}\tilde{q}_{x,r}e^{2\pi i r y/N},
\end{equation}
and similarly for $\mathbf{p}_{x,y}$. 
We set $\kappa_y=\kappa_x=1$ to shorten the notation (indeed they can be absorbed by a canonical transformation).
The Hamiltonian (\ref{eq:complex2d}) then becomes
\begin{equation}
\label{eq:notokay}
H_{CB}=\displaystyle  \sum_{x=1}^{L}\sum_{r=0}^{N-1} \tilde{p}^{\dagger}_{x,r} \tilde{p}_{x,r}+\omega^2_r \tilde{q}^{\dagger}_{x,r} \tilde{q}_{x,r}+(\tilde{q}_{x+1,r}-\tilde{q}_{x,r})^{\dagger}(\tilde{q}_{x+1,r}-\tilde{q}_{x,r}).
\end{equation}
where
\begin{equation}
\label{eq:freqc}
\omega_r^2=\omega_0^2+4\sin^2\frac{\pi r}{N}.
\end{equation}
Because of the additivity of the independent transverse chain modes in Eq. (\ref{eq:notokay}), the entanglement entropy can be computed by using the 1d results, 
as we did for free fermions.

As already discussed in section (\ref{sec:mainREE}), the ground-state reduced density matrices of each 1d chain is obtained by means of corner transfer matrices,
for the bipartition of the infinite chain in two halves. 
Therefore, the entanglement spectrum associated to the $r$-mode/chain along the $y$ direction is given by Eq. \eqref{HCTM}, specialised to the 
frequency $\omega_r$, i.e., the eigenvalues of the entanglement Hamiltonian are now given by
\begin{equation}
\label{eq:eig}
\epsilon^{(r)}_j=\epsilon_r (2j+1),
\end{equation}
where the energy levels are
\begin{equation}
\label{eq:details}
\epsilon_r=\dfrac{\pi I(\sqrt{1-\kappa_r^2})}{I(\kappa_r)}, \qquad \kappa_r=\frac{1}{2}(2+\omega_r^2-\omega_r \sqrt{4+\omega_r^2}).
\end{equation}
The parameter $\kappa_r$ is obtained by solving the equation $\omega^2_r={(1-\kappa_r)^2}/{\kappa_r}$.
 
In our semi infinite strip, each mode gives a contribution to the entanglement entropy which can be computed through the CTM approach. Since they are independent, such contributions simply add up leading to
\begin{equation}
\label{eq:Sbos}
S_n^{2d} =\frac{2}{1-n}\displaystyle \sum_{r=0}^{N-1}  \sum_{j=0}^{\infty} \left( n\log [1-e^{-(2j+1) \epsilon_r}]-\log [1-e^{-(2j+1)n\epsilon_r}]\right).
\end{equation}
This result is valid for arbitrary integer $N$. We can now take the limit of large transverse direction. 
The sum becomes an integral in $\zeta=r/N$, we can write ($\epsilon_r \to \epsilon(\mathrm{\zeta})$)
\begin{equation}\label{eq:SvNreni}
S_n^{2d} =\frac{2 N}{1-n} \displaystyle \int_0^1\, d\zeta\sum_{j=0}^{\infty} \left(  n\log [1-e^{- (2j+1)\epsilon(\zeta)}]-\log [1-e^{-(2j+1)n\epsilon(\zeta)}] \right),
\end{equation}
and in the limit $n \to 1$
\begin{equation}\label{eq:SvNbosons}
S_1^{2d} =2 N \displaystyle \int_0^1\, d\zeta\sum_{j=0}^{\infty} \left( \frac{(2j+1)\epsilon(\zeta)}{e^{(2j+1)\epsilon(\zeta) }-1}-\log [1-e^{-(2j+1)\epsilon(\zeta)}] \right).
\end{equation}

\begin{figure}
\centering
\subfigure
{\includegraphics[width=0.325\textwidth]{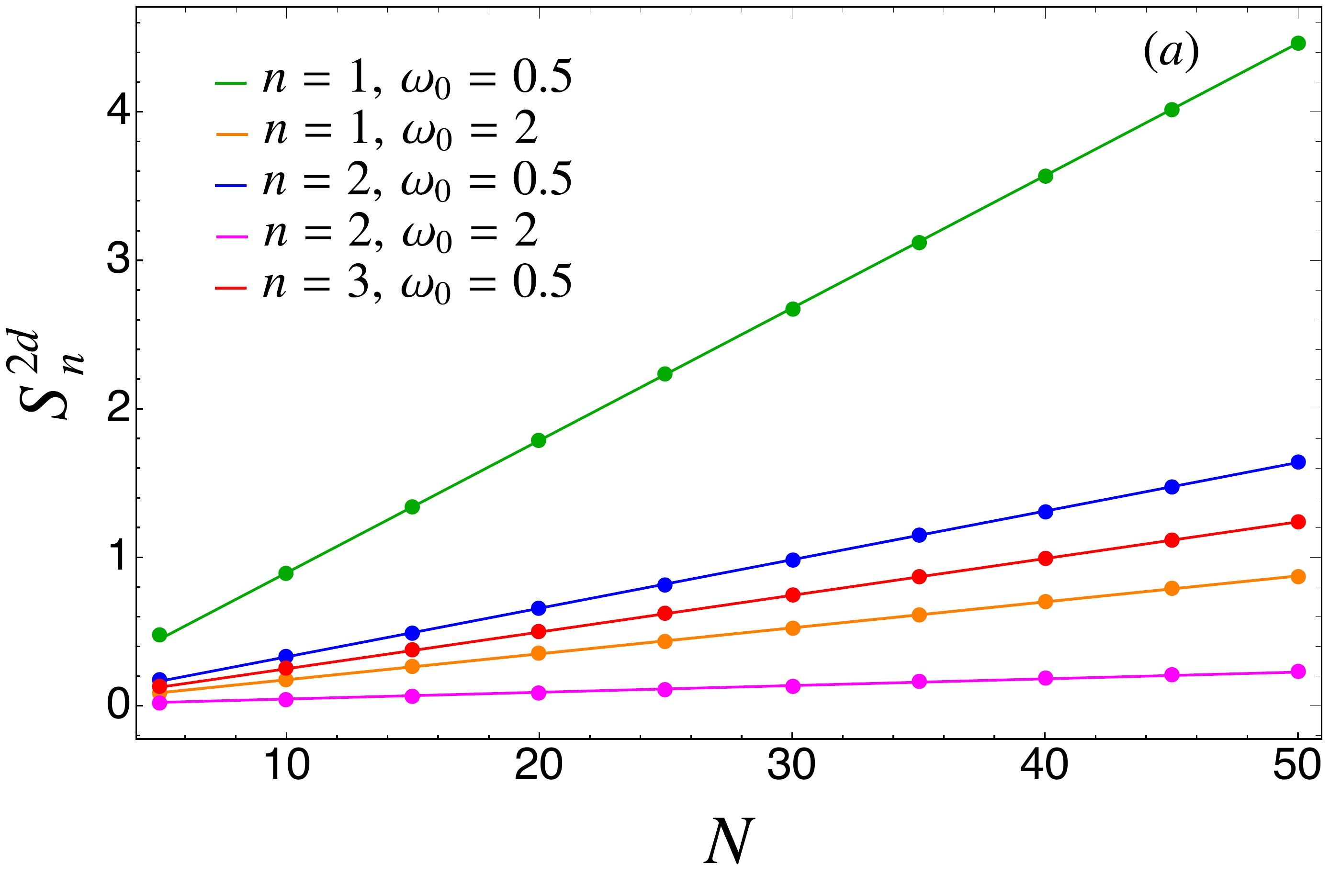}}
\subfigure
{\includegraphics[width=0.325\textwidth]{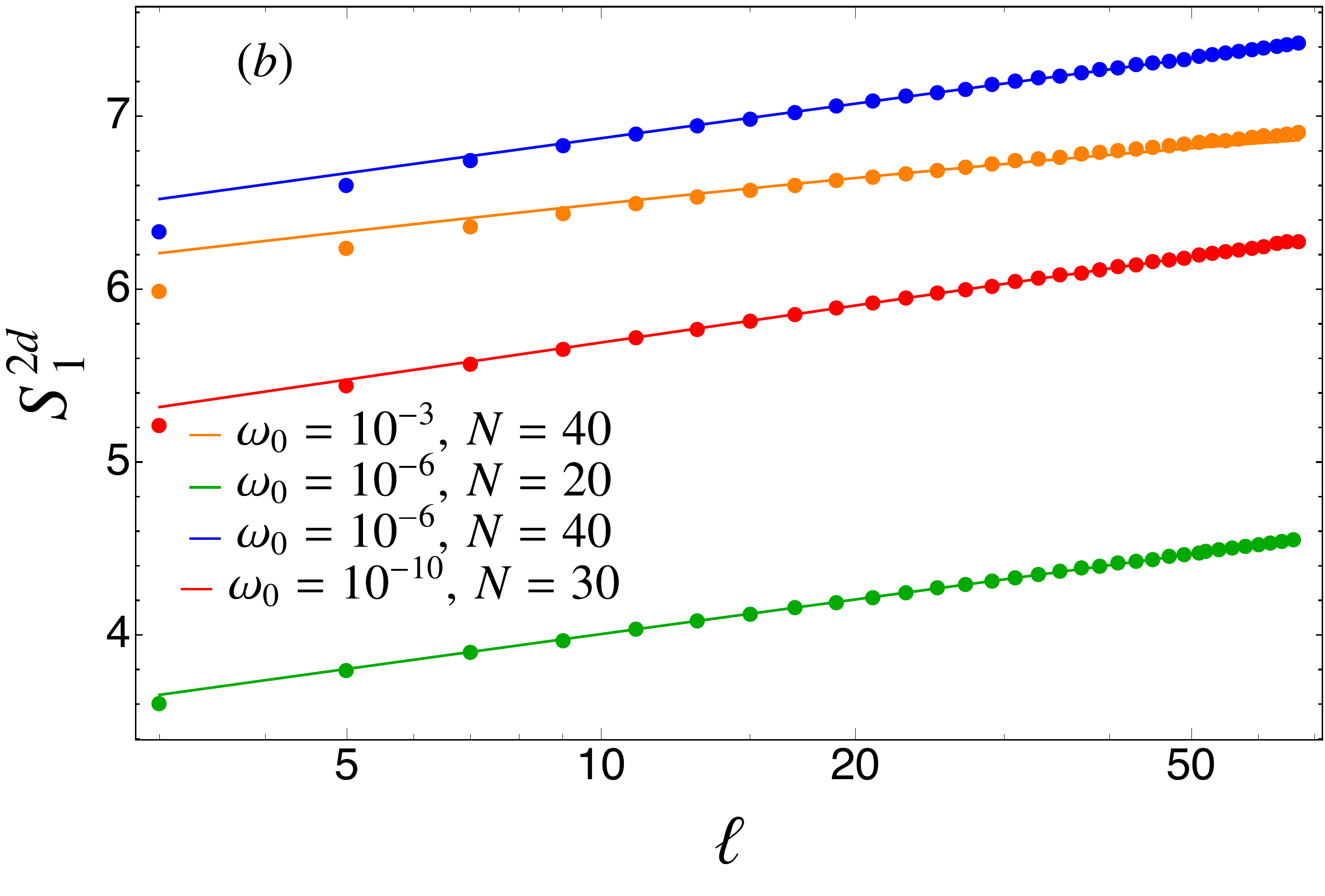}}
\subfigure
{\includegraphics[width=0.325\textwidth]{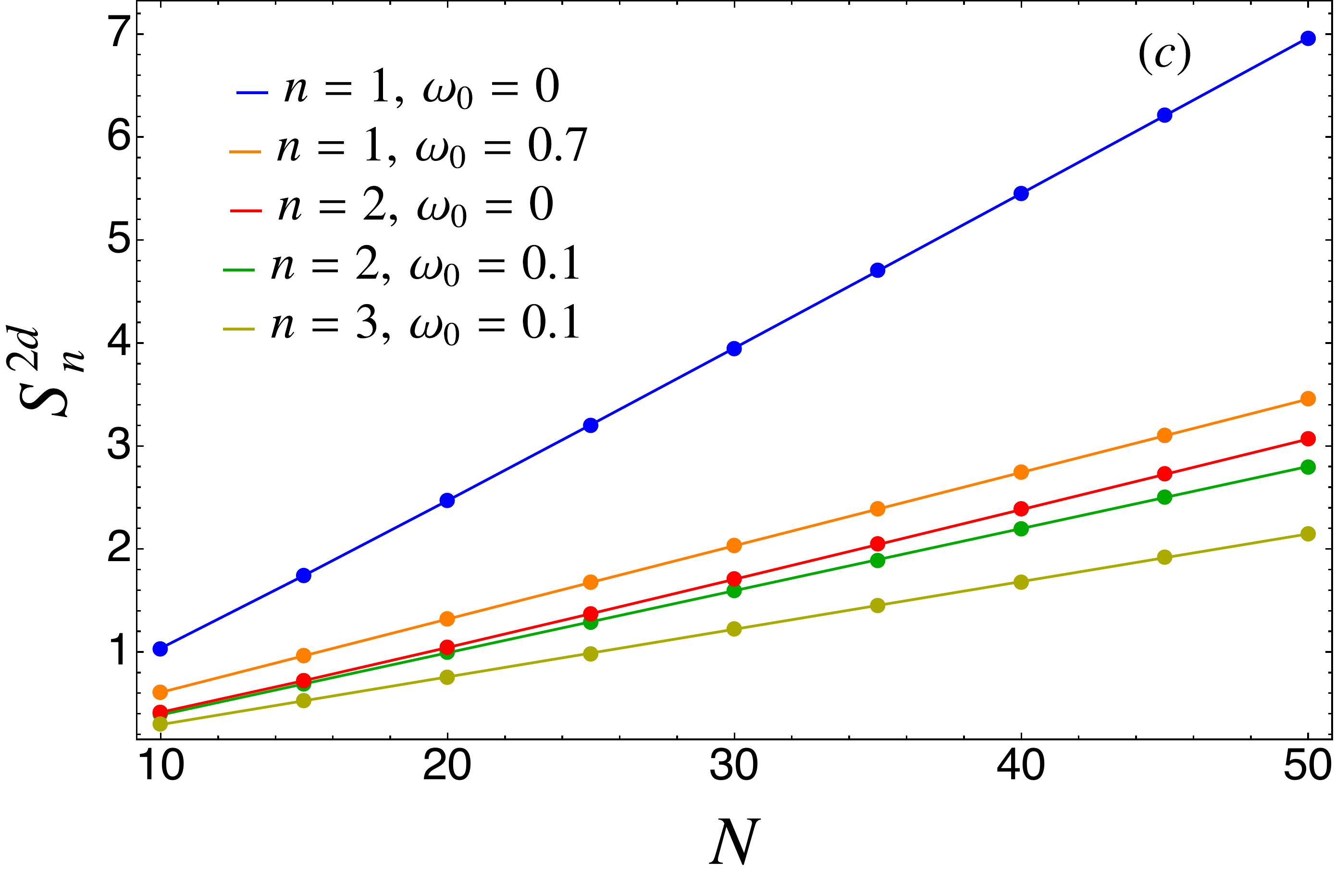}}
\caption{
Entanglement entropies in the 2d lattice of complex oscillators.
(a): $S_n^{2d}$ in the non-critical regime, against the length of the transverse direction $N$, for different $\omega_0$ and $n$ at fixed subsystem size $\ell=50$.
Periodic BC are imposed along $y$.
The symbols correspond to numerical data, while continuous lines are the analytic predictions (\ref{eq:SvNreni}). 
Different colours denote different choices of the parameters $n$ and $\omega_0$.
(b): Entanglement entropy $S_1^{2d}$ in the critical regime $\omega_0 \to 0$. 
Solid lines are the prediction (\ref{eq:sbcritica}) for different $\omega_0$ and $N$. 
The smaller $\omega_0$, the better Eq. (\ref{eq:sbcritica}) works. 
The additive constant $c_1^{1d}$ is numerically extrapolated through a fitting procedure for a chain. 
(c): Same as in (a), but with DBC's along the  transverse direction. The theoretical prediction is Eq. (\ref{eq:Sbosdbc}).
 }\label{fig:renyibosons}
\end{figure}

These results for the entanglement entropies are numerically tested in panel (a) of Figure \ref{fig:renyibosons} using the free-boson techniques reported in the Appendix. 
The considered subsystem is a strip, periodic in the transverse direction (hence of length $N$) and of longitudinal size equal to $\ell$, but such that $\ell$ is much 
larger than the correlation length  of the system (of order $\omega_0^{-1}$) and so the entanglement is just the double of the one for a semi-infinite subsystem.
We have fixed $\ell=50$ which is large enough for the considered values of $\omega_0$. 
Eqs. (\ref{eq:SvNreni}) and (\ref{eq:SvNbosons}) perfectly predict the prefactor of the area-law term, in all cases when the thermodynamic limit along the transverse 
direction is a good approximation. 

We now discuss the critical regime $\omega_0 \to 0$. Here we do not set $\omega_0=0$ from the beginning, but we take a very small $\omega_0$ and then take 
the large $\ell$ limit. The two limits are known to not commute in 1d \cite{CH}. 
Although the technique to obtain the 2d results from 1d one is the same for bosons and fermions, the physics is very different. 
Indeed, while for fermions the $N$ chains  in the Hamiltonian \eqref{eq:Hamiltonian1} are all critical, just with renormalised chemical potentials \eqref{mur},
for free bosons only the zero-mode chain is critical and all the other have a gap given by Eq.~(\ref{eq:freqc}) that does not close as $\omega_0\to0$.
This different behaviour is the origin of the logarithmic multiplicative correction to the area law for massless fermions, while 
massless bosons follow a strict area law, with additive logarithmic corrections. 
While these physical results are well known in the literature (see, e.g., \cite{widom1}) we find their explanation with dimensional reduction particularly clear. 

We can now sum the contributions of the various transverse modes to get the total entanglement entropy. 
For the zero-mode with $r=0$ we take the result from the massive Klein-Gordon theory \cite{CH}. 
Summing up the various contributions, we have for the two-dimensional  lattice complex oscillators
\begin{multline}
\label{eq:sbcritica}
S_n^{2d} = \frac{n+1}{3n}\log\ell +n\log (-\log(\omega_0 \ell))+c_n^{1d}+\\
+\frac{2}{1-n}\displaystyle \sum_{r=1}^{N-1}  \sum_{j=0}^{\infty} \left( n\log [1-e^{-(2j+1) \epsilon_r}]-\log [1-e^{-(2j+1)n\epsilon_r}]\right).
\end{multline}
Here the first line is the zero gapless transverse  mode and the second is the sum over all massive ones. 
The additive constant $c_n^{1d}$ is non-universal and is not predicted by  field theory; we will fix it numerically with a standard fit of the 1d system. 
All the chains with  $r>0$ give a $O(1)$ contribution in $\ell$ since, for large enough $\ell$, it holds $\ell \gg \omega_r^{-1}$;
hence they give rise to an area-law scaling (i.e., $\propto N$).
The panel (b) of Figure \ref{fig:renyibosons} confirms the accuracy of the prediction \eqref{eq:sbcritica} for the critical regime as a function of $N$.

\subsubsection{Some generalisations.}
As in the fermionic case, let us mention that this technique can also be applied when imposing DBC's along the transverse direction, 
i.e., $q_{i,0}=q_{i,N}=p_{i,0}=p_{i,N}=0$. 
Because of the breaking of translational invariance,  we can simply use the Fourier sine transform
\begin{equation}\label{eq:fouriersine}
{\bf q}_{x,y}=\sqrt{\dfrac{2}{N}}\displaystyle \sum_{r=1}^{N-1}\tilde{q}_{x,r}\sin \left(\frac{\pi r y}{N} \right), \qquad \tilde{q}_{x,r}=\sqrt{\dfrac{2}{N}}\displaystyle \sum_{y=1}^{N-1}{\bf q}_{x,y}\sin \left(\frac{\pi r y}{N} \right).
\end{equation}
(and similarly for $\tilde{p}_{x,y}$). The key difference with respect to the periodic case is that the frequencies of the transverse modes are
\begin{equation}
\label{eq:normalfreqdbc}
\omega^2_r=\omega^2_0+4\sin^2\frac{\pi r}{2 N}, \quad r=1,\cdots,N-1.
\end{equation}
Thus, within these BC, the frequencies are \emph{all} different from zero, even for $\omega_0= 0$. The R\'enyi entanglement is
\begin{equation}
\label{eq:Sbosdbc}
S_n^{2d} =\frac{2}{1-n}\displaystyle \sum_{r=1}^{N-1}  \sum_{j=0}^{\infty} \left( n\log [1-e^{-(2j+1) \epsilon_r}]-\log [1-e^{-(2j+1)n\epsilon_r}]\right).
\end{equation}
We can now take the limit of large $N$, similarly to what done in Eq. (\ref{eq:SvNreni}) for $\omega_0>0$, to get
\begin{multline}\label{eq:SvNrenidbc}
S_n^{2d} =\frac{2N}{1-n} \displaystyle \int_0^1\, d\zeta \sum_{j=0}^{\infty} \left(  n\log [1-e^{- (2j+1)\epsilon(\zeta)}]-\log [1-e^{-(2j+1)n\epsilon(\zeta)}] \right) \\
-\frac{2}{1-n} \sum_{j=0}^{\infty} \left(  n\log [1-e^{- (2j+1)\epsilon_0}]-\log [1-e^{-(2j+1)n\epsilon_0}] \right),
\end{multline}
where we need to subtract the contribution from the zero mode, since in Eq. (\ref{eq:Sbosdbc}) the sum starts from $r=1$ rather than $0$.
The accuracy of Eq. (\ref{eq:Sbosdbc}) is checked by numerics in the panel (c) of Figure \ref{fig:renyibosons} in which the agreement is perfect.

\subsection{Symmetry Resolved Entanglement Entropies}\label{sec:bosonsSREE}

Here we compute the contributions to the entanglement entropy coming from the different $U(1)$ symmetry sectors for the 2d lattice of oscillators.
The conserved  charge reduced to the subsystem is just the 2d generalisation of ${Q}_A$ in Eq. \eqref{chargebosonA}. 
To get the 2d results for the strip geometry, we use dimensional reduction and the 1d findings of Ref. \cite{MDC-19-CTM} through the CTM approach.

\begin{figure}
\centering
\subfigure
{\includegraphics[width=0.49\textwidth]{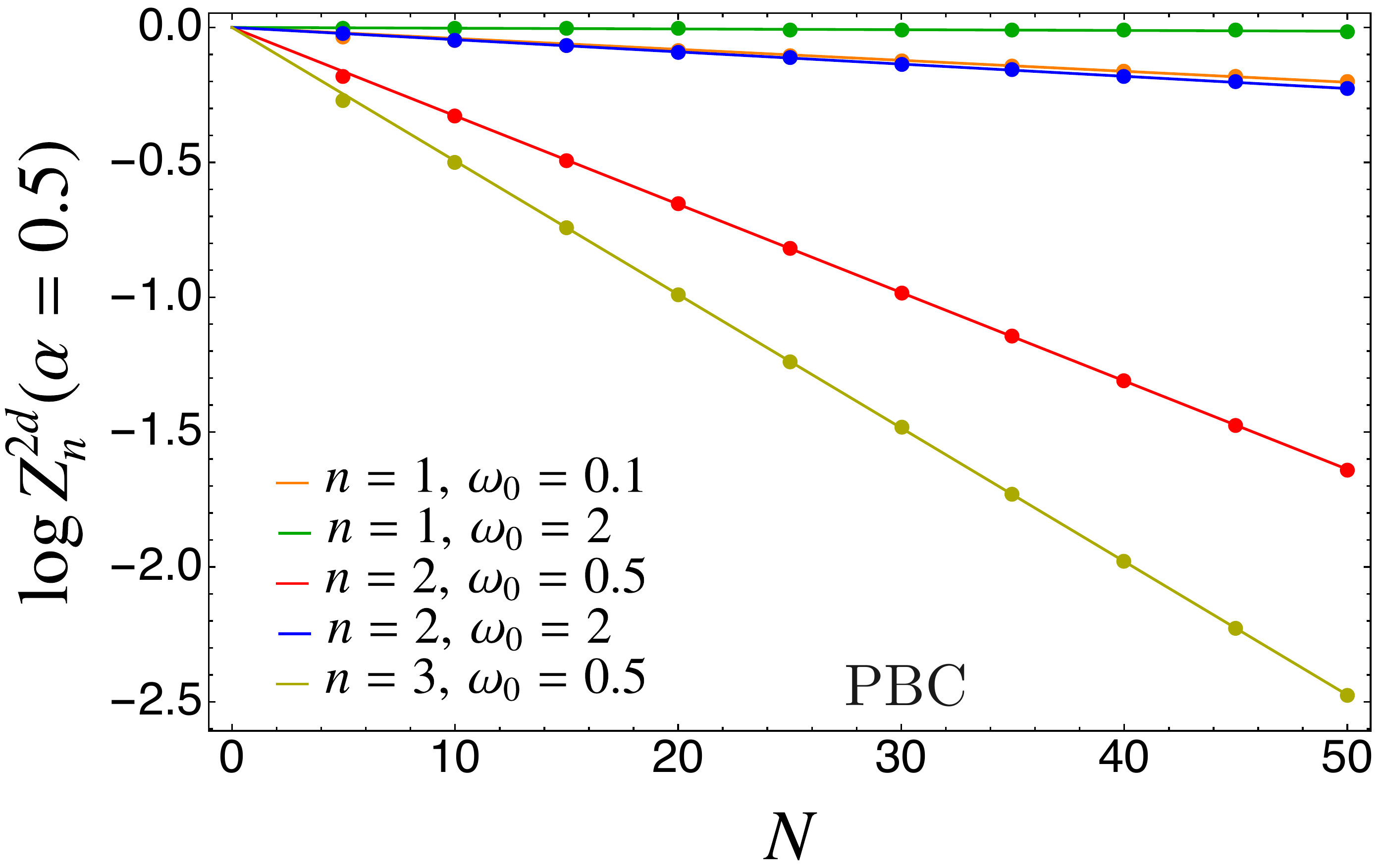}}
\subfigure
{\includegraphics[width=0.48\textwidth]{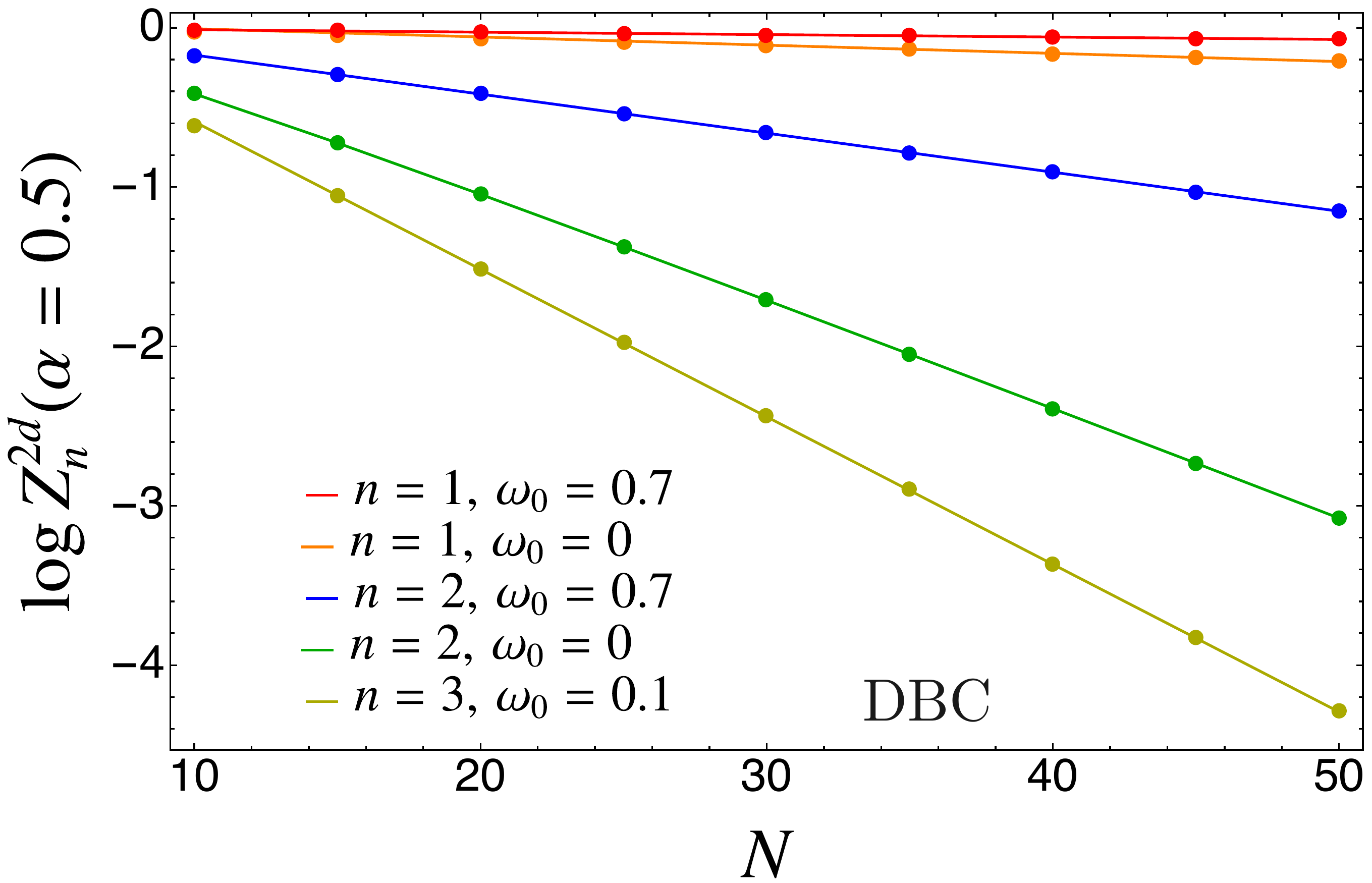}}
\subfigure
{\includegraphics[width=0.49\textwidth]{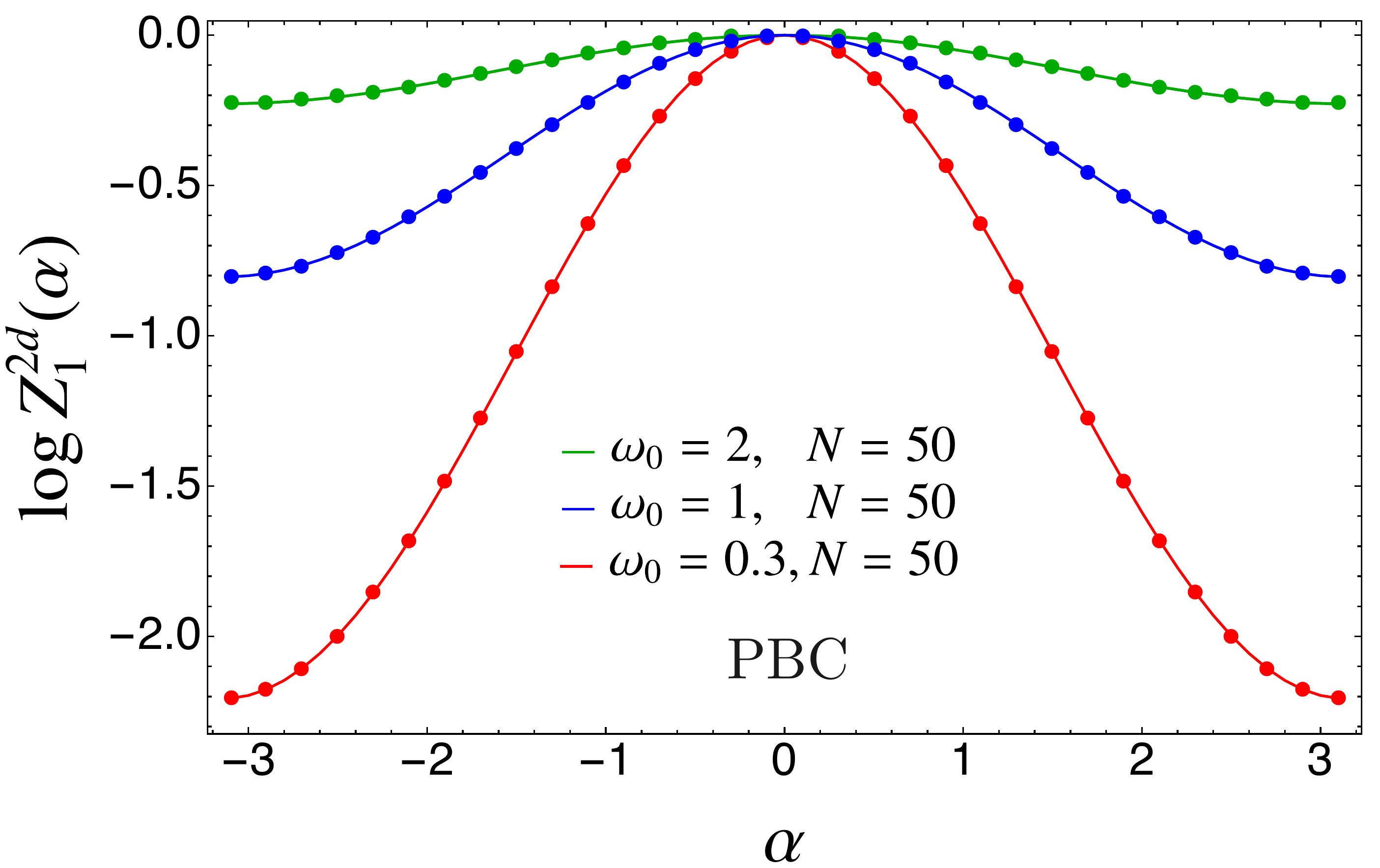}}
\subfigure
{\includegraphics[width=0.48\textwidth]{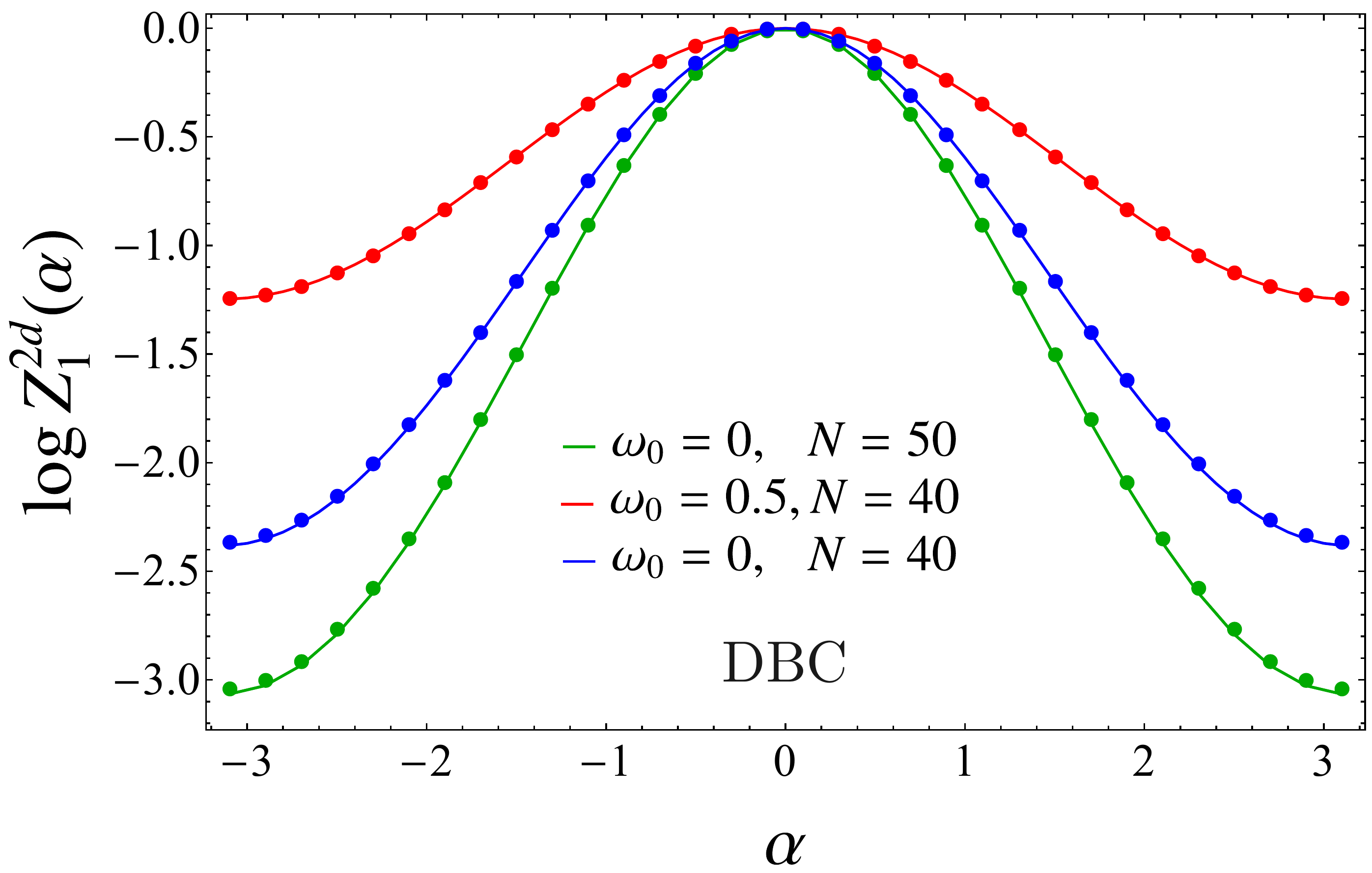}}
\caption{Logarithm of the charged moments $\log Z_n(\alpha)$ for the 2d lattice of oscillators in the off-critical regime. 
Top panels: Plots as a function of $N$, for different $n$ and $\omega_0$, imposing PBC's (left) and DBC's (right) along the transverse direction. 
Bottom panel: Plots against $\alpha$ for different $\omega_0$ and $N$ (again with PBC's (left) and DBC's (right)).
In the right panels, numerical data (symbols) are compared with the analytic predictions (solid lines) of Eq. (\ref{eq:Sbosalphaterm}). 
In the left panels, data are compared with the analytic prediction for DBC's, which is the same as Eq. (\ref{eq:Sbosalpha}), but the sum starts from $r=1$ rather than $r=0$. 
}\label{fig:imrezalphaB}
\end{figure}

\subsubsection{Charged moments.}
For the 2d lattice and for a subsystem being a  periodic  strip, the charged moments  are obtained as a sum of the $N$ independent chains as 
\begin{equation}
\label{eq:Sbosalpha}
\log Z_n^{2d}(\alpha) =\displaystyle \sum_{r=0}^{N-1} \sum_{j=0}^{\infty}\big[ 2n\log [1-e^{-(2j+1) \epsilon_r}]-\log [1-e^{-(2j+1)n\epsilon_r+i\alpha}]-\log [1-e^{-(2j+1)n\epsilon_r-i\alpha}]\big],
\end{equation}
and, taking the limit of large $N$ 
\begin{multline}
\label{eq:Sbosalphaterm}
\log Z_n^{2d}(\alpha) =N \displaystyle  \int_0^1d\zeta \sum_{j=0}^{\infty} \big[ 2n\log [1-e^{-(2j+1) \epsilon(\zeta)}]-\log [1-e^{-(2j+1)n\epsilon(\zeta)+i\alpha}] +  \\
 -\log [1-e^{-(2j+1)n\epsilon(\zeta)-i\alpha}]\big].
\end{multline}
Notice that $\log Z_n^{2d}(\alpha)$ is real and even in $\alpha$.
We plot $\log Z_n^{2d}(\alpha)$  as a function of $\alpha$ and $N$ in Figure \ref{fig:imrezalphaB} (left panels) together with the corresponding numerical data
(for the numerical details see the Appendix). 
The agreement is perfect for all considered values of $n,\alpha,N$, and $\omega_0$. 
We find that for all $\alpha$,  $\log Z_1^{2d}(\alpha)$ is a monotonously increasing function of  the self frequency of the oscillators, $\omega_0$. 
As a function of $\alpha$, they have a single maximum at $\alpha=0$.
The plots as a function of $N$ show that  the integral in Eq.~(\ref{eq:Sbosalphaterm}) well predicts the prefactor of the area-law term of $Z_n(\alpha)$ in the 
massive case, $N \gg \omega_0^{-1}$. To obtain the data in the figure we fix $\ell=50$ which is much larger than the correlation length at all considered $\omega_0$; 
hence, by cluster decomposition, they approach the double of the prediction \eqref{eq:Sbosalphaterm} (see discussion in section \ref{sec:mainREE} below Eq.~\eqref{eq:cftresult}). 
We also analyse the scaling of the charged moments for DBC's along the transverse direction.
The corresponding results are also displayed in the right panels of Figure \ref{fig:imrezalphaB}. 
The computation through the dimensional reduction perfectly works for every $\omega_0$.

\begin{figure}
\centering
\subfigure
{\includegraphics[width=0.49\textwidth]{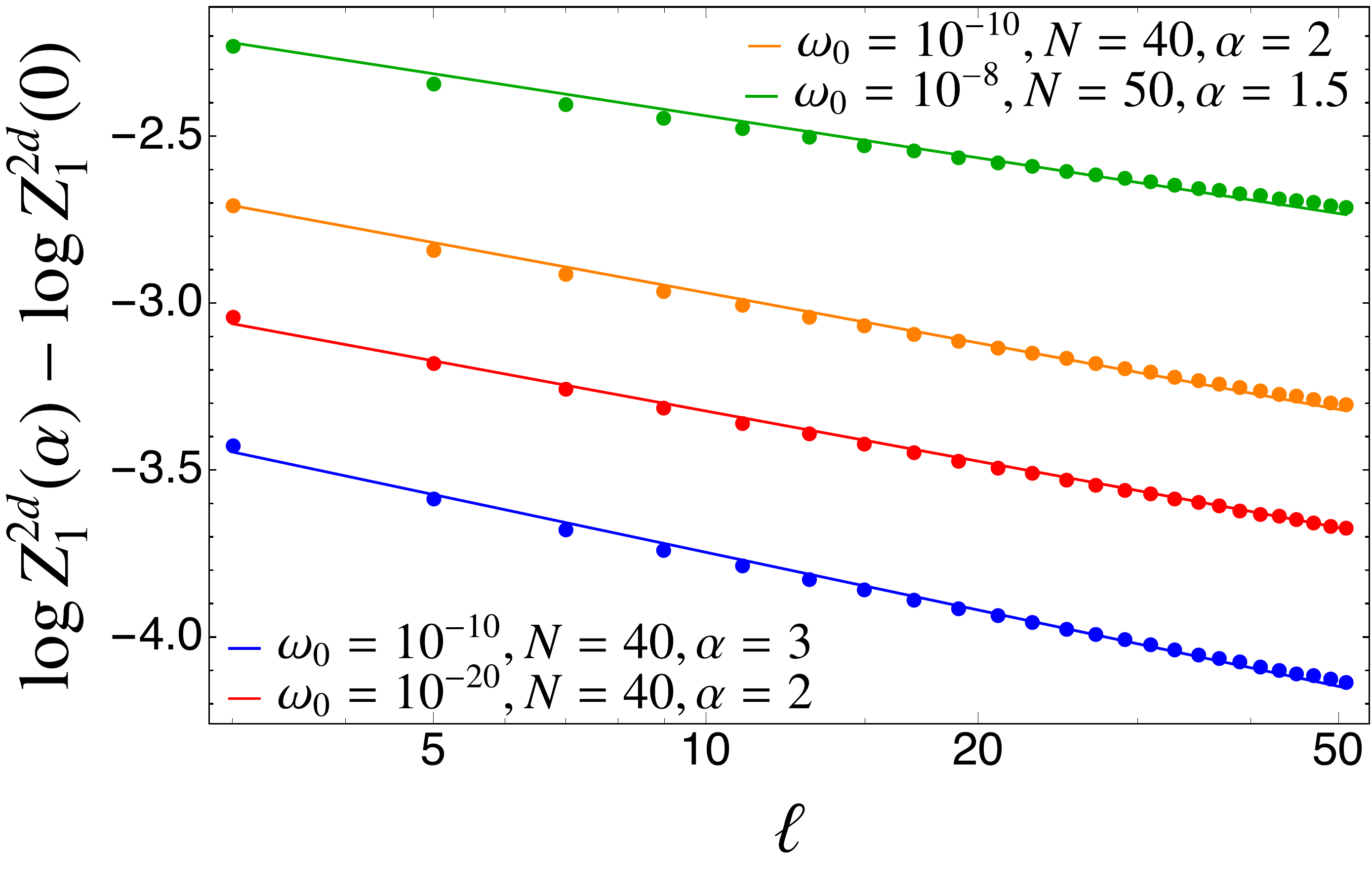}}
\subfigure
{\includegraphics[width=0.49\textwidth]{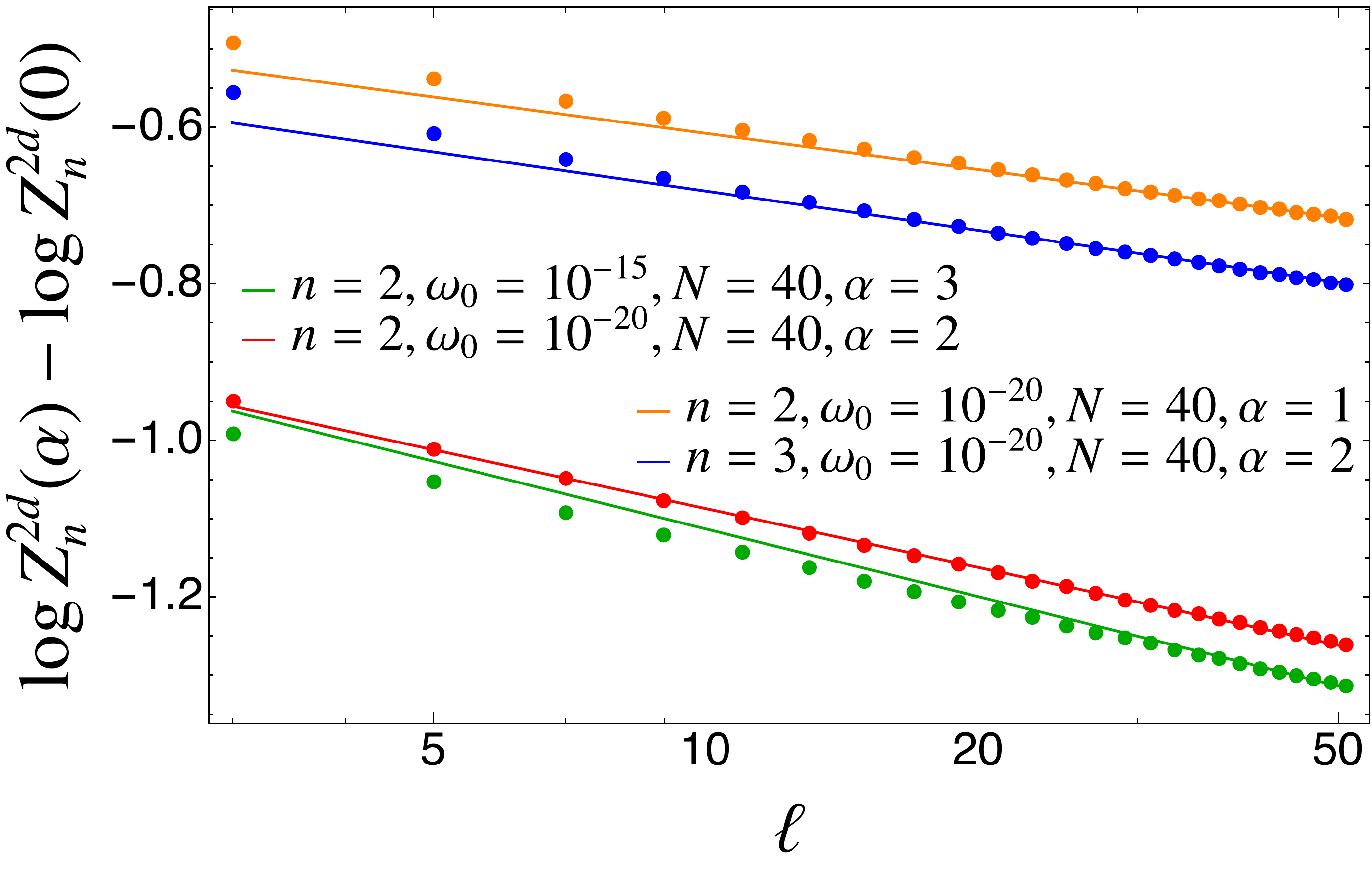}}
\caption{Charged moments $\log( Z^{2d}_n(\alpha)/Z^{2d}_n(\alpha=0))$ of the 2d complex harmonic lattice in the critical regime, as a function of the subsystem size $\ell$, 
for different values of the self-frequency $\omega_0$ and $\alpha$. 
The R\'enyi index is $n=1$ in the left panel and $n= 2, 3$ in the right panel. 
Data (symbols) are compared to the analytic prediction  (\ref{eq:Znalphacrit}) (solid lines). }\label{fig:critica}
\end{figure}

In the critical regime, $\omega_0 \to 0$, the subsystem is a finite strip of longitudinal length $\ell$ and  
the resulting pattern is similar to the case encountered when $\alpha=0$. 
Using Eq. (\ref{eq:crit1}) and assuming that  $\ell \gg \xi$, we obtain 
\begin{multline}
\label{eq:Znalphacrit}
\log \frac{Z_n^{2d}(\alpha)}{Z_n^{2d}(0)}\simeq \frac{2}{n} \left[ \left(\dfrac{\alpha}{2\pi}\right)^2-\dfrac{|\alpha|}{2\pi}\right] \log \ell +\\
 - \displaystyle \sum_{r=1}^{N-1} \sum_{j=0}^{\infty}\big[ \log [1-e^{-(2j+1)n\epsilon_r+i\alpha}]+\log [1-e^{-(2j+1)n\epsilon_r-i\alpha}]\big],
\end{multline}
where the second line is an additive ($\ell$-independent) term that represents the contribution of the chains with $r>0$.
Figure \ref{fig:critica} shows that the agreement of Eq. (\ref{eq:Znalphacrit}) with numerical data is better as $\omega_0$ is smaller, 
i.e., when the approximation of a critical regime is valid. Also it works better for $n$ closer to 1.
Note that we had to consider values of $\omega_0$ much smaller as compared to the same calculation for $\alpha=0$ to fit the numerics with the analytical prediction 
for the critical regime. This is likely due to the lack of a 
more detailed knowledge of the subleading corrections to $Z_n^{(1d)}(\alpha)$ in the critical regime (which instead we have for free fermions). 

\subsubsection{Symmetry resolution.}

\begin{figure}
\centering
\subfigure
{\includegraphics[width=0.49\textwidth]{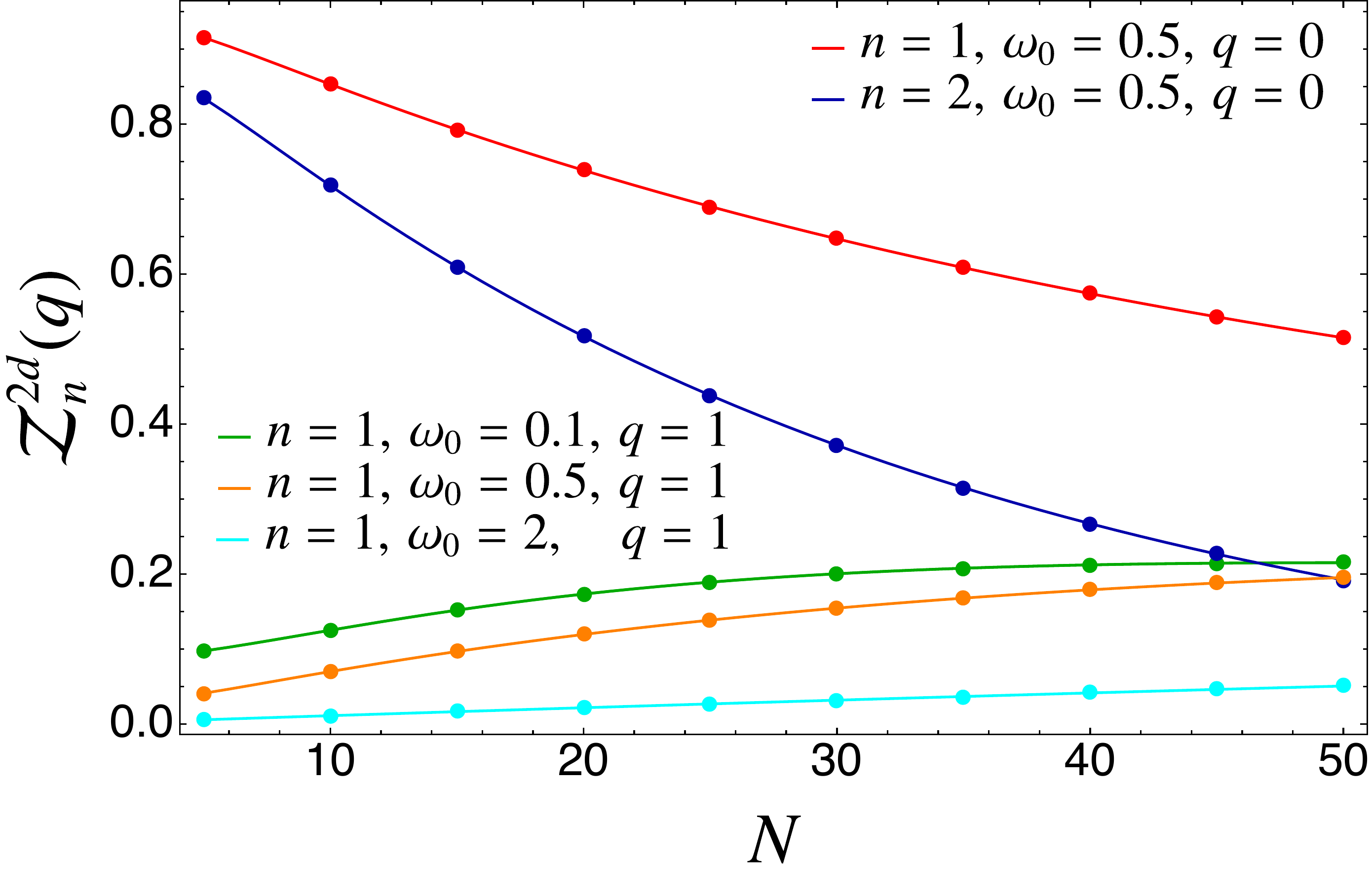}}
\subfigure
{\includegraphics[width=0.49\textwidth]{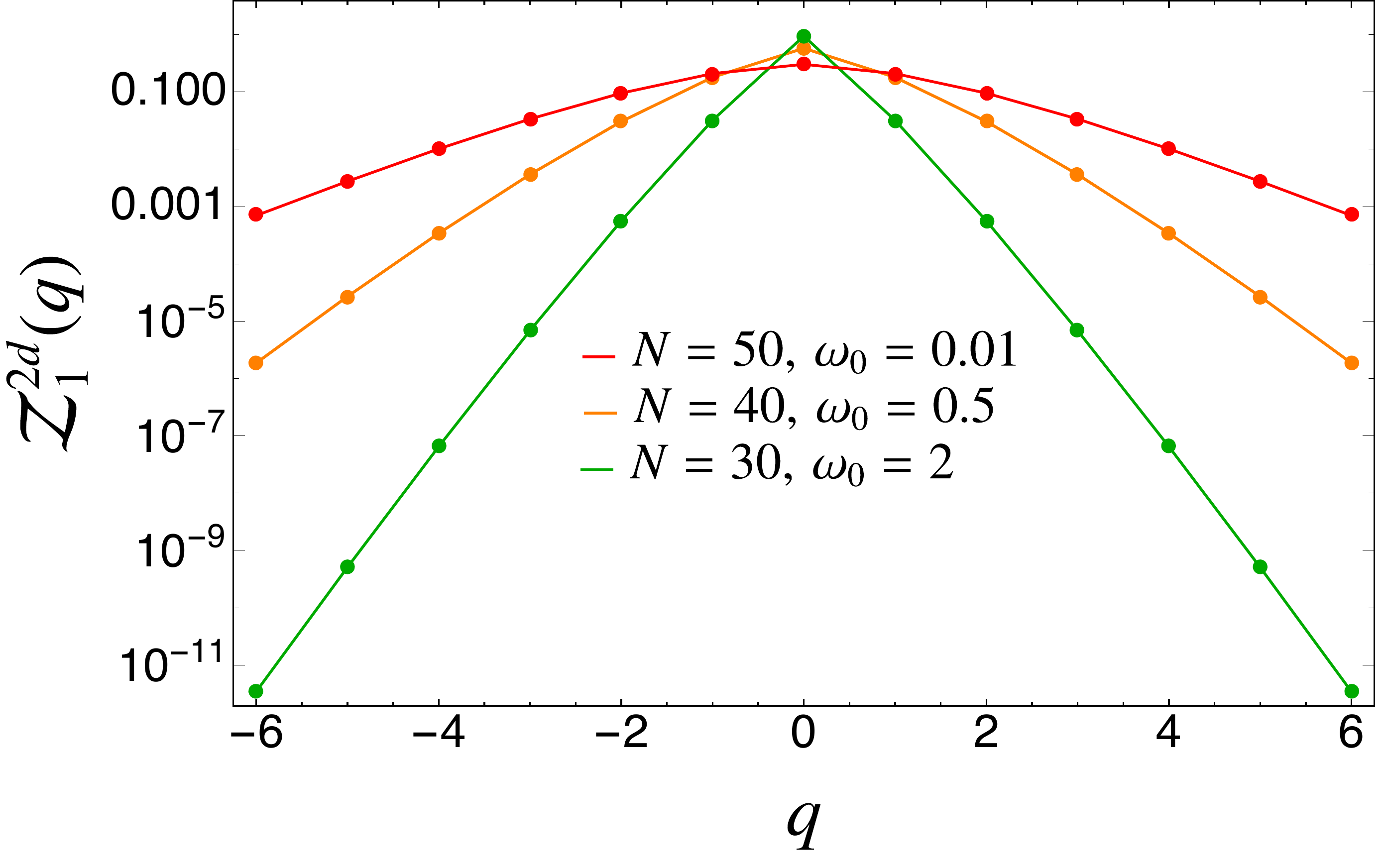}}
\caption{ $\mathcal{Z}_n^{2d}(q)$ in the 2d harmonic lattice for different values of $\omega_0$, $q$, and $n$, as a function of $N$ (left panel) and $q$ (right panel). 
Data (symbols) are compared with the analytic prediction (solid lines) of Eq. (\ref{eq:firstF}).  
$\mathcal{Z}_1^{2d}(q)$ is peaked at $q=0$, which is the average charge in the subsystem. 
Here we do not test the large $N$ result \eqref{eq:asympt}. 
}\label{fig:last}
\end{figure}

In order to get the symmetry resolution it is convenient to first rewrite ${\log}Z^{2d}_{n}(\alpha)$ in the limit $N \to \infty$ as
\begin{equation}
\label{eq:prods}
\log Z^{2d}_n(\alpha)=N \displaystyle \int_0^1 d\zeta \log \left[ \frac{\theta_4(0|e^{-n\epsilon(\zeta)})}{\theta_4(\frac{\alpha}{2}|e^{-n\epsilon(\zeta)})} \displaystyle \prod_{j=1}^{\infty}  \frac{(1-e^{- (2j-1)\epsilon(\zeta)})^{2n}}{(1-e^{-(2j-1)n\epsilon(\zeta)})^2}\right] \equiv N f_n(\alpha),
\end{equation}
where $\theta_4$ is one of the Jacobi theta function
\begin{equation}
\theta_4(z|u)=
\sum_{k=-\infty}^{\infty} (-1)^k\, u^{k^2} e^{2 i k z}.
\end{equation}
Then, the Fourier transform $\mathcal{Z}^{2d}_n(q)$ is 
\begin{equation}
\label{eq:firstF}
\mathcal{Z}^{2d}_n(q)=\displaystyle \int_{-\pi}^{\pi}\dfrac{d\alpha}{2\pi}e^{-iq\alpha} \prod_{r=0}^{N-1} Z^{1d}_{n,r} (\alpha),
\end{equation}
i.e., it is the convolution of the Fourier transforms $\mathcal{Z}^{1d}_{n,r}(q)$ of $Z^{1d}_{n,r} (\alpha)$.
This formula can be easily evaluated for any finite $N$, even very large. 
In order to test its accuracy we  take the Fourier transform of the numerical data for $Z_n^{2d}(\alpha)$ in the previous section 
and compare it with Eq. \eqref{eq:firstF}. 
The results are shown in Figure \ref{fig:last} where the symmetry resolved moments are plotted both against $N$ and $q$ for different values of $\omega_0$.  
The agreement is excellent. Note that $\mathcal{Z}_n(q)$ is peaked at $q=0$, which is the average charge in the subsystem. 

%
For large $N$, we can use Eq.~\eqref{eq:prods} so that Eq.~(\ref{eq:firstF}) can be rewritten as
\begin{equation}
\mathcal{Z}^{2d}_n(q) \simeq\displaystyle \int_{-\pi}^{\pi}\dfrac{d\alpha}{2\pi}e^{-iq\alpha} e^{Nf_n(\alpha)}.
\end{equation}
Given that we are interested in the large $N$ limit, the integral may be performed by saddle point method, with the only maximum of $f_n(\alpha)$ in 
$\alpha=0$, as we can see in Fig. \ref{fig:imrezalphaB}.
Therefore, the integral becomes  
\begin{equation}\label{eq:asympt}
\mathcal{Z}^{2d}_n(q)\simeq e^{N f_n(0)}\displaystyle \int_{-\infty}^{\infty} \frac{d\alpha}{2\pi}e^{-iq\alpha}e^{-\frac{N \alpha^2}{2}  \int_0^1 d\zeta \frac{\theta''_4(0|e^{-n\epsilon(\zeta)})}{4\theta_4(0|e^{-n\epsilon(\zeta)})} } 
= Z^{2d}_n(0) \frac{e^{-\frac{q^2}{2g(n)N}}}{\sqrt{2\pi N g(n)}},
\end{equation} 
where we defined 
\begin{equation}
g(n)\equiv \displaystyle \int_0^1 d\zeta \frac{\theta''_4(0|e^{-n\epsilon(\zeta)})}{4\theta_4(0|e^{-n\epsilon(\zeta)})} .
 \end{equation}
The probability distributions  given by these moments are Gaussian with mean $\bar{q}=0$ and variance that grows as $\sqrt{N}$. 
Unfortunately it is difficult to test Eq. \eqref{eq:asympt} against numerical calculations because we would need rather large values of $N$.
We instead checked that indeed Eq. \eqref{eq:firstF} converges for large $N$ to \eqref{eq:asympt}. 
Anyhow, we will show the corresponding plot only for the symmetry resolved entropies below.

\begin{figure}
\centering
\subfigure
{\includegraphics[width=0.495\textwidth]{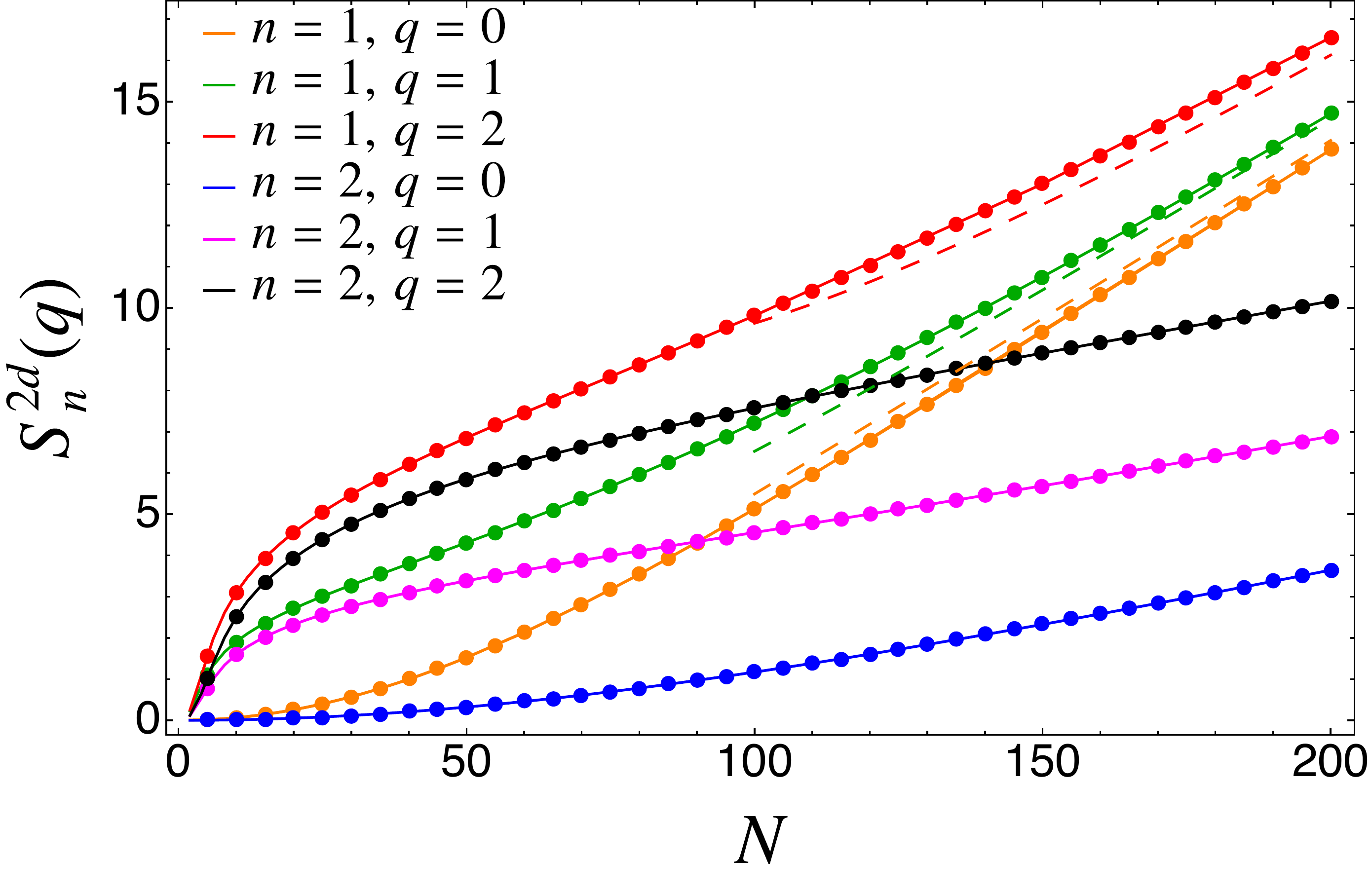}}
\subfigure
{\includegraphics[width=0.495\textwidth]{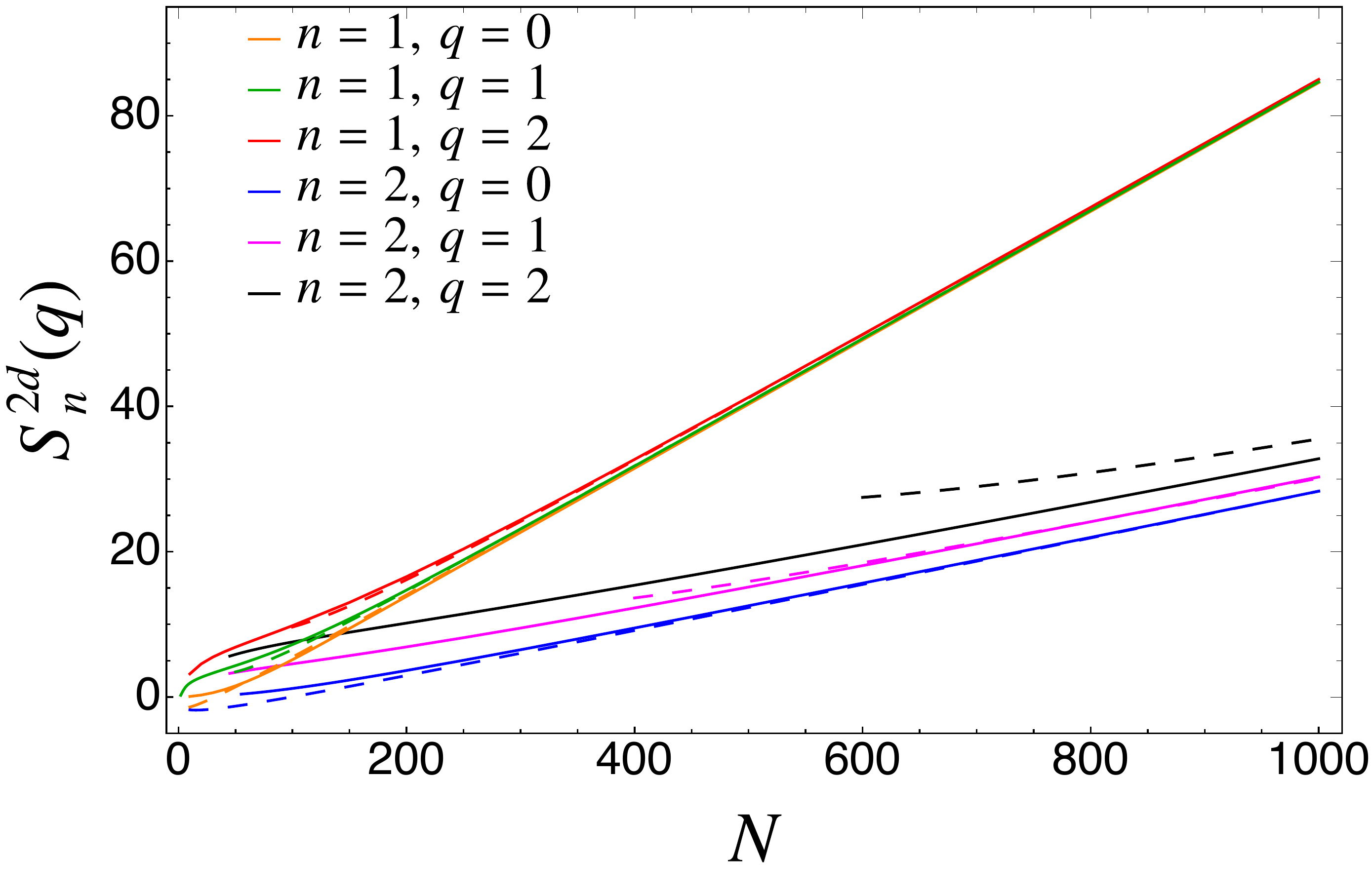}}
\caption{Left: Numerical results for the symmetry resolved entropies $S_n^{2d} (q)$ of the 2d lattice of complex oscillators. 
The numerical data for $q =0, 1,2,$ $n = 1,2$ and $\omega_0 = 0.5$ are compared with the  predictions (\ref{eq:res}) and (\ref{eq:res1})
finding perfect agreement. For these relatively small values of $N$, equipartition of entanglement does not hold, 
even though for $n=1$ data start approaching the asymptotic behaviour described by Eq. (\ref{eq:asympt3}) and plotted through the dashed lines.
Right: We plot the analytic prediction for the symmetry resolved entropy (\ref{eq:res}) (full lines) valid at any $N$ together with its asymptotic expansion 
\eqref{eq:asympt2} (dashed lines), showing that for large $N$ equipartition of entanglement is recovered. 
Notice that, as $n$ and $q$ grow, equipartition occurs at much larger values of $N$.
}\label{fig:last1}
\end{figure}

The last step now is to use Eq. (\ref{eq:firstF}) to calculate the symmetry resolved entropies
\begin{equation}
\label{eq:res}
S^{2d}_n(q)=\dfrac{1}{1-n}\log \left[\dfrac{\mathcal{Z}^{2d}_n(q)}{\mathcal{Z}^{2d}_1(q)^n} \right],
\end{equation}
whose limit $n\rightarrow 1$ is the symmetry resolved von Neumann entropy
\begin{equation}
\label{eq:res1}
S^{2d}_1(q)\simeq -\dfrac{\partial_n \mathcal{Z}_n^{2d}(q)|_{n=1}}{\mathcal{Z}_1^{2d}(q)}+\log \mathcal{Z}_1^{2d}(q).
\end{equation}
Using the previously obtained $\mathcal{Z}^{2d}_n(q)$, we have analytic predictions valid for any $N$.
In Figure \ref{fig:last1} we test the accuracy of this prediction for the entropy in each symmetry sector for $N$ as large as $200$. 
The figure clearly shows that {\it for these relatively small values of $N$}, the equipartition of the entanglement does not hold, even though for $n=1$ data start 
becoming parallel to each other, suggesting a possible onset of equipartition.

In order to understand if and how equipartition is attained at larger values of $N$, we work out the large $N$ limit. 
In the limit $N \to \infty$  plugging Eq. (\ref{eq:asympt}) into Eq. (\ref{eq:res}), we obtain
\begin{equation}
\label{eq:asympt1}
S^{2d}_n(q)=\dfrac{1}{1-n}\log \frac{Z_n^{2d}(0)}{(Z_1^{2d}(0))^n}e^{-\frac{q^2}{2N} \left(\frac{1}{g(n)}-\frac{n}{g(1)}\right)}\frac{( 2\pi N g(1))^{n/2}}{(2 \pi Ng(n))^{1/2}}. 
\end{equation}
The first ratio in Eq. (\ref{eq:asympt1}) just gives the total R\'enyi entropy of order $n$, while the non-trivial dependence on $n$ of $g(n)$ is responsible for the breaking of the equipartition of the entanglement. After some algebra, we obtain
\begin{equation}
\label{eq:asympt2}
S^{2d}_n(q)=S_n^{2d}-\frac{1}{2} \log (2\pi N)-\frac{1}{2(1-n)}\log \frac{g(n)}{g(1)^n}-\frac{q^2}{2(1-n)N}\left(\frac{1}{g(n)}-\frac{n}{g(1)}\right),
\end{equation}
whose limit $n \to 1$ is 
\begin{equation}
\label{eq:asympt3}
S^{2d}_1(q)=S_1^{2d}-\frac{1}{2} \log (2\pi N)-\frac{q^2}{2N}\frac{g'(1)+g(1)}{g(1)^2}+\frac{1}{2}\left(\frac{g'(1)}{g(1)}-\log g(1) \right).
\end{equation}
Hence, we have shown that the leading terms in the expansion for large $N$ satisfy the equipartition of entanglement.
The first term breaking it is at order $1/N$ and has an amplitude proportional to $q^2$.

Unfortunately, as already mentioned, it is difficult to test numerically the validity of Eq. \eqref{eq:asympt2} because it requires too large value of $N$. 
A posteriori, the reason of this peculiar behaviour is easily understood from Eqs. \eqref{eq:asympt2} and \eqref{eq:asympt3}: 
the prefactor of the equipartition breaking term multiplying $q^2/N$ is $-103.485\dots$ for $n=1$ and $-1793.66\dots$ for $n=2$,
very large in both cases. 
Hence, we should get to values of $N$ of order of thousands in order to see equipartition and this is not simply done numerically. 
What instead we can easily do is to test that for large $N$ the analytic prediction (\ref{eq:res}) tends indeed to the predicted asymptotic behaviour \eqref{eq:asympt2}.
This is shown in the left of Figure \ref{fig:last1} where we see that very large values of $N$ are required to recover the asymptotic behaviour, especially for large values 
of $q$ and $n$. Hence equipartition is attained for larger and larger values of $N$ as $q$ and $n$ grow, as very clear from the figure.

\section{Conclusions}
\label{concl}
In this work we exploited \emph{dimensional reduction} for the computation of R\'enyi and symmetry resolved entropies of two-dimensional systems of free fermions and 
bosons with a translational invariant geometry in the transverse direction. 

In section \ref{sec:fermiREE}, we computed the R\'enyi entropies of a 2d non-relativistic free fermionic system, which show the expected logarithmic violation 
of the area law. The latter transparently follows as a result of having $N$ independent 1d critical systems.   
We then proceeded to compute the symmetry-resolved entanglement for the same model (section \ref{sec:fermiSREE}). 
We first obtained an exact asymptotic expression for the charged moments  (cf. Eq. (\ref{eq:try2exact})) and then moved to the truly symmetry resolved entropies 
given by the Fourier transform of the charged ones, Eq. (\ref{eq:Ftrasform}). 
We found that leading terms for large subsystems satisfy entanglement equipartition (as in 1d \cite{xavier}) and we identified the first subleading correction breaking it. 
It turns out that the exact knowledge of non-universal subleading constants is fundamental for a proper description of the symmetry resolved entanglement entropies
while the sole leading term known from Widom conjecture does not provide accurate quantitative results. 

We then considered a 2d lattice of complex oscillators (lattice version of Klein Gordon theory).
We computed the R\'enyi entropies and the symmetry resolved entanglement  in section \ref{sec:bosonsREE} and \ref{sec:bosonsSREE}, respectively. 
We found an area-law behaviour of the entropies regardless of the model being massive or massless. 
Such behaviour, which is very different from the fermionic case, is already well known and its origin is very clear in dimensional reduction approach: 
only one 1d transverse mode is massless, while all other acquire a non zero mass so that the total entanglement is the sum of finite terms without logarithmic violations. 
The massless mode only provides a subleading additive logarithmic term. More generically, the different number of transverse modes corresponding to critical chains, which determine the logarithmic violation of the area law, can be associated to the dispersion relation of the model (for example free fermions with a Dirac dispersion on
a honeycomb lattice have no logarithmic correction  \cite{honeycomb}).
We then moved to the computation of the symmetry resolved entanglement which are fully characterised through the charged moments and their Fourier transform.
Importantly we found that only the leading terms in the expansion for large transverse  size $N$ satisfy entanglement equipartition which is violated for finite $N$,
even in the limit of large longitudinal subsystem length. 

The different structure of the entanglement equipartition in 2d bosonic and fermionic systems clearly shows how such intriguing phenomenon is related to the 
gaussianity of the probability distribution of the conserved $U(1)$ charge, which generically follows from the central limit theorem emerging from the 
large number of elementary constituents. Yet, there are important counterexamples, like the 1d free boson \cite{MDC-19-CTM}, which affect also the physics of some 2d systems 
we considered here. 
Understanding the fine details of entanglement equipartition, such as the precise conditions for its validity and the form of the first subleading term breaking it, remains 
an important open issue.

Having understood how dimensional reduction works for the symmetry resolved entanglement in 2d free theories is also the starting point for studying interacting ones, 
e.g. along the lines of Refs. \cite{konik,dsy-12,p-16} for the total entanglement, but a lot of challenging work is still necessary to get results in this direction.

\section*{Acknowledgments}
We thank Xhek Turkeshi and Giuseppe Di Giulio for useful discussions and collaborations on related topics. 
We thank an anonymous referee for spotting a typo that was the cause of a relevant error in the first version of this manuscript.
PC and SM acknowledge support from ERC under Consolidator grant number 771536 (NEMO).

\begin{appendices}

\section{Numerical tools}\label{app:a}
In this appendix we describe how the numerical data reported in the main text have been obtained. \

For free fermions, as already explained in section \ref{sec:fermiREE}, the correlation matrix restricted to the subsystem $A$ and corresponding to the $m$-th mode is
 \begin{equation}
 \label{eq:corelation1app}
  C_{k^{(m)}_y}(i,j)=\dfrac{\sin k_{m}^F (i-j)}{\pi (i-j)}.
 \end{equation}
Denoting the eigenvalues of the matrix $ C_{k^{(m)}_y}$ by $\varepsilon_{i}^{(m)}$ (with $i \in [1,\ell]$), then simple algebra leads to the moments of 
$ \rho^A_{k_y^{(m)}}$ \cite{correlation}
\begin{equation}
\mathrm{Tr} \rho^{A,n}_{k_y^{(m)}}=\displaystyle \prod_{i=1}^{\ell} [(\varepsilon_{i}^{(m)})^n+(1-\varepsilon_{i}^{(m)})^n],
\end{equation}
and, equivalently, to the R\'enyi entropies 
\begin{equation}
\label{eq:app1}
S^{1d}_{n,m}=\displaystyle \sum_{i=1}^{\ell}e_n(\varepsilon_{i}^{(m)}), \qquad e_n(x)\equiv \frac{1}{1-n} \log \left[ x^n+(1-x)^n \right].
\end{equation}
Once we diagonalise each block of the correlation matrix $\mathbf{C}$ of the entire system, we only have to sum Eq. (\ref{eq:app1}) over all modes. 

The $\alpha$-dependent moments $Z^{1d}_{n,m}(\alpha)$ for the $m$-th mode can be
also easily written in terms of the eigenvalues of the correlation matrix with a simple modification of the above formulas, i.e., \cite{goldstein}
\begin{equation}\label{eq:Zalphanumerics}
Z^{1d}_{n,m}(\alpha)=\displaystyle \prod_{i=1}^{\ell} [(\varepsilon_{i}^{(m)})^n e^{i\alpha}+(1-\varepsilon_{i}^{(m)})^n].
\end{equation}
Then we proceed as for the moments by taking the product of all the independent contributions of the transverse modes.

This approach is equally applicable to a system of coupled oscillators. 
We report results for real oscillators,  the complex case is just the combination of two real ones.
The factorisation of the Hilbert space is such that we can study the eigenvalues of the correlation matrices associated to each tranverse mode.
Let us denotes as $X_{k^{(m)}_y}$ and $P_{k^{(m)}_y}$ the matrices of the correlators of positions and momenta of the $m$-th 
mode (i.e. $X_{ij}=\braket{q_{i,m}q_{j,m}}$ and $P_{ij}=\braket{p_{i,m}p_{j,m}}$). 
Let us also denote by $\sigma_{i,m}$, (with $i \in [1, \ell]$) the eigenvalues of $\sqrt{X_{k^{(m)}_y} P_{k^{(m)}_y}}$. 
The reduced density matrix of $A$ can be written as \cite{holevo1,holevo2}
\begin{equation}
\label{eq:rho_A lambda_n}
\mathrm{Tr} \rho^{A,n}_{k_y^{(m)}}=\displaystyle \prod_{i=1}^{\ell} \frac{1}{\left[\left(\sigma_{i,m}+\frac{1}{2}\right)^n+\left(\sigma_{i,m}-\frac{1}{2}\right)^n\right]},
\end{equation}
and, equivalently, the R\'enyi entropies
\begin{equation}
S^{1d}_{n,m}= \frac{-1}{1-n} \displaystyle \sum_{i=1}^{\ell}\log \left[\left(\sigma_{i,m}+\frac{1}{2}\right)^n+\left(\sigma_{i,m}-\frac{1}{2}\right)^n\right].
\end{equation}
In presence of $\alpha$, the above formula generalises as
\begin{equation}
\label{eq:logZ n alfa lattice} 
Z^{1d}_{n,m}(\alpha) 
=
\prod_{i=1}^\ell
\frac{1}{
\left(
\sigma_{i,m}+\frac{1}{2}
\right)^n
-
e^{i\alpha}
\left(
\sigma_{i,m}-\frac{1}{2}
\right)^n
}.
\end{equation}
Again summing the contribution over $N$ modes we get results for the 2d lattice. 
\section{Anisotropic case}\label{app:b}

The computations done in section \ref{sec:fermions} simplify considerably when studying a setting in which all the $N$ transverse modes correspond to critical chains. 
An example is given by the anisotropic fermionic tight-binding model on a two-dimensional square lattice described by the Hamiltonian
\begin{equation}
\label{eq:Hamiltoniananis}
H_{FF}=-\frac{1}{2}\sum_{i,j}(J_x c^{\dagger}_{i+1,j}c_{i,j}+J_y c^{\dagger}_{i,j+1}c_{i,j})+h.c. +\mu\sum_{i}c^{\dagger}_{i,i}c_{i,i},
\end{equation}
with $J_x=2J_y$. For simplicity, we set $J_y=1$. The Fourier transform along the compact direction $y$ leads to 
Eq. ($\ref{eq:Hamiltonian2}$) and the regime in which all transverse modes are critical occurs when $|\mu_{r}/J_x|=|\mu_{r}/2|<1$. 
We then choose $|\mu|<1$ to ensure this constraint for all modes. In the following sections we highlight the simplifications arising compared to the isotropic case of
section \ref{sec:fermions}. Indeed the factor $f_N(\mu)$ in Eq.\,(\ref{eq:try1}) reduces to $1$ and does not depend on $\mu$.

\subsection{R\'enyi and Entanglement Entropies}\label{sec:fermiREE}

From the structure of the Hamiltonian in Eq. ($\ref{eq:Hamiltonian1}$), the R\'enyi entropies decompose as 
 \begin{equation}
 \label{eq:entropy1app}
 S^{2d}_n\left( \bigotimes_{r=0}^{N-1} \rho^n_{k_y^{(r)},A} \right)=\sum_{r=0}^{N-1}S^{1d}_{n,r}
 \end{equation}
 where $S^{1d}_{n,r}$ is given in Eq. (\ref{eq:entropy1}).

\begin{figure}[t]
\centering
{\includegraphics[width=0.45\textwidth]{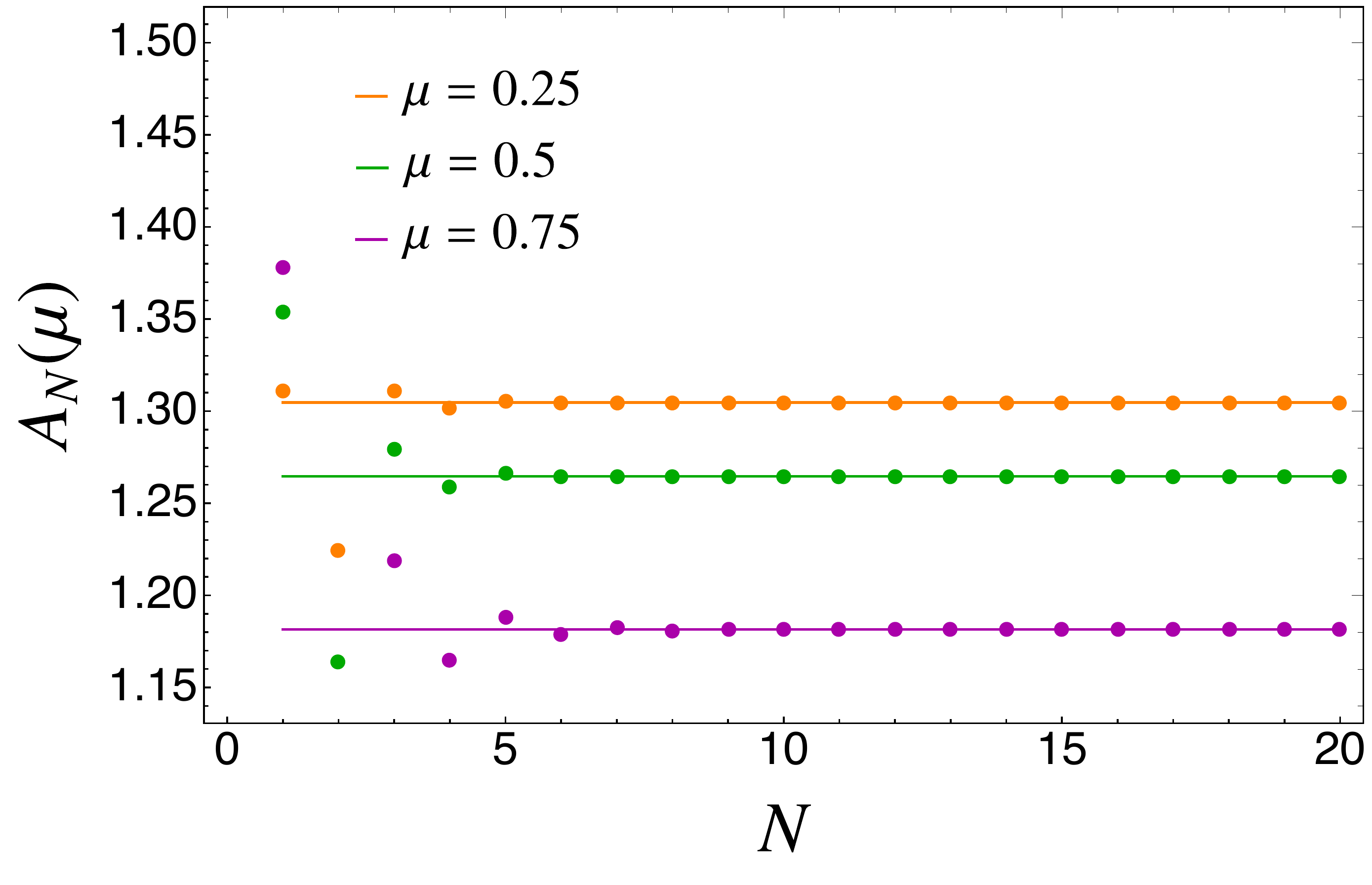}}
\caption{The function $A_N(\mu)$ in Eq. \eqref{ANmapp} as a function of length $N$ of the transverse direction for three different values of chemical potential $\mu$.
For all $\mu$, the curves quickly approach $A_\infty(\mu)$ reported as full lines. 
}\label{fig:ANmapp}
\end{figure}

In this setting, the Fermi momentum of each transverse mode-chain  is 
 \begin{equation}
 \label{eq:fermiapp}
 \sin k_r^F=\sqrt{1-\left(\dfrac{\mu}{2}-\dfrac{1}{2}\cos \left(\dfrac{2\pi r}{N} \right) \right)^2}.
 \end{equation}
Plugging this relation into Eq. (\ref{eq:entropy1app}) we get 
 \begin{equation}
\label{eq:try1app}
S^{2d}_n =\dfrac{N}{6}\Big(1+\dfrac1n\Big)\log (2\ell)+N\Upsilon_n+\dfrac{1}{12}\Big(1+\dfrac1n\Big)
\sum_{r=1}^N\log \left[ 1-\left(\dfrac{\mu}{2}-\dfrac{1}{2}\cos\left(\dfrac{2\pi r}{N}\right)\right)^2\right]. 
\end{equation}
For any given $N$, the sum over $r$ can been performed using elementary trigonometric identities which lead to 
\begin{equation}
S^{2d}_n  =\dfrac{N}{6}\Big(1+\dfrac1n\Big) \log \frac{\ell}{2}+ \dfrac{1}{12} \Big(1+\dfrac1n\Big)\log [4(T_N(2-\mu)-(-1)^N)(T_N(2+\mu)-1)]+N\Upsilon_n,
\end{equation}
where $T_N$ is the $N$-th Chebyshev polynomial. This formula is valid for any finite $N$. 
It is useful to define a quantity analogous to Eq.\,(\ref{ANm}), i.e.
\begin{equation}
A_N(\mu) = \dfrac{1}{2N}\log [4 (T_N(2-\mu)-(-1)^N)(T_N(2+\mu)-1)], 
\label{ANmapp}
\end{equation}
so that we have  
\begin{equation}
S^{2d}_n  =\dfrac{N}{6}\Big(1+\dfrac1n\Big) \Big(\log \frac{ \ell}{2}+ A_N(\mu)\Big) +N\Upsilon_n,
\end{equation}
The function $A_N(\mu)$ is plotted as function of $N$ in Figure \ref{fig:ANmapp}: as $N$ increases, it approaches an asymptotic value more quickly than in the isotropic case (see Figure \ref{fig:ANm}).
In the limit of large $N$, the sum in Eq. \eqref{eq:try1app} turns into an integral
\begin{equation}
\label{eq:sumtointegralapp}
\dfrac{1}{6}\sum_{r=1}^N\log \left| 1-\left(\dfrac{\mu}{2}-\dfrac{1}{2}\cos\left(\dfrac{2\pi r}{N}\right)\right)^2\right| \rightarrow \dfrac{N}{12 \pi} \displaystyle \int_{0}^{2\pi} dx\, \log \left( 1-\dfrac{1}{4}\left( \mu - \cos (x) \right)^2 \right).
\end{equation}
which, can be performed (e.g., using the Taylor expansion of the logarithm and resumming the series). 
The final result is (easier than the analogous in Eq.\,(\ref{eq:integral}))
\begin{equation}
\label{eq:integralapp}
N A_\infty(\mu)=\dfrac{N}{2} \log \left[ \left(2+\mu +\sqrt{\mu^2+4\mu+3} \right)  \left(2-\mu+\sqrt{\mu^2-4\mu+3} \right) \right].
\end{equation}
Hence the total entropy for large $N$ is 
\begin{equation}
\label{eq:integral2app}
S^{2d}_n =\dfrac{N}{6}\Big(1+\dfrac1n\Big) \log \frac{\ell}2+ \dfrac{N}{6} \Big(1+\dfrac1n\Big)  A_\infty(\mu) +N\Upsilon_n ,
\end{equation}
which shows the expected logarithmic correction as a consequence of the fact that we are dealing with $N$ critical chains

As for the isotropic setting, the same approach is straightforwardly adapted to the computation of the entanglement entropies  in the case of  DBC's along 
the transverse direction ($y$-axis), where the only final difference is that, in Eq. (\ref{eq:entropy1app}), we have to sum over $N-1$ modes, rather than $N$.

Moreover, also in this case the same strategy applies when the total system is a finite block of $L$ sites along the $x$-direction with PBC's, by replacing $\ell$ with $\frac{L}{\pi} \sin \frac{\pi\ell }{L}$.

\begin{figure}[t]
\centering
\subfigure
{\includegraphics[width=0.325\textwidth]{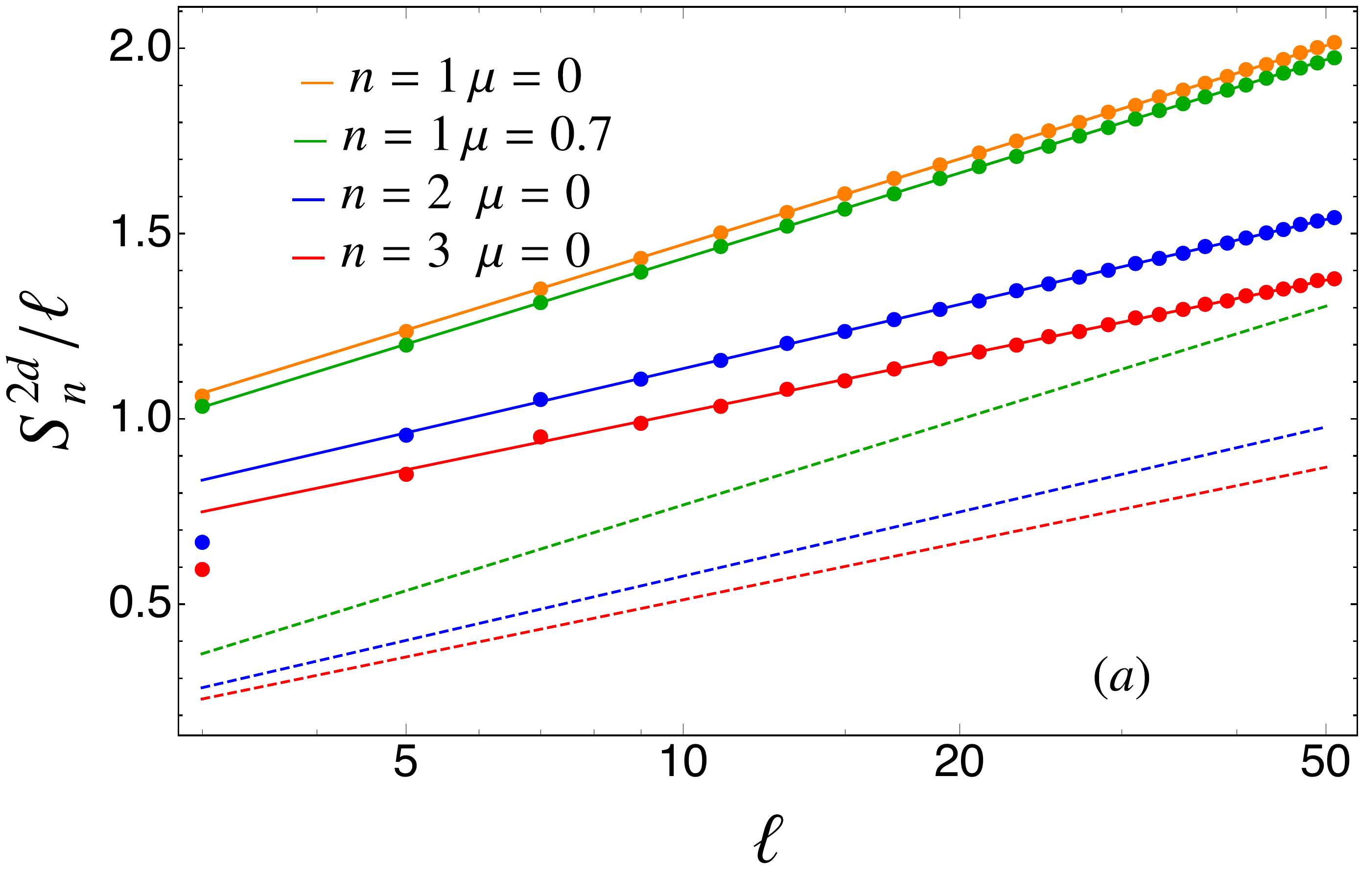}}
\subfigure
{\includegraphics[width=0.325\textwidth]{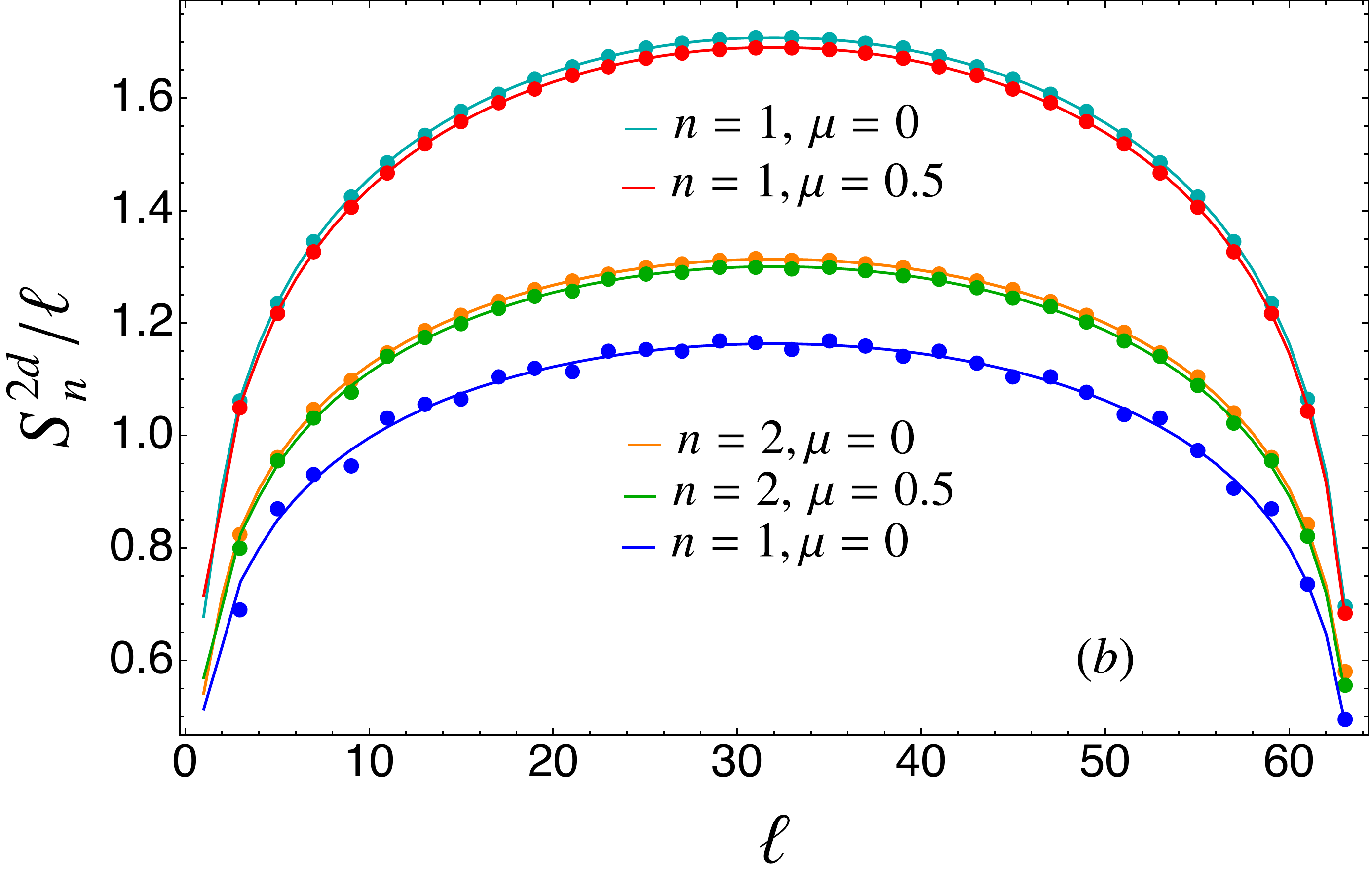}}
\subfigure
{\includegraphics[width=0.325\textwidth]{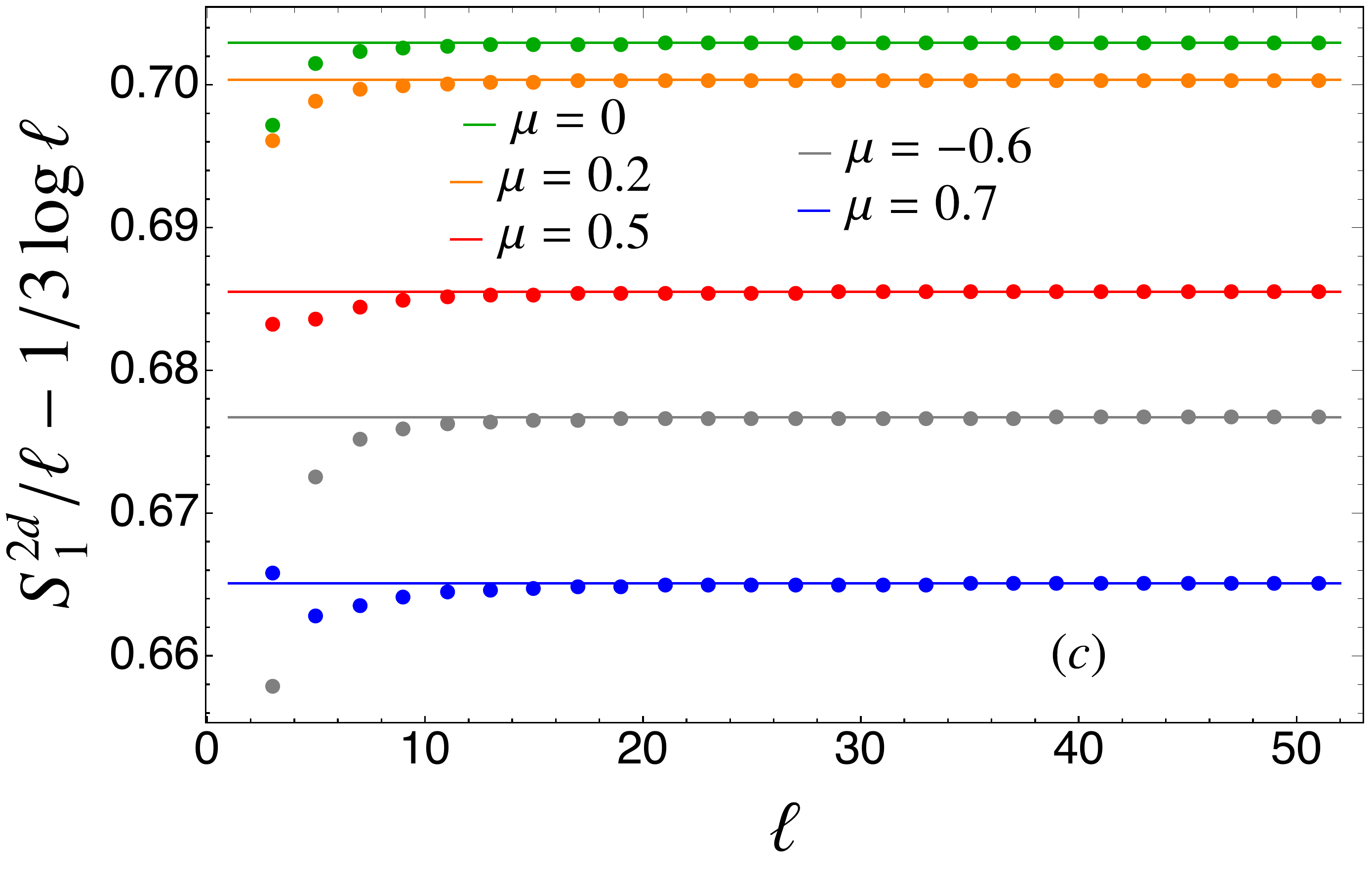}}
\caption{Leading scaling behaviour of the R\'enyi entropies $S^{2d}_n$ of 2d free fermions both for infinite (a) and finite system size $L$ (b) 
in the longitudinal direction.  
In the transverse direction, we fix the periodic size $N$ to equal $\ell$, the subsystem length in the longitudinal direction. 
The numerical results (symbols) for different values of $\mu$ and $n$ are reported as function of $\ell$. 
They match well the theoretical prediction of Eq.  (\ref{eq:integral2app}); 
the dashed lines in (a) are the leading behaviour $\propto \ell \log\ell$ which is clearly not enough accurate. 
The non-universal coefficient proportional to the area, $2\ell$, in Eq. (\ref{eq:integralapp}) is well captured by the numerics, as highlighted in (c).
}\label{fig:primaapp}
\end{figure}

The results for the total entropies are checked against exact numerical calculations in Figure~\ref{fig:primaapp}, where we report the numerical data of the R\'enyi entropies for different values of the  index $n$ and chemical potential $\mu$, 
both for infinite (panel (a)) and finite (panel (b)) system size. 
It is evident that the analytical results correctly describe the data, not only through the sole leading universal behaviour $\propto \ell \log \ell$, but, above all, through to the subleading terms $\propto  \ell$, as showed in the panel (c) of the same figure. 
As in the isotropic setting, the subleading oscillating corrections for $n\neq 1$ are described by the sum over the oscillating contributions given by each mode in Eq. (\ref{eq:oscsn1d}).

\subsection{Symmetry Resolved Entanglement Entropies}\label{sec:fermiSREE}
As already discussed in section \ref{sec:fermiSREE}, the same dimensional reduction technique can further be used to compute the symmetry resolved entanglement entropies of a system in which (in the current setting) all transverse modes correspond to 1d critical chains. 
Let us start with the computation of the charged moments.
The factorisation of Eq.\,(\ref{eq:fact}) allows us to write
\begin{equation}
 \label{eq:entropyapp}
 \log Z_n^{2d}(\alpha)=\sum_{r=0}^{N-1}\log Z^{1d}_{n,r}(\alpha),
 \end{equation}
and, using the explicit 1d result Eq. (\ref{eq:1d}), the  sum is performed as 
\begin{multline}
\label{eq:try2exactapp}
\log Z^{2d}_n(\alpha)\simeq i \bar{q} \alpha
-\left[\frac{1}{6}\left(n-\frac{1}{n} \right) +\frac{2}{n}\left(\frac{\alpha}{2\pi} \right)^2\right] \left( \log \frac{\ell}{2}  + A_N(\mu)\right)N +N \Upsilon(n,\alpha).
\end{multline}
The first purely  imaginary term in Eq. (\ref{eq:try2exactapp}) is the average number of particle within $A$, for large $N$ explicitly given by 
\begin{equation}
\label{eq:approxapp}
\bar{q}= \frac{N \ell  }{2\pi^2}\displaystyle \int_0^{2\pi}dx\arccos \left(\dfrac{\mu-\cos x}{2} \right) .
\end{equation}
It is extensive in the subsystem volume ($N \ell$), as it should, and at half-filling, $\mu=0$, it reproduces the simple result $\bar{q}=N \ell/2$.

\begin{figure}
\centering
\subfigure
{\includegraphics[width=0.45\textwidth]{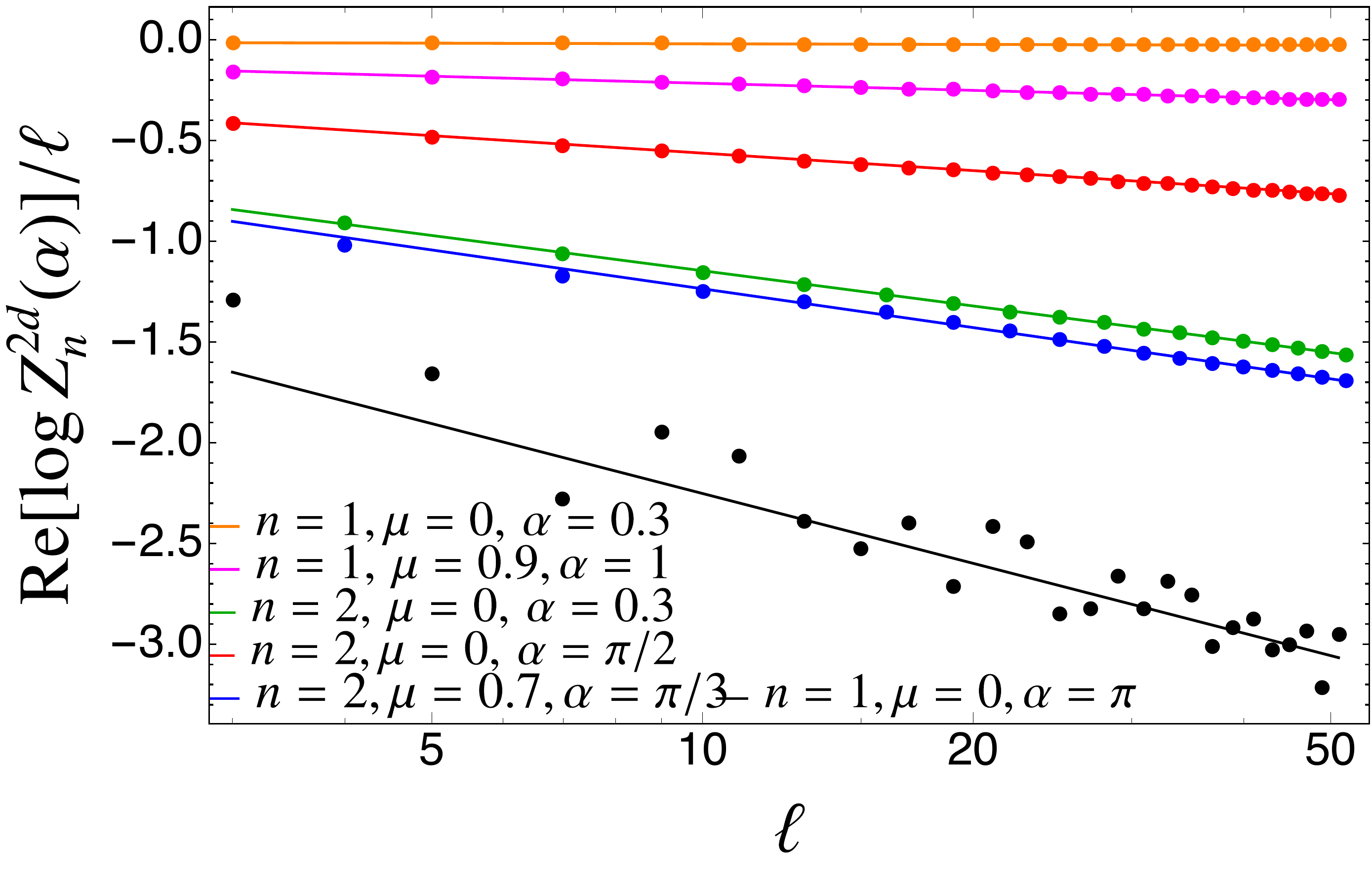}}
\subfigure
{\includegraphics[width=0.44\textwidth]{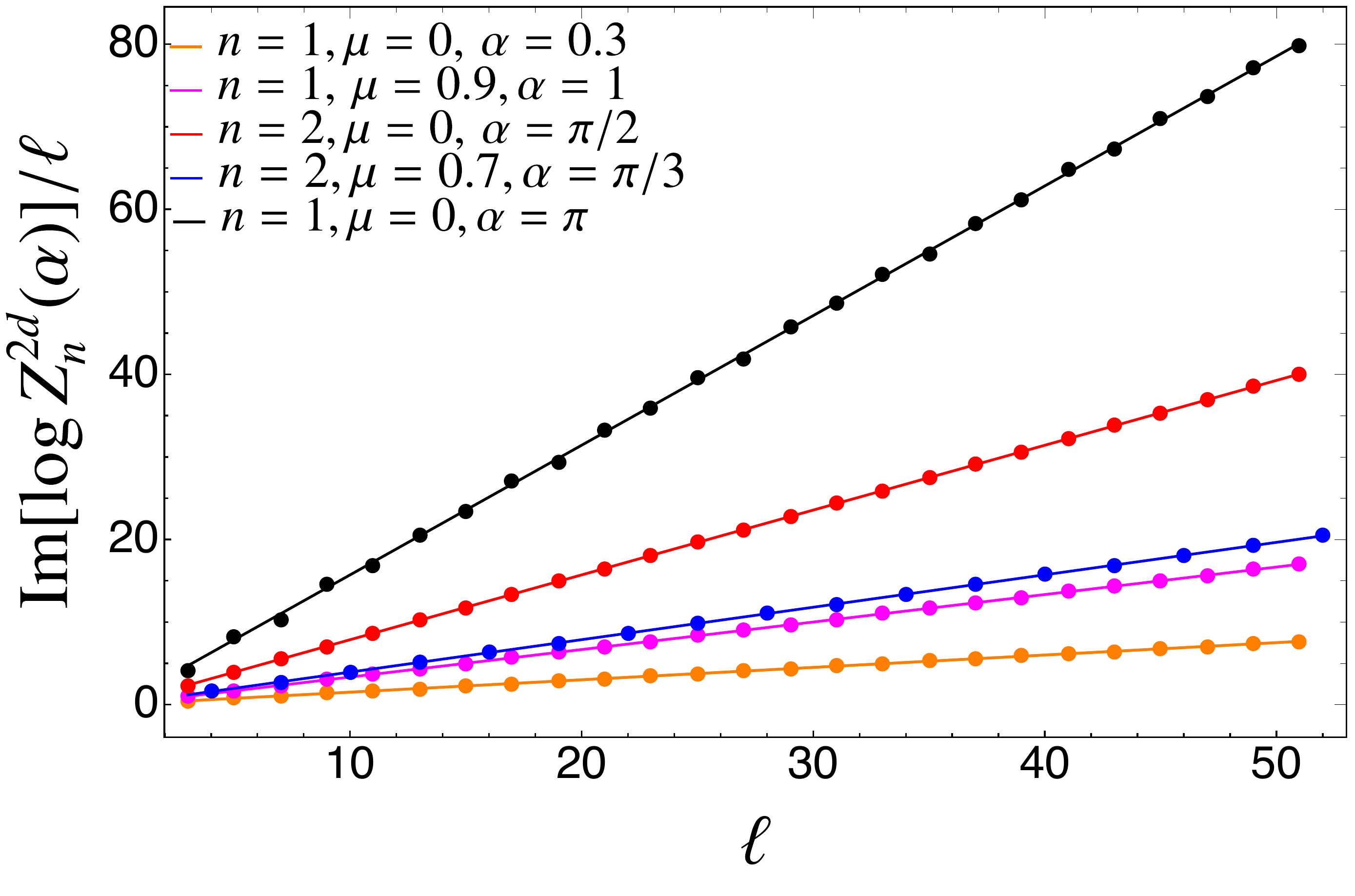}}
\subfigure
{\includegraphics[width=0.45\textwidth]{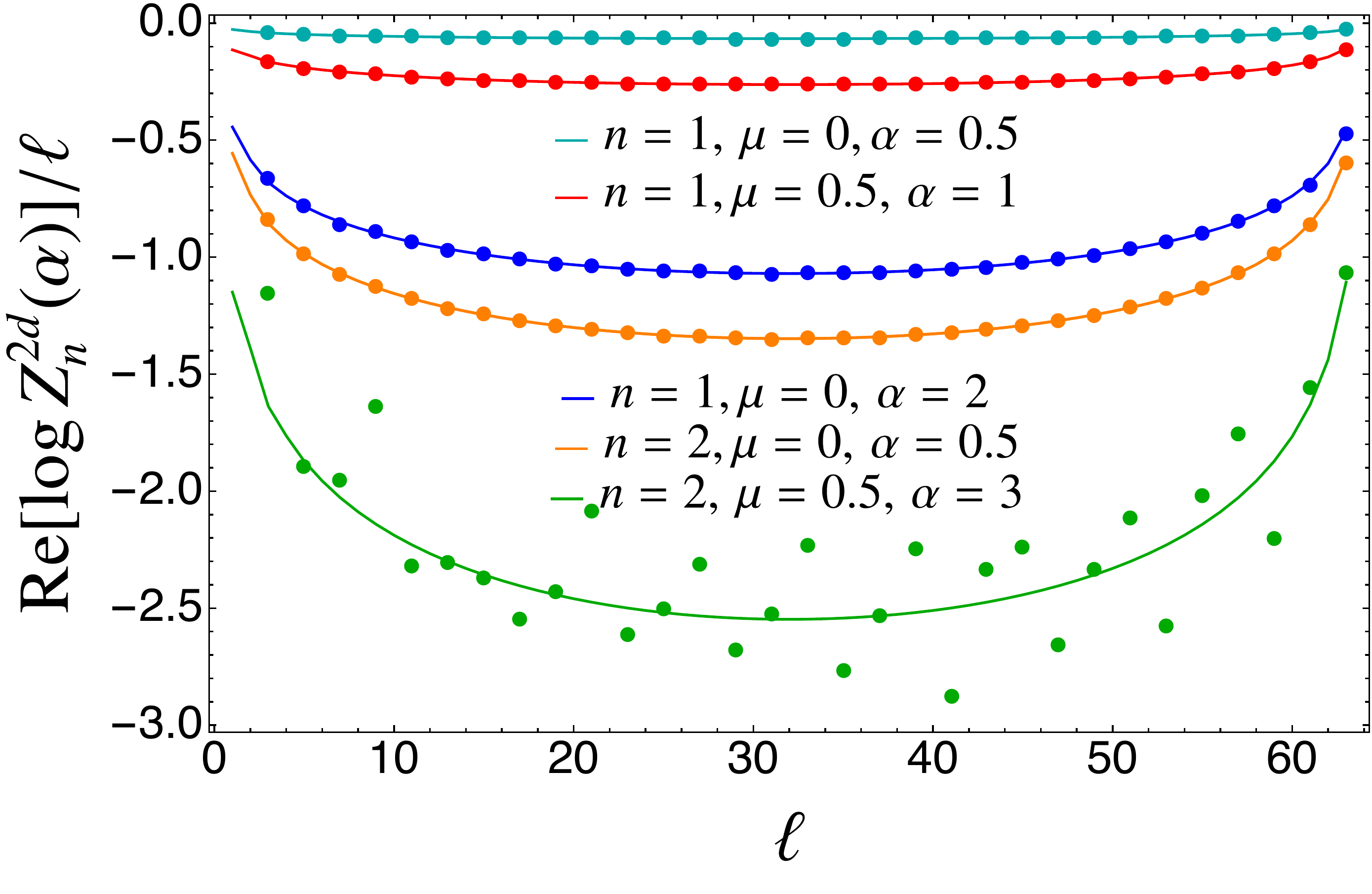}}
\subfigure
{\includegraphics[width=0.44\textwidth]{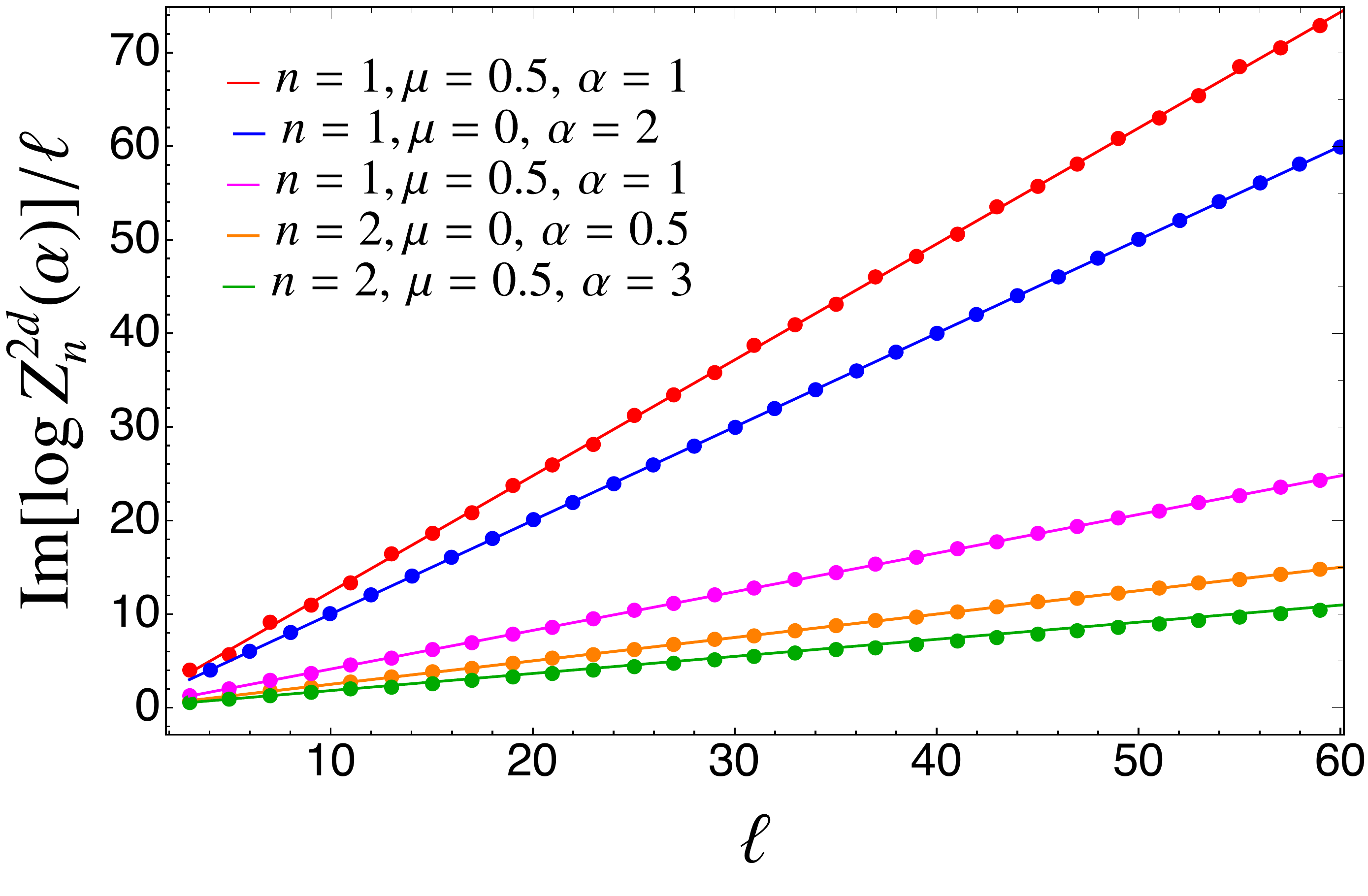}}
\caption{{\bf Top panels}: Leading scaling behaviour of the real and imaginary part of the charged moments $\log Z^{2d}_n(\alpha)$ in 2d free fermionic model 
for an infinite cylinder with transverse length  $N=\ell$, equal to the subsystem length in the longitudinal direction. 
The numerical results (symbols) for several values of $\alpha$ and $n$ are reported as function of $\ell$ for different $\mu$'s. 
Different colours represent different choices of the parameters $n, \alpha, \mu$. 
The corresponding analytic predictions (continuous lines), Eq. (\ref{eq:try2exactapp}), is also reported. 
{\bf Bottom panels}:  The same quantity is studied for a periodic system 
in both directions. The longitudinal circumference is $L=64$ while the transverse one is equal to $\ell$.
}
\label{fig:imrezalphaFapp}
\end{figure}

Keeping only the terms up to $O(\alpha^2)$, we rewrite (\ref{eq:try2exactapp}) in the compact form as: 
\begin{equation}
\label{eq:compactapp}
\log Z^{2d}_n(\alpha)\simeq\log Z^{2d}_n(0)+i\bar{q}\alpha-\alpha^2( \mathcal{B}_n \log \ell +\mathcal{C}_n) N,
\end{equation}
with  $\mathcal{B}_n$ given in Eq.\,(\ref{eq:istr}) and
\begin{equation}
\label{eq:istrapp}
\begin{split}
& \mathcal{C}_n=\dfrac{1}{2\pi^2n} \left(A_N(\mu)-\log2  \right)-\gamma(n).
\end{split}
\end{equation}

In Figure \ref{fig:imrezalphaFapp} we report the numerical data both for the real and the imaginary part of $\log Z^{2d}_n(\alpha)$ for different values of $n$ and $\alpha$ with the theoretical prediction in Eq. (\ref{eq:try2exactapp}) showing that the analytical result correctly describes the data as long as $|\alpha| < \pi$. In the same figure, we also test the usual generalisation of the geometry of a torus using the finite size form with the chord length.
Also in this case, the subleading oscillatory behaviour is easily obtained by the sum over contributions for each mode given by Eq. (\ref{eq:osc1d}).

\begin{figure}
\centering
\subfigure
{\includegraphics[width=0.45\textwidth]{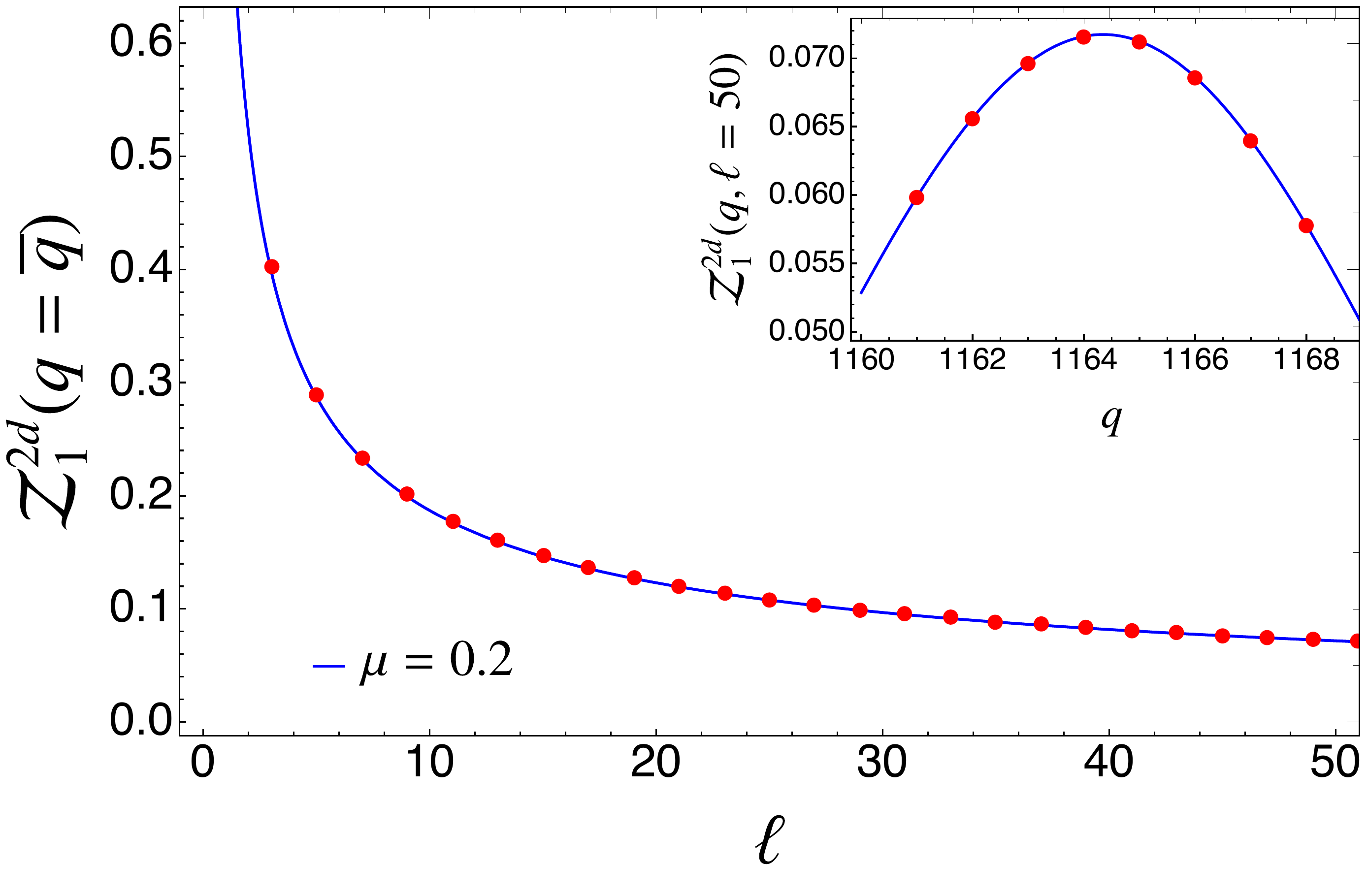}}
\subfigure
{\includegraphics[width=0.45\textwidth]{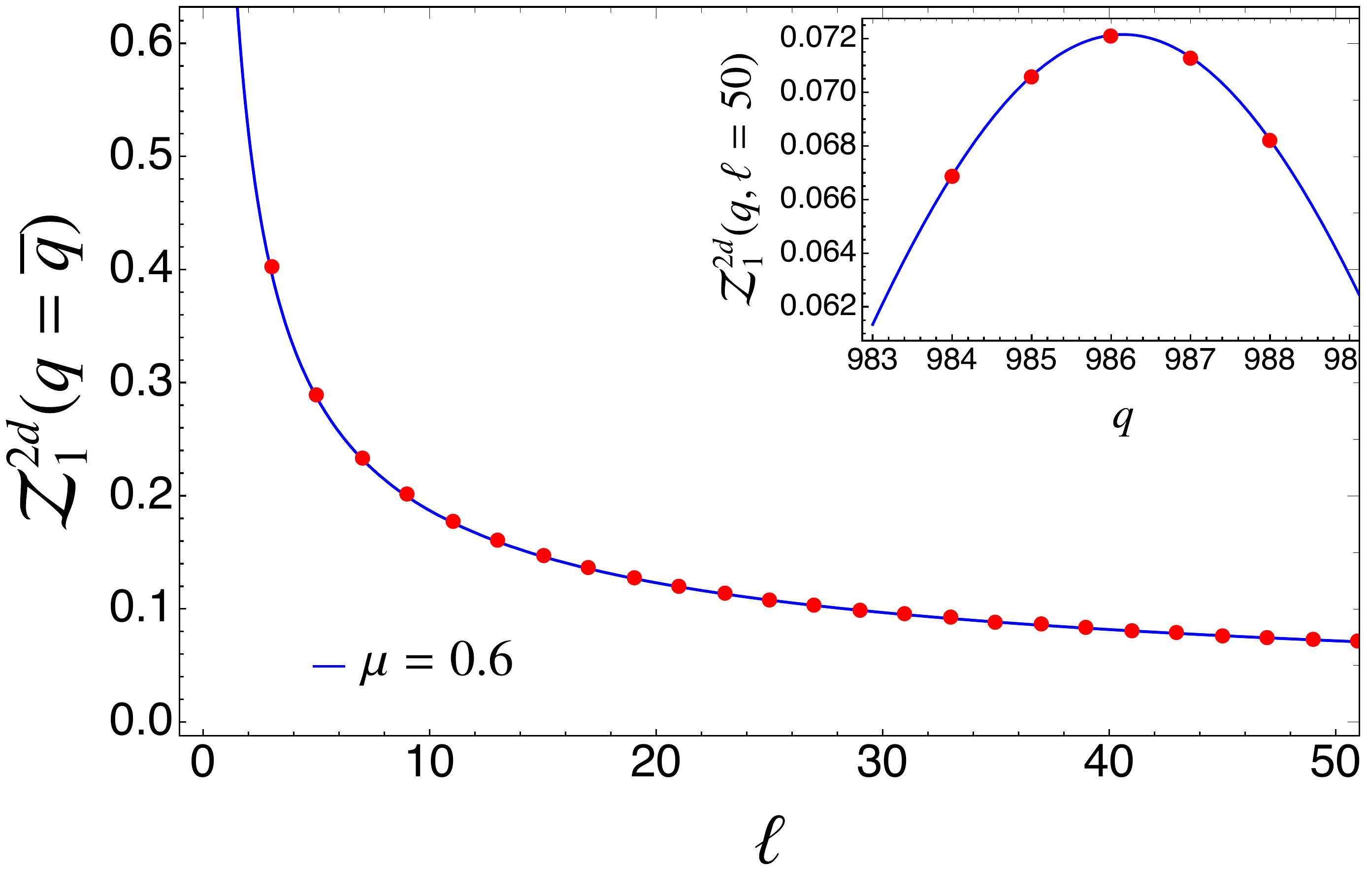}}
\caption{The probability $\mathcal{Z}_1^{2d}(q)$ for 2d free fermions with chemical potential $\mu=0.2$ (left) and $\mu=0.6$ (right). 
The red symbols are the numerical values and blue lines are the analytical prediction (\ref{eq:gaussian}).
In the main frame $\mathcal{Z}_1^{2d}(q=\bar{q})$ is shown as a function of $\ell$, whereas in the inset we fix $\ell=50$ and $\mathcal{Z}_1^{2d}(q)$ 
is plotted as a function of $q$. }
\label{fig:znqfermionapp}
\end{figure}

We now can compute the Fourier transform $\mathcal{Z}^{2d}_n(q)$ of the charged moments, i.e. 
\begin{equation}
\label{eq:Ftrasformapp}
\mathcal{Z}^{2d}_n(q) = \displaystyle \int_{-\pi}^{\pi} \dfrac{d\alpha}{2\pi}e^{-iq\alpha}Z^{2d}_n(\alpha)\simeq
Z_n^{2d}(0)\displaystyle \int_{-\pi}^{\pi} \dfrac{d\alpha}{2\pi}e^{-i(q-\overline{q})\alpha-\alpha^2 b_n},
\end{equation}
where the coefficient of the quadratic term is 
\begin{equation}
\label{eq:varianceRapp}
b_n= \mathcal{B}_n N \log \ell +\mathcal{C}_n N.
\end{equation}
By means of the saddle point approximation we get a Gaussian distribution function with mean  $\bar{q}$ and variance 
growing as $\sqrt{N \log \ell}$, i.e.
\begin{equation}
\label{eq:gaussianapp}
\mathcal{Z}_n^{2d}(q)\simeq 
Z_n^{2d}(0) e^{-\frac{(q-\overline{q})^2}{4(\mathcal{B}_n N\log \ell +\mathcal{C}_n N)}}\sqrt{\dfrac{1}{4\pi(\mathcal{B}_n N\log \ell +\mathcal{C}_nN)}},
\end{equation}
whose accuracy is checked in Figure \ref{fig:znqfermionapp}.

Finally, the asymptotic behaviour of the symmetry resolved entanglement, given by Eq.\,(\ref{eq:def}), is
\begin{multline}
\label{eq:simmresolvedapp}
S^{2d}_n(q)=S^{2d}_n -\dfrac{1}{2} \log \left( \dfrac{2N}{\pi} \left(  \log(\ell/2)+ \delta_n + A_\infty(\mu)  \right)\right)+\dfrac{\log n}{2(1-n)}+\\ 
(q-\overline{q})^2\pi^4\dfrac{n}{1-n}\dfrac{ (\gamma(1)-n\gamma(n))}{N[ \log (\ell/2) + \kappa_n + A_\infty(\mu) ]^2}+\cdots,
\end{multline}
where $S^{2d}_n$ is the total R\'enyi entropy, $\delta_n$ and $\kappa_n$ are respectively defined in Eq.s\,(\ref{eq:deltan}) and \ref{eq:kappan}.
\begin{figure}
\centering
\subfigure
{\includegraphics[width=0.45\textwidth]{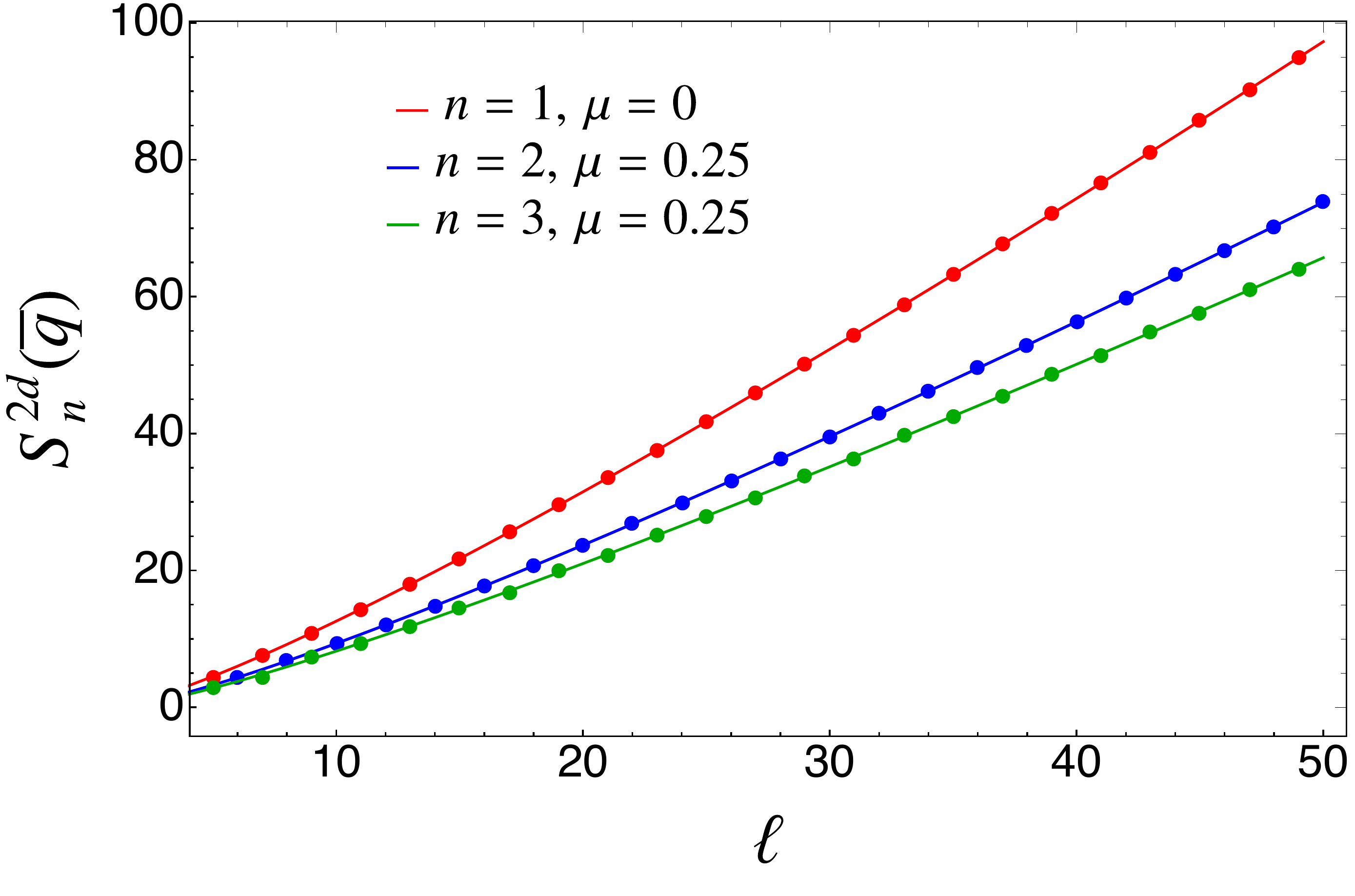}}
\subfigure
{\includegraphics[width=0.45\textwidth]{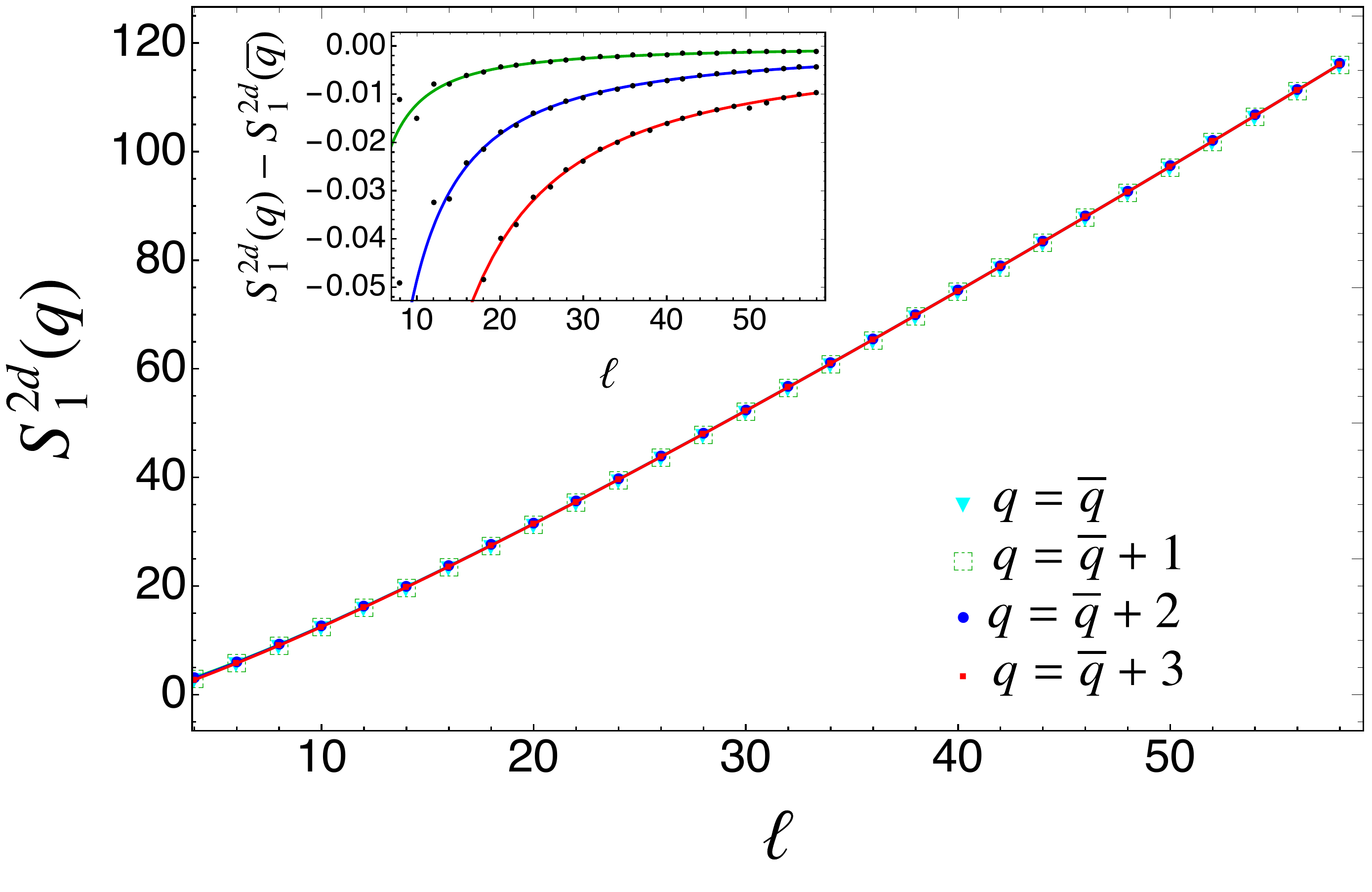}}
\subfigure
{\includegraphics[width=0.45\textwidth]{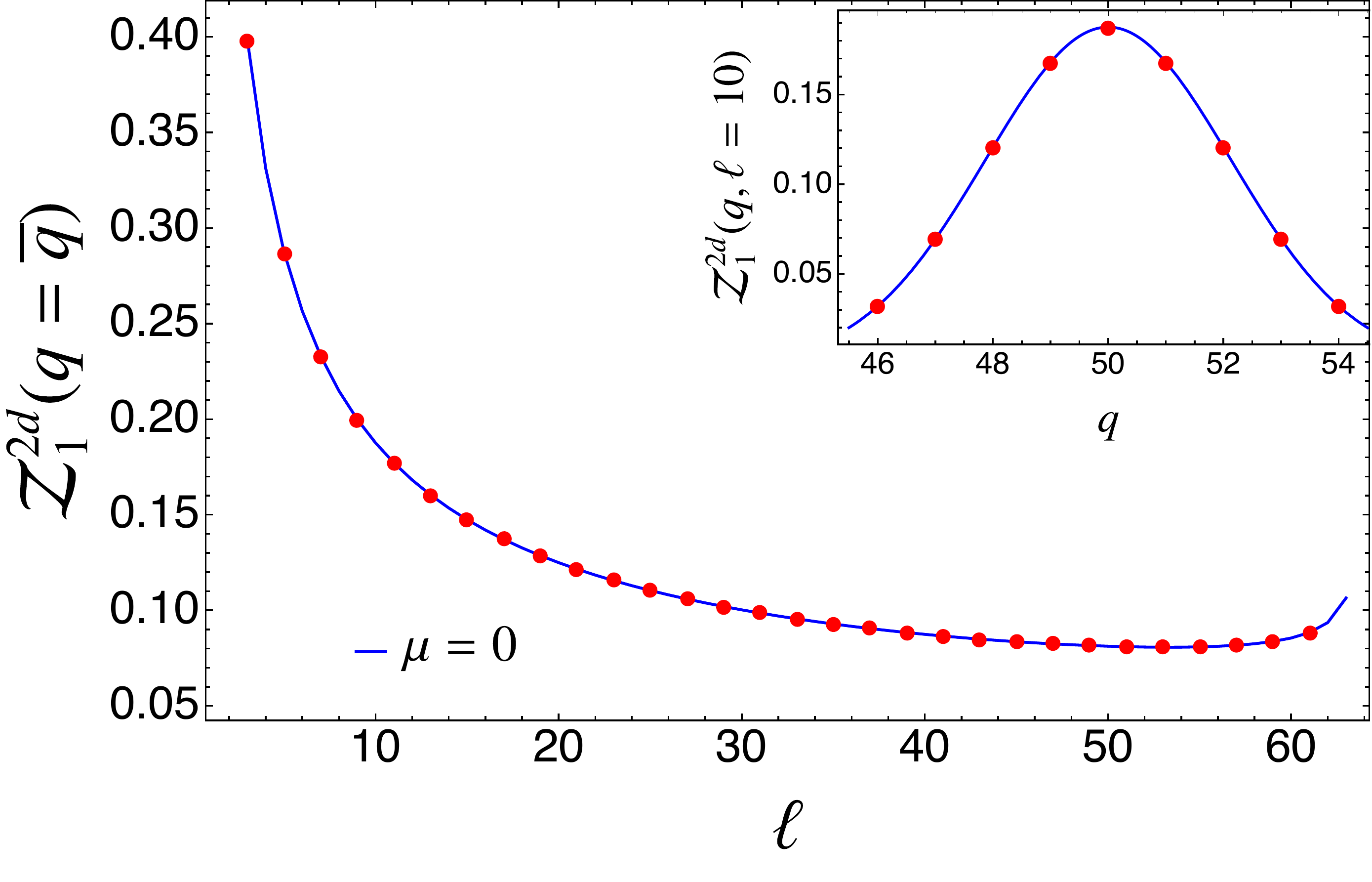}}
\subfigure
{\includegraphics[width=0.44\textwidth]{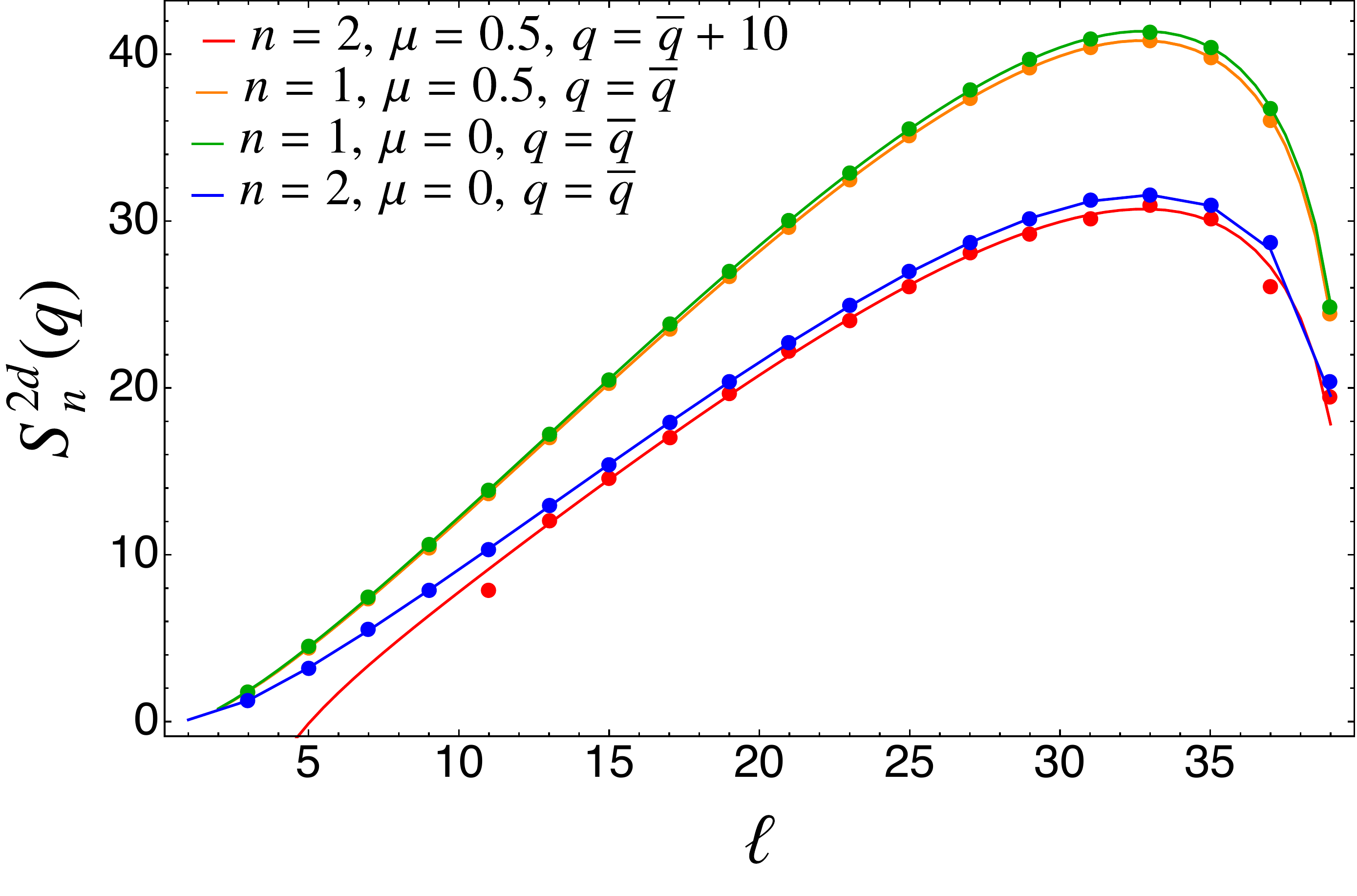}}
\caption{ {\bf Top panels}: Symmetry resolved R\'enyi entanglement entropies $S_n^{2d} (q)$ of 2d free fermions for $n=1,2,3$ and different values of $\mu$. We fix the transverse direction $N=\ell$, equal to the length of the subsystem in the longitudinal direction.
In the left panels the numerical data (symbols) of 2d free fermions for $q = \bar{q}$ are compared with the theoretical predictions of Eqs. (\ref{eq:simmresolvedapp}). 
In the right panel we show four values of $q$ (namely $q-\bar q=0,1,2,3$).
The data are almost coinciding on this scale, so in the inset we report their difference which is perfectly captured by the theoretical prediction. {\bf Bottom panels}: Left panel: $\mathcal{Z}_1^{2d}(q=\bar{q})$ of 2d free fermions for $\mu=0$ and finite torus of longitudinal length $L=64$. 
The inset shows $\mathcal{Z}_1^{2d}(q)$ for $\ell=10$ as a function of $q$. 
Right panel: Symmetry resolved R\'enyi entanglement entropies $S_n^{2d} (q)$ of 2d free fermions for $n=1,2$ and different $\mu$'s and $q$. 
}
\label{fig:symmresfermionsapp}
\end{figure}

\end{appendices}


\newpage
\newpage

\section*{References}
\begin{thebibliography}{130}


\bibitem{eisert-2010}
J.~Eisert, M.~Cramer, and M.~B.~Plenio, \emph{Area laws for the entanglement entropy}, 
\href{http://dx.doi.org/10.1103/RevModPhys.82.277}{Rev. Mod. Phys. {\bf 82}, 277 (2010).}

\bibitem{bombelli}
L. Bombelli, R. K. Koul, J. Lee, and R. D. Sorkin, \emph{A Quantum Source of Entropy for Black Holes}, 
\href{http://dx.doi.org/10.1103/PhysRevD.34.373}{Phys. Rev.  D {\bf 34}, 373 (1986).}
\bibitem{sredniki}
M. Srednicki, \emph{Entropy and area}, 
\href{http://dx.doi.org/10.1103/PhysRevLett.71.666}{Phys. Rev. Lett. {\bf 71}, 666 (1993).} 

\bibitem{intro1}
L.~Amico, R.~Fazio, A.~Osterloh, and V.~Vedral,
\emph{Entanglement in many-body systems}, 
\href{http://dx.doi.org/10.1103/RevModPhys.80.517}{Rev. Mod. Phys. {\bf 80}, 517 (2008)}.

\bibitem{intro2}
P.~Calabrese, J.~L. Cardy, and B.~Doyon,
\emph{Entanglement entropy in extended quantum systems}, 
\href{http://dx.doi.org/10.1088/1751-8121/42/50/500301}{J. Phys. A {\bf 42}, 500301 (2009)}.

\bibitem{intro3}
N.~Laflorencie,
{\it Quantum entanglement in condensed matter systems}, 
\href{http://dx.doi.org/10.1016/j.physrep.2016.06.008}{Phys. Rep. {\bf 643}, 1 (2016)}.


\bibitem{hlw-94}
C. Holzhey, F. Larsen, and F. Wilczek, {\it Geometric and renormalized entropy in conformal field theory}, 
\href{http://dx.doi.org/10.1016/0550-3213(94)90402-2}{Nucl. Phys. B {\bf 424}, 443 (1994)}.

\bibitem{vidal}
G. Vidal, J. I. Latorre, E. Rico, and A. Kitaev, {\it Entanglement in quantum critical phenomena}, 
\href{http://dx.doi.org/10.1103/PhysRevLett.90.227902}{Phys. Rev. Lett. {\bf 90}, 227902 (2003)};\\
J. I. Latorre, E. Rico, and G. Vidal,
{\it Ground state entanglement in quantum spin chains},
Quant. Inf. Comp. {\bf 4}, 048 (2004). 

\bibitem{cc-04}
P. Calabrese and J. Cardy, {\it Entanglement entropy and quantum field theory}, 
\href{http://dx.doi.org/10.1088/1742-5468/2004/06/P06002}{J.  Stat. Mech. P06002 (2004)}.

\bibitem{cc-09}
P. Calabrese and J. Cardy, {\it Entanglement entropy and conformal field theory}, 
\href{http://dx.doi.org/10.1088/1751-8113/42/50/504005}{J. Phys. A {\bf 42}, 504005 (2009)}.



\bibitem{widom2}
D. Gioev and I. Klich,
{\it Entanglement entropy of fermions in any dimension and the Widom conjecture,} 
\href{https://doi.org/10.1103/PhysRevLett.96.100503}{Phys. Rev. Lett. {\bf 96}, 100503 (2006)}

\bibitem{widom3}
M. M. Wolf, {\it Violation of the Entropic Area Law for Fermions},  
\href{https://journals.aps.org/prl/abstract/10.1103/PhysRevLett.96.010404}{Phys. Rev. Lett. {\bf 96}, 010404 (2006)}.

\bibitem{widom4}
W. Li, L. Ding, R. Yu, T. Roscilde, and S. Haas,
{\it Scaling Behavior of Entanglement in Two- and Three-Dimensional Free Fermions,} 
\href{https://doi.org/10.1103/PhysRevB.74.073103}{Phys. Rev. B {\bf 74}, 073103 (2006)}.

\bibitem{mint2}
P.~Calabrese, M.~Mintchev, and E.~Vicari.
\newblock {Entanglement entropies in free fermion gases for arbitrary dimension}, 
\href{https://arxiv.org/abs/1110.6276} {EPL {\bf{97}}, 20009 (2012)}.

\bibitem{widom5}
R. C. Helling, H. Leschke and W. L. Spitzer,
{\it A special case of conjecture by Widom with implications to fermionic entanglement entropy}, 
\href{https://doi.org/10.1093/imrn/rnq085}{Int. Math. Res. Not. {\bf 2011}, 1451 (2011)}.


\bibitem{fz-07}
S. Farkas and Z. Zimboras, {\it The von Neumann entropy asymptotics in multidimensional fermionic systems},
\href{http://dx.doi.org/10.1063/1.2800167}{J. Math. Phys. {\bf 48}, 102110 (2007)}.

\bibitem{s-12} 
B. Swingle,  {\it Renyi entropy, mutual information, and fluctuation properties of Fermi liquids}, 
\href{https://doi.org/10.1103/PhysRevB.86.045109}{Phys. Rev. B {\bf 86}, 045109 (2012)}. 

\bibitem{cmv-12}
P. Calabrese, M. Mintchev and E. Vicari, {\it Exact relations between particle fluctuations and entanglement in Fermi gases},
\href{http://dx.doi.org/10.1209/0295-5075/98/20003}{EPL {\bf 98}, 20003 (2012).}

\bibitem{ryu}
M. T. Tan and S. Ryu,
{\it Particle Number Fluctuations, R\'enyi and Symmetry-resolved Entanglement Entropy in Two-dimensional Fermi Gas from Multi-dimensional Bosonization}, 
\href{https://arxiv.org/abs/1911.01451}{arxiv:1911.01451}.

\bibitem{widom1}
T. Barthel, M.-C. Chung, and U. Schollw\"ock,
{\it Entanglement scaling in critical two-dimensional fermionic and bosonic systems,} 
\href{https://journals.aps.org/pra/abstract/10.1103/PhysRevA.74.022329} {Phys. Rev. A {\bf 74}, 022329 (2006)}.


\bibitem{suggest}
M. C. Chung and I. Peschel,
{\it Density-matrix spectra for two-dimensional quantum systems,} 
\href{https://journals.aps.org/prb/abstract/10.1103/PhysRevB.62.4191}{Phys. Rev. B {\bf 62}, 4191 (2000)}.

\bibitem{areaLawplenio2}
M. B. Plenio, J. Eisert, J. Dreissig, and M. Cramer,
{\it Entropy, entanglement, and area: analytical results for harmonic lattice systems,} 
\href{https://journals.aps.org/prl/abstract/10.1103/PhysRevLett.94.060503}{Phys. Rev. Lett. {\bf 94}, 060503 (2005)}.

\bibitem{areaLawplenio1}
M. Cramer, J. Eisert, M.\,B. Plenio, and J. Dreissig,
{\it An entanglement-area law for general bosonic harmonic lattice systems,} 
\href{https://journals.aps.org/pra/abstract/10.1103/PhysRevA.73.012309} {Phys. Rev. A {\bf 73}, 012309 (2006)}.


\bibitem{areaLawplenio3}
M. Cramer, J. Eisert, and M.B. Plenio,
{\it Statistics dependence of the entanglement entropy,} 
\href{https://journals.aps.org/prl/abstract/10.1103/PhysRevLett.98.220603} {Phys. Rev. Lett. {\bf 98}, 220603 (2007)}

\bibitem{CH2d}
H. Casini and M. Huerta,
{\it Universal terms for the entanglement entropy in 2+1 dimensions}, 
\href{https://doi.org/10.1016/j.nuclphysb.2006.12.012}{Nucl. Phys. B {\bf 764}, 183 (2007)}.

\bibitem{mfs-09}
M. A. Metlitski, C. A. Fuertes, and S. Sachdev, {\it Entanglement Entropy in the O(N) model},
\href{http://dx.doi.org/10.1103/PhysRevB.80.115122}{Phys. Rev. B {\bf 80}, 115122 (2009)}.

\bibitem{s-10}
B. Swingle, {\it Entanglement Entropy and the Fermi Surface},
\href{https://doi.org/10.1103/PhysRevLett.105.050502}{Phys. Rev. Lett. {\bf 105}, 050502 (2010)}.

\bibitem{dsy-12}
W. Ding, A. Seidel, and K. Yang, {\it Entanglement Entropy of Fermi Liquids via Multi-dimensional Bosonization},
\href{https://doi.org/10.1103/PhysRevX.2.011012}{Phys. Rev. X {\bf 2}, 011012 (2012)}.
  
\bibitem{fendley}
H. Ju, A. B. Kallin, P. Fendley, M. B. Hastings, and R. G. Melko, {\it Entanglement scaling in two-dimensional gapless systems},  
\href{https://doi.org/10.1103/PhysRevB.85.165121}{Phys. Rev. B {\bf 85}, 165121 (2012)}.

\bibitem{twist}
X. Chen, W. Witczak-Krempa, T. Faulkner, and E. Fradkin,
{\it Two-cylinder entanglement entropy under a twist,} \href{https://doi.org/10.1088/1742-5468/aa668a}{J. Stat. Mech.  043104 (2017)}.

\bibitem{konik}
A. J. A. James and R. M. Konik, {\it Understanding the entanglement entropy and spectra of 2d quantum systems through arrays of coupled 1d chains}, 
\href{https://doi.org/10.1103/PhysRevB.87.241103}{Phys. Rev. B {\bf 87}, 241103 (R) (2013)}.

\bibitem{kdmsa-13}
F. Kolley, S. Depenbrock, I. P. McCulloch, U. Schollw\"ock, and V. Alba, {\it Entanglement spectroscopy of SU(2)-broken phases in two dimensions}, 
\href{https://doi.org/10.1103/PhysRevB.88.144426}{Phys. Rev. B {\bf 88}, 144426 (2013)}.

\bibitem{sjfm-13}
J.-M. Stephan, H. Ju, P. Fendley, and R. G. Melko, {\it Entanglement in gapless resonating-valence-bond states},
\href{https://doi.org/10.1088/1367-2630/15/1/015004}{New J. Phys. {\bf 15}, 015004 (2013)}.

\bibitem{im-13}
S. Inglis and R. G. Melko, {\it Entanglement at a two-dimensional quantum critical point: a $T=0$ projector quantum Monte Carlo study},
\href{https://doi.org/10.1088/1367-2630/15/7/073048}{New J. Phys. {\bf 15},  073048 (2013)}.

\bibitem{mt-13}
J. McMinis and N. M. Tubman, {\it  Renyi entropy of the interacting Fermi liquid}, 
\href{https://doi.org/10.1103/PhysRevB.87.081108}{Phys. Rev. B {\bf 87}, 081108(R) (2013)}.

\bibitem{dp-15}
J. E. Drut and W. J. Porter, {\it Hybrid Monte Carlo approach to the entanglement entropy of interacting fermions},
\href{https://doi.org/10.1103/PhysRevB.92.125126}{Phys. Rev. B {\bf 92}, 125126 (2015)}.

\bibitem{p-16}
T. Palmai, {\it Entanglement entropy from the truncated conformal space}, \href{https://doi.org/10.1016/j.physletb.2016.06.012}{Phys. Lett. B {\bf 759}, 439 (2016)}.

\bibitem{wws-16}
S. Whitsitt, W. Witczak-Krempa,  and S. Sachdev, {\it Entanglement entropy of the large N Wilson-Fisher conformal field theory}, 
\href{http://dx.doi.org/10.1103/PhysRevB.95.045148}{Phys. Rev. B {\bf 95}, 045148 (2017)}.






\bibitem{lr-14}
N. Laflorencie and S. Rachel, {\it Spin-resolved entanglement spectroscopy of critical spin chains and Luttinger liquids},
\href{http://dx.doi.org/10.1088/1742-5468/2014/11/P11013}{J. Stat. Mech. (2014) P11013}.

\bibitem{goldstein}
M. Goldstein and E. Sela, {\it Symmetry-Resolved Entanglement in Many-Body Systems}, 
\href{https://journals.aps.org/prl/abstract/10.1103/PhysRevLett.120.200602} {Phys. Rev. Lett. {\bf 120}, 200602 (2018)}.

\bibitem{goldstein1}
M. Goldstein and E. Sela, {\it Imbalance Entanglement: Symmetry Decomposition of Negativity}, 
\href{https://journals.aps.org/pra/abstract/10.1103/PhysRevA.98.032302} {Phys. Rev. A {\bf 98}, 032302 (2018)}.

\bibitem{goldstein2}
N. Feldman and M. Goldstein, {\it Dynamics of Charge-Resolved Entanglement after a Local Quench}, 
\href{http://dx.doi.org/10.1103/PhysRevB.100.235146}{Phys. Rev. B {\bf 100}, 235146 (2019)}.

\bibitem{xavier}
J. C. Xavier, F. C. Alcaraz, and G. Sierra, {\it Equipartition of the entanglement entropy}, 
\href{https://journals.aps.org/prb/abstract/10.1103/PhysRevB.98.041106} {Phys. Rev. B {\bf  98}, 041106 (2018)}.

\bibitem{MDC-19-CTM}
S. Murciano, G. Di Giulio, and P. Calabrese, {\it Symmetry resolved entanglement in gapped integrable systems: a corner transfer matrix approach}, 
\href{https://arxiv.org/pdf/1911.09588}{arXiv:1911.09588}.


\bibitem{riccarda}
R. Bonsignori, P. Ruggiero, and P. Calabrese, \textit{Symmetry resolved entanglement in free fermionic systems},
\href{https://doi.org/10.1088/1751-8121/ab4b77}{J. Phys. A  \textbf{52}, 475302 (2019)}.
 
\bibitem{SREE2dG}
S. Fraenkel and M. Goldstein,
{\it Symmetry resolved entanglement: Exact results in 1d and beyond}, 
\href{https://doi.org/10.1088/1742-5468/ab7753}{J. Stat. Mech.  033106 (2020)}.

\bibitem{fis}
A. Lukin, M. Rispoli, R. Schittko, M. E. Tai, A. M. Kaufman, S. Choi, V. Khemani, J. Leonard, and M. Greiner, 
{\it Probing entanglement in a many-body localized system}, \href{https://dx.doi.org/10.1126/science.aau0818}{Science {\bf 364}, 6437 (2019)}.


\bibitem{korepin}
B. Q. Jin and V. E. Korepin,
 {\it Quantum Spin Chain, Toeplitz Determinants and Fisher-Hartwig Conjecture}, 
 \href{http://dx.doi.org/10.1023/B:JOSS.0000037230.37166.42}{J. Stat. Phys. {\bf 116}, 79 (2004)}.
 
 \bibitem{ce-10}
P.~Calabrese and F.~H.~L. Essler,
{\it Universal corrections to scaling for block entanglement in spin-1/2 XX chains}, 
\href{http://dx.doi.org/10.1088/1742-5468/2010/08/P08029}{J. Stat. Mech. P08029 (2010)}.






\bibitem{baxter}
R. J Baxter, {\it Exactly solved models in statistical mechanics.} Academic Press, San
Diego (1982). 

\bibitem{solvable}
I.~{Peschel}, {\it Entanglement in solvable many-particle models}, \href{https://dx.doi.org/10.1007/s13538-012-0074-1}{ Braz. J. Phys. {\bf 42}, 267 (2012)}.
 


 
\bibitem{nishino}
T. Nishino, {\it Density Matrix Renormalization Group Method for 2d Classical Models}, 
\href{http://dx.doi.org/10.1143/JPSJ.64.3598}{  J. Phys. Soc. Jpn. {\bf 74}, 3598 (1995)}.

\bibitem{nishino1}
T. Nishino and K. Okunishi, {\it Density Matrix and Renormalization for Classical Lattice Models}, 
\href{http://dx.doi.org/10.1007/BFb0104638}{Lect. Notes Phys. {\bf 478}, 167 (1997).}

 \bibitem{Gaussian}
I. Peschel and T. T. Truong, {\it Corner Transfer Matrices for the Gaussian Model},
\href{http://dx.doi.org/10.1002/andp.19915030116}{Ann. Physik (Leipzig) {\bf 48}, 185 (1991)}.
 

\bibitem{Peschel}
I.~{Peschel} and M. C. {Chung},  {\it Density Matrices for a Chain of Oscillators}, 
 \href{https://dx.doi.org/10.1088/0305-4470/32/48/305}{J. Phys. A {\bf 32},  8419 (1999)}.
 
 \bibitem{peschel1}
 I. Peschel, M. Kaulke, and O. Legeza, {\it Density-matrix spectra for integrable models},  
 \href{https://dx.doi.org/10.1002/(SICI)1521-3889(199902)8:2<153::AID-ANDP153>3.0.CO;2-N}{Ann. Physik (Leipzig) {\bf 8}, 153 (1999)}.
 
 \bibitem{albaES}
V. Alba, M. Haque, and A. M. L\"auchli, {\it Boundary-Locality and Perturbative Structure of Entanglement Spectra in Gapped Systems}, 
\href{https://link.aps.org/doi/10.1103/PhysRevLett.108.227201}{Phys. Rev. Lett. {\bf 108}, 227201  (2012)}.

\bibitem{ccd-08}
J. L. Cardy,  O. A. Castro-Alvaredo, and B. Doyon,
{\it Form factors of branch-point twist fields in quantum integrable models and entanglement entropy},   
\href{http://dx.doi.org/10.1007/s10955-007-9422-x}{J. Stat. Phys. {\bf 130}, 129 (2008)}. 

\bibitem{cd-09}
O. A. Castro-Alvaredo and B. Doyon, 
{\it Bi-partite entanglement entropy in massive 1+1-dimensional quantum field theories}, 
\href{http://dx.doi.org/10.1088/1751-8113/42/50/504006}{J. Phys. A {\bf 42}, 504006 (2009)}.


\bibitem{nc-10}
M.~A. Nielsen and I.~L. Chuang, \textit{{Quantum computation and quantum  information}}.
 \href{http://dx.doi.org/10.1017/CBO9780511976667}{Cambridge University Press, Cambridge, UK, 10th anniversary~ed. (2010)}.


\bibitem{wv-03}
H. M. Wiseman and J. A. Vaccaro, {\it Entanglement of Indistinguishable Particles Shared between Two Parties}, 
\href{https://dx.doi.org/10.1103/PhysRevLett.91.097902}{ Phys. Rev. Lett. {\bf 91}, 097902 (2003)}.


\bibitem{SREE2d}
H. Barghathi, C. M. Herdman, and A. Del Maestro, {\it R\'enyi generalization of the operational entanglement entropy}, 
\href{http://dx.doi.org/10.1103/PhysRevLett.121.150501}{Phys. Rev. Lett. {\bf 121}, 150501 (2018)}.

\bibitem{delmaestro2}
H. Barghathi, E. Casiano-Diaz, and A. Del Maestro, {\it Operationally accessible entanglement of one-dimensional spinless fermions}, 
\href{https://journals.aps.org/pra/abstract/10.1103/PhysRevA.100.022324}{Phys. Rev. A {\bf 100}, 022324 (2019)}.

\bibitem{kusf-20}
M. Kiefer-Emmanouilidis, R. Unanyan, J. Sirker, and M. Fleischhauer, {\it Bounds on the entanglement entropy by the number entropy in non-interacting fermionic systems},
\href{https://arxiv.org/pdf/2003.03112.pdf}{ArXiv:2003.03112}.

\bibitem{kusf-20b}
M. Kiefer-Emmanouilidis, R. Unanyan, J. Sirker, and M. Fleischhauer, {\it Evidence for unbounded growth of the number entropy in many-body localized phases},
\href{https://arxiv.org/pdf/2003.04849.pdf}{ArXiv:2003.04849}.


\bibitem{crc-20}
L. Capizzi, P. Ruggiero, and P. Calabrese, {\it Symmetry resolved entanglement entropy of excited states in a CFT}, \href{https://arxiv.org/abs/2003.04670}{arXiv:2003.04670}.





\bibitem{ccgm-20}
P. Calabrese, M. Collura, G. Di Giulio, and S. Murciano, {\it Full counting statistics in the gapped XXZ spin chain},
\href{https://iopscience.iop.org/article/10.1209/0295-5075/129/60007}{EPL {\bf 129}, 6 (2020)}.

\bibitem{xhek}
X. Turkeshi, P. Ruggiero, V. Alba, and P. Calabrese,
 {\it Entanglement equipartition in critical random spin chains}, 
\href{https://arxiv.org/abs/2005.03331}{arXiv:2005.03331} (2020).


\bibitem{CFH}
H. Casini, C. D. Fosco, and M. Huerta, 
{\it Entanglement and alpha entropies for a massive Dirac field in two dimensions}, 
\href{https://iopscience.iop.org/article/10.1088/1742-5468/2005/07/P07007}{J. Stat. Mech. P07007 (2005)}.

\bibitem{CFH2}
R. E. Arias, D.D. Blanco, H. Casini, 
{\it Entanglement entropy as a witness of the Aharonov-Bohm effect in QFT}, 
\href{https://doi.org/10.1088/1751-8113/48/14/145401}{J. Phys. A {\bf 48}, 145401 (2015)}.

\bibitem{ch-rev}
H. Casini and M. Huerta, {\it Entanglement entropy in free quantum field theory},
\href{https://doi.org/10.1088/1751-8113/42/50/504007}{J. Phys. A {\bf 42}, 504007 (2009)}.

\bibitem{ssr-17}
H. Shapourian, K. Shiozaki, and S. Ryu, {\it Partial time-reversal transformation and entanglement negativity in fermionic systems}, 
\href{https://doi.org/10.1103/PhysRevB.95.165101}{Phys. Rev. B {\bf 95}, 165101 (2017)}.

\bibitem{neg2}
 H. Shapourian, P. Ruggiero, S. Ryu, and P. Calabrese, {\it Twisted and untwisted negativity spectrum of free fermions}, 
 \href{http://dx.doi.org/10.21468/SciPostPhys.7.3.037}{ 	SciPost Phys. {\bf 7}, 037 (2019)}.
 
\bibitem{clss-19}
E. Cornfeld, L. A. Landau, K. Shtengel, and E. Sela, 
{\it Entanglement spectroscopy of non-Abelian anyons: Reading off quantum dimensions of individual anyons},
\href{http://dx.doi.org/10.1103/PhysRevB.99.115429}{Phys. Rev. B {\bf 99}, 115429 (2019)}.

\bibitem{cms-13}
P. Caputa, G. Mandal, and R. Sinha, {\it Dynamical entanglement entropy with angular momentum and U(1) charge},
\href{https://dx.doi.org/10.1007/JHEP11(2013)052}{JHEP 11 (2013) 052}.

\bibitem{d-16}
J. S. Dowker,  {\it Conformal weights of charged R\'enyi entropy twist operators for free scalar fields in arbitrary dimensions},
\href{https://doi.org/10.1088/1751-8113/49/14/145401}{J. Phys. A {\bf 49}, 145401 (2016)};\\
J. S. Dowker,  {\it Charged R\'enyi entropies for free scalar fields},
\href{https://doi.org/10.1088/1751-8121/aa6178}{J. Phys. A {\bf 50}, 165401 (2017)}.

\bibitem{matsuura}
A. Belin, L.-Y. Hung, A. Maloney, S. Matsuura, R. C. Myers, and T. Sierens, {\it Holographic charged R\'enyi entropies}, 
\href{https://doi.org/10.1007/JHEP12(2013)059}{JHEP {\bf 12} (2013) 059}.



\bibitem{SREE}
P. Caputa, M. Nozaki, and T. Numasawa, {\it Charged Entanglement Entropy of Local Operators}, 
\href{https://dx.doi.org/10.1103/PhysRevD.93.105032}{Phys. Rev. D {\bf 93}, 105032 (2016)}.



\bibitem{correlation}
I. Peschel, {\it Calculation of reduced density matrices from correlation functions},
 \href{https://iopscience.iop.org/article/10.1088/0305-4470/36/14/101} {J. Phys. A {\bf 36}, L205 (2003)}.
 
\bibitem{pe-09} 
I. Peschel and V. Eisler, {\it Reduced density matrices and entanglement entropy in free lattice models},  
\href{https://iopscience.iop.org/article/10.1088/1751-8113/42/50/504003} {J. Phys. A {\bf 42}, 504003 (2009)}.

\bibitem{sublead}
J.~Cardy and P.~Calabrese,
 {\it Unusual Corrections to the Scaling in Entanglement Entropy}, \href{http://dx.doi.org/10.1088/1742-5468/2010/04/P04023}{J. Stat. Mech. (2010) P04023}.

\bibitem{pc2010}
P. Calabrese, M. Campostrini, F. Essler and B. Nienhuis,
{\it Parity effects in the scaling of block entanglement in gapless spin chains}
\href{http://dx.doi.org/10.1103/PhysRevLett.104.095701}{Phys. Rev. Lett. {\bf 104}, 095701 (2010)}.

\bibitem{CMV-11}
P. Calabrese, M. Mintchev, and E. Vicari, {\it The entanglement entropy of 1d systems in continuous and homogenous space}, 
\href{https://dx.doi.org/10.1088/1742-5468/2011/09/P09028} {J. Stat. Mech. P09028 (2011)}.


\bibitem{CH}
H. Casini and M. Huerta, {\it Entanglement and alpha entropies for a massive scalar field in two dimensions}. 
\href{https://doi.org/10.1088/1742-5468/2005/12/P12012}{J. Stat. Mech. P12012 (2005)}.

\bibitem{honeycomb}
W. L. You,
{\it The scaling of entanglement entropy in a honeycomb lattice on a torus, }\href{https://doi.org/10.1088/1751-8113/47/25/255301}
{J. Phys. A. {\bf 47}, 255301 (2014)}

\bibitem{holevo1}
A. Holevo, M. Sohma,  and O. Hirota, {\it Capacity of quantum Gaussian channels},
\href{https://doi.org/10.1103/PhysRevA.59.1820}{Phys. Rev. A {\bf 59}, 1820 (1999)}.

\bibitem{holevo2}
 A. Holevo, {\it Probabilistic and Statistical Aspects of Quantum Theory}, 
 \href{https://link.springer.com/book/10.1007/978-88-7642-378-9}{Publications of the Scuola Normale Superiore (2011)}.




\end{thebibliography}
\end{document}